\documentclass[aps,prd,floatfix,nofootinbib,superscriptaddress,preprint,11pt]{revtex4-1}
\usepackage{amssymb}
\usepackage{amsmath}
\usepackage{amsfonts}
\usepackage{graphicx}
\usepackage{color}
\usepackage{xspace}
\usepackage{ulem}
\usepackage{amsmath,amssymb,amsfonts,latexsym,cancel}
\usepackage{mathrsfs}
\usepackage{xcolor}
\usepackage{cancel}
\usepackage{anysize}
\usepackage{wrapfig}
\usepackage{slashed}
\usepackage{lipsum}
\usepackage{multirow}
\usepackage{gensymb}

\def\ket#1{\left\vert #1\right\rangle}
\def\bra#1{\left\langle #1\right\vert}

\newcommand\ptwiddle[1]{\mathord{\mathop{#1}\limits^{\scriptscriptstyle(\sim)}}}

\allowdisplaybreaks

\newcommand{\AddrUCL}{
Department of Physics and Astronomy, University College London,\\
London WC1E 6BT, United Kingdom
}

\newcommand{\AddrTUM}{
Physik Department T70, Technische Universit\"at M\"unchen,\\ James-Franck-Stra{\ss}e, D-85748 Garching, Germany
}

\begin{document}

\title{Probing New Physics with Long-Range Neutrino Interactions:\\ An Effective Field Theory Approach}

\author{Patrick D. Bolton}
\email{patrick.bolton.17@ucl.ac.uk}\affiliation{\AddrUCL}
\author{Frank F. Deppisch}
\email{f.deppisch@ucl.ac.uk}\affiliation{\AddrUCL}
\author{Chandan Hati}
\email{c.hati@tum.de}\affiliation{\AddrTUM}

\begin{abstract}
\noindent
We investigate forces induced by the exchange of two light neutrinos between Standard Model (SM) fermions in the presence of effective operators parametrising physics beyond the SM. We first set up a general framework in which we derive the long-range potential mediated by weakly interacting neutrinos in the SM, retaining both spin-independent and spin-dependent terms. We then derive neutrino-mediated potentials when there are vector, scalar and tensor non-standard interactions present as well as an exotic neutrino magnetic moment. Examining the phenomenology of such long-range potentials in atomic scale laboratory experiments, we derive upper bounds on the Wilson coefficients of the effective operators and compare these to those from processes such as charged lepton flavour violation.
\end{abstract}

\maketitle

\vspace{-0.5em}
\section{Introduction}
\vspace{-0.5em}
\label{sec:intro}
Along with the four known types of interaction in nature -- the electromagnetic, strong, weak and gravitational -- it is possible that there exist additional forces. The exchange of a new spin-zero or spin-1 boson between two fermions may give rise to some exotic new force, e.g., an axion (spin-zero), a dark photon (spin-1) or a light $Z'$ are among the potential candidates for such a scenario which might lead to an exotic fifth force, which are being actively searched for at several experiments~\cite{Ni:1999di, Heckel:2006ww, Baessler:2006vm, Hammond:2007jm, Heckel:2008hw, Vasilakis:2008yn, Serebrov:2009ej, Ignatovich:2009zz, Serebrov:2009pa, Karshenboim:2010cg, Karshenboim:2010cj, Petukhov:2010dn, Karshenboim:2011dx, Hoedl:2011zz, Raffelt:2012sp, Yan:2012wk, Tullney:2013wqa, Chu:2012cf, Bulatowicz:2013hf, Mantry:2014zsa, Hunter:2013gba, Salumbides:2013dua, Leslie:2014mua, Arvanitaki:2014dfa, Stadnik:2014xja, Afach:2014bir, Leefer:2016xfu, Ficek:2016qwp, Ji:2016tmv, Crescini:2017uxs, Dzuba:2017puc, Delaunay:2017dku, Stadnik:2017hpa, Rong:2017wzk, Safronova:2017xyt, Ficek:2018phf, Rong:2018yos, Dzuba:2018anu, Kim:2017yen,Poddar:2019zoe, Babu:2019iml}. Some such forces between elementary particles scale, at large distances, with an inverse power of the distance between the particles; they are often referred to as long-range forces. They may arise from the exchange of a massless mediator between two particles -- the inverse-square Coulomb interaction between charged particles being the most common example, arising due to an exchange of a single photon. The prospects of discovering a new long-range force coupling to ordinary matter is highly intriguing from both the theoretical and experimental points of view. In electroweak (EW) theory, the neutrinos are the lightest particles and can be considered as nearly massless on the scale of atoms. The exchange of a single neutrino (and in general any fermion) will change the angular momentum of the exchanging particles involved and therefore cannot give rise to a force between stable matter. However, the exchange of two neutrinos can keep the quantum numbers of the exchanging particles unchanged, and can potentially lead to a long-range force. Historically, the idea of a long-range force mediated by the exchange of two neutrinos was conceived a long time ago~\cite{Feynman:1963ax, Feinberg:1968zz, Feinberg:1989ps, Hsu:1992tg, Fischbach:1996qf}. The first explicit computation of the two-neutrino exchange force using the four-Fermi approximation was performed at the leading order in Ref.~\cite{Feinberg:1968zz} to obtain a potential of the form $V(r) = G_F^2 /(4\pi^3 r^5)$, where $G_F$ is the Fermi constant. The neutral-current effects were included in Ref.~\cite{Feinberg:1989ps} and the velocity dependence up to first order was included in Ref.~\cite{Hsu:1992tg}. However, in all these calculations neutrinos were assumed to be massless and of a single flavour.

An interesting series of discussions in the literature resulted also from Ref.~\cite{Fischbach:1996qf}, where it was reported that if neutrinos were massless, the two-neutrino exchange force between neutrons can lead to a large self-energy in a neutron star system through many-body interactions, which far exceeds the order of magnitude of the rest mass of the system itself. It was noted that such a situation can be avoided if the neutrinos are massive, shortening the range of the relevant interaction. In Ref.~\cite{Smirnov:1996vj}, however, it was argued that the creation and subsequent capture of low-energy neutrinos in the star will fill a degenerate Fermi neutrino sea that can block the free propagation of the neutrinos that are responsible for the neutrino force. In Ref.~\cite{Abada:1996nx}, a calculation of the self-energy using a different technique was reported to obtain a negligible contribution. In Ref.~\cite{Fischbach:1996ep} it was stated that the two-neutrino exchange force can be repulsive leading to repulsion among neutrinos instead of filling up the Fermi sea in the neutron star. In Ref.~\cite{Kiers:1997ty}, on the other hand, it was argued that the neutrino self-energy does not threaten the stability of the neutron star, to which Ref.~\cite{Fischbach:2018rfp} differs\footnote{This issue is not the subject of investigation of this work and is only included to provide a comprehensive historical overview.}.

In Ref.~\cite{Grifols:1996fk} the two-neutrino exchange potential was calculated for massive Majorana neutrinos, which was further improved in Ref.~\cite{Lusignoli:2010gw} to include the effects of flavour mixing for the spin-independent part of the force. In a recent work~\cite{Thien:2019ayp}, one can also find a detailed inclusion of flavour mixing effects for the spin-independent part of the neutrino exchange force. In Ref.~\cite{Asaka:2018qfg}, the second order EW effects have been discussed which is usually ignored in the effective (Fermi theory) approach and the relevant EW second-order shifts are calculated for muonium energy levels. While the first-order EW contribution to the hyperfine splitting in $1S$ muonium is found to be of order 65~Hz, the second-order corrections are found suppressed by two orders of magnitude, therefore making any new physics corrections of the first-order EW contribution more relevant than higher-order effects in the EW theory.

Recently, atomic and nuclear systems have attracted substantial attention as probes of SM physics and beyond~\cite{Fichet:2017bng, Stadnik:2017yge, Arcadi:2019uif, Ghosh:2019dmi, Segarra:2020rah}, with excellent improvements in the experimental precision and a promising future prospect for further improvements on experimental measurements~\cite{Strasser:2019fbk}. In particular, in Ref.~\cite{Stadnik:2017yge}, it was explored how the long-range neutrino exchange force can be probed using atomic and nuclear spectroscopy. In Ref.~\cite{Segarra:2020rah}, the possibility of distinguishing between the Dirac versus Majorana nature of neutrinos was discussed in the context of a violation of the weak equivalence principle, while the same has been explored using the Casimir-like force induced by neutrinos between two plates or a point particle and a plate in Ref.~\cite{Costantino:2020bei}. Given the significant amount of progress already made in the literature, it seems desirable to have a robust and systematic analysis of all the possible operator realisations for the long-range neutrino exchange force with equal emphasis on spin-independent and spin-dependent parts. The latter part is particularly relevant given the atomic and nuclear spectroscopy provides sensitivity to both types of long-range neutrino exchange forces, as discussed in Ref.~\cite{Stadnik:2017yge}. Furthermore, a clear distinction and comparison of the Dirac versus Majorana neutrino cases and possible connection of the relevant short-range operators with other relevant observables are also expedient.

In the present work, we consider an effective field theory (EFT) approach to analyse the long-range potential induced by the exchange of two neutrinos in a systematic way, including all the possibilities for the relevant four-fermion contact interactions including the usual SM vector and axial-vector interactions, the scalar and pseudo-scalar interactions and tensor interactions. The effects of flavour mixing are kept completely general and the possibility of having a right-handed current for the neutrinos are also considered in view of many SM extensions pointing towards such a possibility, see e.g., Refs.~\cite{Pati:1973rp, Pati:1973uk, Pati:1974yy, Mohapatra:1974gc, Senjanovic:1975rk,Davidson:1978pm, Mohapatra:1980qe, Keung:1983uu, Dhuria:2015wwa, Deppisch:2017vne, Hati:2017aez, Cepedello:2018zvr, Bolton:2019bou, Hati:2018tge}. We present both spin-independent and spin-dependent results for the long-range potential induced by the exchange of two neutrinos and also analyse the effects for a considerable neutrino magnetic moment. We also discuss the possibility of probing spin-independent and spin-dependent components of the long-range potential using state-of-the-art atomic and nuclear spectroscopy experiments. In particular, the muonium atom currently provides the most precise probe providing access to physics at the scale of tens of GeV and is sensitive to the spin-dependent components of the long-range potential, which has prospects of further improvement at J-PARC Muon Science Facility (MUSE) with new high-intensity muon beam~\cite{Strasser:2019fbk}. In view of the relevant effective operators also inducing charged lepton flavour violating (cLFV) observables, subject to very tight constraints from ongoing and upcoming experiments, we also compare the relevant constraints and comment on their possible complementarity in view of an EFT approach. We also comment on other particle physics probes of these operators, e.g. electron-$\nu$ and nucleon-$\nu$ scattering, beta decays and $ee\to\nu\nu\gamma$ at LEP.

This paper is organised as follows. In Sec.~\ref{sec:model} we introduce the low-energy EFT formalism and its connection to the SM gauge-invariant EFT. We also discuss the current bounds on the relevant Wilson coefficients from probes such as cLFV processes,  electron-neutrino scattering, nucleon-neutrino scattering, beta decays and LEP data. In Sec.~\ref{sec:derivepotential} we outline the derivation of a potential associated with the exchange of a virtual particle between two fermions, taking into account their spins. In Sec.~\ref{sec:effectivepotential} we derive potentials induced by the exchange of two neutrinos between SM charged- and neutral-current interactions in addition to other non-standard vector, scalar and tensor interactions. We conclude this section by discussing and comparing the potentials in these scenarios. In Sec.~\ref{sec:spec} we discuss the prospect of probing beyond the SM effective operators using atomic spectroscopy measurements. We summarise the current experimental measurements and use them to derive limits on the various Wilson coefficients for the Dirac and Majorana neutrino scenarios. In Sec.~\ref{sec:mag}, we discuss the effects of non-vanishing electromagnetic properties of neutrinos and derive the relevant long-range potentials. We also derive the relevant limits on the neutrino electric and magnetic dipole moments using the currently available experimental data. Finally, in Sec.~\ref{sec:con}, we make our concluding remarks.

\section{Effective General Neutrino Interactions}
\label{sec:model}
\subsection{Low energy EFT}
\label{sec:LEFT}
In order to study the effects of new physics interactions of neutrinos in the context of neutrino-mediated long-range potentials, we first need to specify the relevant effective field theory (EFT) framework. If the non-standard interactions are the result of some new physics at a high energy scale $\Lambda_{\text{NP}}$, the general impact of such interactions is to induce operators containing all possible permutations of SM fields respecting the global  and gauge symmetries present at a lower scale $\mu\ll \Lambda_{\text{NP}}$, where $\Lambda_{\text{NP}} $ is the cut-off scale of validity for the EFT. This can be written as a series of higher dimension ($d\geq 5$) non-renormalisable operators,
\begin{align}
\mathcal{L}_{\mathrm{eff}}=\mathcal{L}_{\mathrm{SM}}+\sum_{i} \sum_{d \geq 5} \frac{C^{(d)}_{i}}{\Lambda_{\text{NP}}^{d-4}} \,\mathcal{O}_{i}^{(d)}\,,
\end{align}
where $\mathcal{L}_{\mathrm{SM}}$ is the SM Lagrangian, $\mathcal{O}_{i}^{(d)}$ are dimension-$d$ combinations of SM fields, $C^{(d)}_{i}$ are the associated dimensionless Wilson coefficients and the index $i$ sums over all invariant combinations of fields. It can be seen that higher-dimension operators are suppressed by the factor $\Lambda_{\text{NP}}^{4-d}$. For $\mathcal{O}(1)$ coefficients $C^{(d)}_{i}$, the scale $\Lambda_{\text{NP}}$ corresponds to the mass of new physics mediators. 

At energies below the EW scale -- relevant for the long-range exchange of two neutrinos -- the SM gauge group is broken and the operators $\mathcal{O}_{i}^{(d)}$ must be invariant under $SU(3)_{c}\times U(1)_{\mathrm{em}}$. This is the so-called low energy effective field theory (LEFT) which has been studied in detail, for example, in Refs.~\cite{Jenkins:2017jig,Jenkins:2017dyc,Dekens:2019ept}. In those works a complete basis of operators up to dimension-six is given along with their associated anomalous dimensions, needed to compute the running of the operators from the scale $\mu$ up to $\Lambda_{\text{NP}}$ via the renormalisation group (RG) equations. Also given are the matching conditions between the LEFT operators and the EFT respecting the SM gauge group (SMEFT) valid at the scale $\Lambda_{\text{NP}}$. A complete basis and set of anomalous dimensions has also been computed in the SMEFT up to dimension-six~\cite{Buchmuller:1985jz,Grzadkowski:2010es,Jenkins:2013zja,Jenkins:2013wua,Alonso:2013hga}. However, in general the operators considered in the LEFT can be lepton number violating and all such LEFT operators with $d\geq 6$ require SMEFT operators with odd dimension higher than six. The only SMEFT operator at dimension-five is the well-known LNV Weinberg operator \cite{Weinberg:1979sa}
\begin{align}
\label{eq:weinberg}
\mathcal{L}^{(5)}_{\mathrm{eff}}=\frac{C^{(5)}_{\rho\sigma}}{\Lambda_{\text{NP}}} (\bar{l^{c}_\rho} \tilde{H}^*)(\tilde{H}^{\dagger} l_\sigma)+\mathrm{h.c.}\,,
\end{align}
where $l_\rho$ ($\rho=e,\mu,\tau$) and $H$ are the leptonic and Higgs $SU(2)_{L}$ doublets, respectively, $l_\rho^{c} = \mathbb{C}\bar{l}_\rho^{T}$ with the charge-conjugation matrix $\mathbb{C}$ and $\tilde{H}=i\sigma_{2}H^{*}$, where $\sigma_2$ is the second Pauli matrix.

Because the LEFT and SMEFT are both constructed out of the SM field content they contain only the left-handed neutrino field $\nu_{L}$, either explicitly for the former and contained in the lepton doublet $l$ for the latter. Technically no assumption is made about the nature of massive neutrinos -- whether they are Dirac fermions and have a right-handed component $\nu_{R}$ or are self-conjugate Majorana fermions, $\nu_{R}=\nu_{L}^c = \mathbb{C}\bar{\nu}_{L}^{T}$. After the EW symmetry breaking the Weinberg operator in Eq.~\eqref{eq:weinberg} generates a Majorana neutrino mass term -- however, if neutrinos are Dirac and lepton number is strictly conserved then the coefficient $C^{(5)}$ must vanish.\footnote{In the Dirac case the $\nu_{R}$ field is neglected in the SMEFT because it would imply a $V+A$ interaction arising, for example, in a left-right symmetric model.}

The non-standard neutrino interactions relevant to long-range neutrino exchange are those that contain a neutral-current (NC) for the neutrinos $(\bar{\nu}\Gamma\nu)$ and the interacting fermions $(\bar{f}\Gamma f)$, where $\Gamma$ is a product of gamma matrices. We include the right-handed component $\nu_{R}$ so that the light neutrinos can be either Dirac or Majorana. The lowest dimension operators containing both neutral neutrino and fermion currents are at dimension-six. In order to compare these with the low-energy Fermi limit of the SM weak interactions, we normalise the Wilson coefficients with respect to the Fermi constant $G_{F}$. There are ten different Lorentz-invariant operators in the resulting effective Lagrangian,
\begin{align}
\begin{aligned}
\label{eq:Leff1}
\mathcal{L}_\mathrm{eff}^{\bar \nu \nu\bar f f} = \frac{4G_F}{\sqrt{2}}
\Big[&
c^{LL}_{\alpha\beta;\rho\sigma} (\bar f_{\alpha L} \gamma_\mu  f_{\beta L})(\bar \nu_{\rho L}
\gamma^\mu \nu_{\sigma L})+c^{RL}_{\alpha\beta;\rho\sigma} (\bar f_{\alpha R} \gamma_\mu  f_{\beta R})(\bar \nu_{\rho L}
\gamma^\mu \nu_{\sigma L}) 
  \\
&+c^{LR}_{\alpha\beta;\rho\sigma} (\bar f_{\alpha L} \gamma_\mu  f_{\beta L})(\bar{\nu}_{\rho R}
\gamma^\mu \nu_{\sigma R})
+c^{RR}_{\alpha\beta;\rho\sigma} (\bar f_{\alpha R} \gamma_\mu  f_{\beta R})(\bar \nu_{\rho R}
\gamma^\mu \nu_{\sigma R})\\
&+g^{LL}_{\alpha\beta;\rho\sigma}(\bar f_{\alpha R} f_{\beta L})(\bar{\nu}_{\rho R}
\nu_{\sigma L})+g^{RL}_{\alpha\beta;\rho\sigma} (\bar f_{\alpha L} f_{\beta R})(\bar{\nu}_{\rho R}
\nu_{\sigma L})  \\
&+g^{LR}_{\alpha\beta;\rho\sigma}(\bar f_{\alpha R} f_{\beta L})(\bar \nu_{\rho L}
\nu_{\sigma R}) +g^{RR}_{\alpha\beta;\rho\sigma} (\bar f_{\alpha L} f_{\beta R})(\bar \nu_{\rho L}
\nu_{\sigma R})\\
&+ h^{LL}_{\alpha\beta;\rho\sigma} (\bar f_{\alpha R}
\sigma^{\mu\nu}f_{\beta L})(\bar{\nu}_{\rho R} \sigma_{\mu\nu}
\nu_{\sigma L}) +  h^{RR}_{\alpha\beta;\rho\sigma} (\bar f_{\alpha L}
\sigma^{\mu\nu}f_{\beta R})(\bar \nu_{\rho L} \sigma_{\mu\nu}
\nu_{\sigma R})
\Big]\,,
\end{aligned}
\end{align}
where $f=(\ell,u,d)$ and the fields are in the flavour eigenstate basis with $\alpha,\beta=e, \mu, \tau$, $\alpha,\beta=u, c, t$, $\alpha,\beta=d, s, b$, respectively. Likewise, $\rho,\sigma$ label the neutrino flavours, $\rho,\sigma=e, \mu, \tau$. SM weak interactions induce the operators with coefficients $c^{LL}$ and $c^{RL}$ in the first line. For charged leptons ($f=\ell$) both the charged-current (CC) and NC weak interactions contribute to $c^{LL}$ (through a Fierz transformation of the CC term), while only the NC interaction contributes to $c^{RL}$. For quarks ($f=u, d$) just the NC interaction contributes to $c^{LL}$ and $c^{RL}$. For low energies relevant to long-range neutrino exchange, however, quarks are contained within non-relativistic nucleons, themselves contained within nuclei. Quark currents can be matched to non-relativistic nucleon currents using heavy baryon chiral EFT as detailed in Refs.~\cite{Jenkins:1990jv,Bishara:2016hek,Altmannshofer:2018xyo} and at the end of this subsection. Finally, all other operators in Eq.~\eqref{eq:Leff1} require the presence of $\nu_{R}$ and must be generated by new physics. It should be noted that the notation in Eq.~\eqref{eq:Leff1} is somewhat similar to the basis often used for the non-standard neutrino interactions, see e.g., Ref.~\cite{Bischer:2019ttk}. We clarify the relevant relations with other commonly used bases in Appendix~\ref{sec:compare}.

If neutrinos are Majorana fermions, thus $\nu_{R}=\nu_L^c$ in the above, then the following additional symmetry relations can be found between the coefficients in Eq.~\eqref{eq:Leff1} under the exchange of $\rho$ and $\sigma$, 
\begin{equation}
\label{eq:Majorana_relations}
\begin{aligned}
\begin{split}
& c^{LL}_{\alpha\beta;\rho\sigma}=-c^{LR}_{\alpha\beta;\sigma\rho}~,\\
& g^{LL}_{\alpha\beta;\rho\sigma}=\phantom{-}g^{LL}_{\alpha\beta;\sigma\rho}~,\\
& g^{LR}_{\alpha\beta;\rho\sigma}=\phantom{-}g^{LR}_{\alpha\beta;\sigma\rho}~, \\
& h^{LL}_{\alpha\beta;\rho\sigma}=-h^{LL}_{\alpha\beta;\sigma\rho}~,\\
\end{split}
~~~
\begin{split}
&c^{RL}_{\alpha\beta;\rho\sigma}=-c^{RR}_{\alpha\beta;\sigma\rho}~,\\
&g^{RL}_{\alpha\beta;\rho\sigma}=\phantom{-}g^{RL}_{\alpha\beta;\sigma\rho}~,\\
&g^{RR}_{\alpha\beta;\rho\sigma}=\phantom{-}g^{RR}_{\alpha\beta;\sigma\rho}~,\\
&h^{RR}_{\alpha\beta;\rho\sigma}=-h^{RR}_{\alpha\beta;\sigma\rho}~,
\end{split}
\end{aligned}
\end{equation}
reducing the number of degrees of freedom by effectively eliminating the operators with coefficients $c^{LR}$ and $c^{RR}$. Note that, in the Majorana case ($\nu_{R}=\nu_L^c$) both $\nu_{L}$ and $\nu_R$ will be lepton number violating (LNV) and will give rise to a number of LNV observables, which are subject to strong constraints from experiments searching for neutrinoless double beta decay, LNV meson decays and LNV collider searches \cite{Ali:2007ec,Abada:2017jjx, Bolton:2019pcu,  Bolton:2019wta, Abada:2019bac, Deppisch:2018eth, Das:2017hmg, Dev:2019rxh, Atre:2005eb, Atre:2009rg, Cai:2017mow, Deppisch:2019ldi, Deppisch:2019kvs, Rodejohann:2019quz, Dolinski:2019nrj, Deppisch:2020mxv, Graf:2018ozy, Chun:2019nwi}.

So far we have kept the coefficients $c^{XY}$, $g^{XY}$ and $h^{XX}$ ($X, Y = L, R$) in the flavour basis of neutrino and fermion fields. The relevant coefficients in the mass eigenstate basis should therefore contain the relevant elements of the Cabibbo–Kobayashi–Maskawa (CKM) and
Pontecorvo–Maki–Nakagawa–Sakata (PMNS) mixing matrices. We will follow the convention that the down-type quark and lepton Yukawa matrices $Y_{d}$ and $Y_{\ell}$ are diagonal and the diagonalisation of the up-type quark and neutrino Yukawa matrices (in the Dirac case) proceed via the bi-unitary transformations
\begin{align}
V\cdot Y'_{u} \cdot\tilde{V}^{\dagger}=Y_{u}~,~~U^{\dagger}\cdot Y'_{\nu}\cdot \tilde{U}=Y_{\nu}\,,
\end{align}
where $V$, $\tilde{V}$, $U$, $\tilde{U}$ rotate the left and right-handed up-type quark and neutrino fields according to
\begin{align}
u'_{\alpha L}=[V^{\dagger}]_{\alpha \beta}u_{\beta L}~,~~u'_{\alpha R}=[\tilde{V}^{\dagger}]_{\alpha \beta}u_{\beta R}~,~~\nu'_{\alpha L}=U_{\alpha i}\nu_{i L}~,~~\nu'_{\alpha R}=\tilde{U}_{\alpha i}\nu_{i R}~,
\end{align}
where for clarity the primed and unprimed fields denote flavour and mass eigenstates respectively -- the neutrino mass eigenstates are also labelled with the index $i$. The matrices $V$ and $U$ then correspond to the CKM and PMNS matrices appearing in the SM charged-current,
\begin{align}
j_{W}^{\mu}=2\bar\nu_{L}U^{\dagger}\gamma^{\mu}\ell_{L}+2\bar u_{L}\gamma^{\mu}V d_{L}~.
\end{align}
The matrices $\tilde{V}$ and $\tilde{U}$ do not appear in any SM interaction -- while $u_{R}$ is present in the SM NC, the form of the current $\bar{u}_{R}\gamma^{\mu}u_{R}$ cancels $\tilde{V}$, provided it is unitary. $\nu_{R}$ is not present at all in the SM and so $\tilde{V}$ and $\tilde{U}$ are usually taken to be unphysical. On the other hand, they will appear for some of the operators in Eq.~\eqref{eq:Leff1} by rotating the fields to the mass basis. 

One can choose to define the coefficients $c^{XY}$, $g^{XY}$ and $h^{XX}$ ($X, Y = L, R$) in the mass basis by absorbing the CKM and PMNS matrix elements into the coefficients in the flavour basis. For example, the Wilson coefficient $c^{LL}$ in the mass eigenstate basis is given by
\begin{align}
\label{eq:LHweaktomass}
c^{LL}_{\alpha\beta;ij} = \sum_{\gamma,\delta}\sum_{\rho,\sigma}c^{LL}_{\gamma\delta;\rho\sigma}(V_{\gamma\alpha} V^*_{\delta \beta})U_{\rho i}^*U_{\sigma j}^{}\, ,
\end{align}
where the $V_{\gamma\alpha} V^*_{\delta \beta}$ factor is only present for $f=u$. On the other hand the Wilson coefficient $c^{RR}$ is written in the mass basis as
\begin{align}
\label{eq:RHweaktomass}
c^{RR}_{\alpha\beta;ij} = \sum_{\gamma,\delta}\sum_{\rho,\sigma}c^{RR}_{\gamma\delta;\rho\sigma}(\tilde{V}_{\gamma\alpha} \tilde{V}^*_{\delta \beta})\tilde{U}_{\rho i}^*\tilde{U}_{\sigma j}^{}\,,
\end{align}
which contains rotation matrices for the right-handed neutrino fields (and possible right-handed up-type quark fields). We immediately see a redundancy in Eq.~\eqref{eq:RHweaktomass} because there is more than one unknown parameter on the right-hand side. The unknown mixing angles and phases in the mixing matrices $\tilde{V}$ and $\tilde{U}$ can instead be absorbed back into the parameters of the matrix $c^{RR}$ in the flavour basis -- this is equivalent to setting $\tilde{V} = \tilde{U} = \mathbb{I}$ from the outset. However, we will see that $\tilde{U}$ may contain information about the presence of additional sterile states in the model.

In the SM, the values of the coefficients $c^{LL}$ and $c^{RL}$ are given in the mass basis as
\begin{align}
\label{eq:SMcLLcRL}
\begin{aligned}
\begin{split}
&c^{LL}_{\ell\ell';ij} = U^*_{\alpha i} U^{}_{\alpha j} +g^{\ell}_L \delta_{ij}~,\\
&c^{LL}_{uu';ij} = g^{u}_L \delta_{ij}~,\\
&c^{LL}_{dd';ij} = g^{d}_L \delta_{ij}~,\\
\end{split}
~
\begin{split}
&c^{RL}_{\ell\ell';ij} = g^{\ell}_R \delta_{ij}\,,\\
&c^{RL}_{uu';ij} = g^{u}_R\delta_{ij}\,,\\
&c^{RL}_{dd';ij} = g^{d}_R\delta_{ij}\,,
\end{split}
\end{aligned}
\end{align}
where $g_{L}^{\ell} = -1/2+s_{W}^2$, $g_{R}^{\ell} = s_{W}^2$, $g_{L}^{u} = 1/2-2s_{W}^2/3$, $g_{R}^{u} = -2s_{W}^2/3$, $g_{L}^{d} = -1/2+s_{W}^2/3$, $g_{R}^{d} = s_{W}^2/3$ for $s_{W}^2=\sin^2\theta_{W}$ and $\theta_{W}$ is the weak mixing angle. In models such as the type-I seesaw that can introduce additional mass eigenstates, therefore making the $3\times 3$ PMNS mixing matrix non-unitary, it is easy to replace $\delta_{ij} \rightarrow C_{ij}$, where
\begin{align}
C_{ij}=\sum_{\alpha}U^*_{\alpha i} U^{}_{\alpha j}\, .
\end{align}
Here $U^{}_{\alpha j}$ corresponds to the generalised PMNS mixing matrix.

To go from effective coefficients at the quark level (e.g. $c^{LL}_{uu;ij}$ and $c^{LL}_{dd;ij}$) to the level of non-relativistic nucleons one must make use of the heavy baryon chiral EFT -- viable when the relevant momentum exchange is below the cut-off scale of the EFT, $\Lambda_{\mathrm{ChEFT}}\sim\mathcal{O}(1~\mathrm{GeV})$. At leading order in the EFT, the light pseudoscalar masses are of order $m_{\pi}\sim\mathcal{O}(q)$ and the neutrinos in the loop only interact with a single nucleon. Interactions of neutrinos with more than one nucleon (for example in deuterium) are suppressed by powers of $q/\Lambda_{\mathrm{ChEFT}}$. 

Following the approach of Ref. \cite{Altmannshofer:2018xyo}, the coefficients for effective operators containing nucleon currents can be written in terms of the quark-level coefficients as
\begin{align}
\begin{aligned}
\label{eq:nucleoncoefficients}
c^{LL}_{\mathcal{N}\mathcal{N};ij}&=\frac{1}{2}\sum\limits_{q}\bigg\{F_{1}^{q/\mathcal{N}}(q^2)\big(c^{LL}_{qq;ij}+c^{RL}_{qq;ij}\big)+F_{A}^{q/\mathcal{N}}(q^2)\big(c^{LL}_{qq;ij}-c^{RL}_{qq;ij}\big)\bigg\}\,,\\
c^{RL}_{\mathcal{N}\mathcal{N};ij}&=\frac{1}{2}\sum\limits_{q}\bigg\{F_{1}^{q/\mathcal{N}}(q^2)\big(c^{LL}_{qq;ij}+c^{RL}_{qq;ij}\big)-F_{A}^{q/\mathcal{N}}(q^2)\big(c^{LL}_{qq;ij}-c^{RL}_{qq;ij}\big)\bigg\}\,,
\end{aligned}
\end{align}
where the sum is over $q=u,d,s$ and $F_{1}^{q/\mathcal{N}}(q^2)$ and $F_{A}^{q/\mathcal{N}}(q^2)$ are the NC vector and axial vector form factors for the quark $q$ within the nucleon or nucleus $\mathcal{N}$, respectively. For the proton the following linear combinations at zero-momentum exchange are given in the SM by
\begin{align}
\begin{aligned}
c^{LL}_{pp;ij}+c^{RL}_{pp;ij}\equiv g^{p}_{V}\delta_{ij}\,~~
c^{LL}_{pp;ij}-c^{RL}_{pp;ij}\equiv g^{p}_{A}\delta_{ij}\,,
\end{aligned}
\end{align}
where $g^{p}_{V}\approx(2g_{V}^u+g_{V}^d)=(1/2-2s_{W}^2)$ and $g^{p}_{A}\approx(2g_{A}^u+g_{A}^d)g_{A}=g_A/2$. Here we have used $F_{1}^{u/p}(0)=2$, $F_{1}^{d/p}(0)=1$, $F_{A}^{u/p}(0)=2g_{A}$ and $F_{A}^{d/p}(0)=g_{A}$ and neglected the small contribution from non-valence quarks. Likewise the SM values for the deuteron are
\begin{align}
\begin{aligned}
c^{LL}_{DD;ij}+c^{RL}_{DD;ij}\equiv g^{D}_{V}\delta_{ij}\,,~~
c^{LL}_{DD;ij}-c^{RL}_{DD;ij}\equiv g^{D}_{A}\delta_{ij}\,,
\end{aligned}
\end{align}
where $g_{V}^{D}\approx(3g_{V}^u+3g_{V}^d)=-2s_{W}^2$ and $g_{A}^{D}\approx F_{A}^{s/D}(0)g_{A}^d$. We have used that the vector form factors for the valence quarks in the deuteron are $F_{1}^{u/D}(0)=F_{1}^{d/D}(0)=3$. The equivalent axial form factors vanish, $F_{1}^{u/D}(0)=F_{1}^{d/D}(0)=0$, and the main contribution arises from strange quarks. The strange quark contribution computed in the chiral EFT is given by
\begin{align}
F_{A}^{s/D}(0)\approx 2\Delta s\left(1-\frac{g_{A}^2m_{D}m_{\pi}^2}{4\pi f_{\pi}^2(m_{\pi}+2\gamma)}\right)-\frac{8\gamma(\mu-\gamma)^2}{m_{D}\mu^2}\sim -0.09\,,
\end{align}
for $\gamma=\sqrt{m_{D}E_{D}}$ \cite{Kaplan:1998sz}. Here $\Delta s$ is the strange axial moment of the deuteron, $m_{D}$ is the deuteron mass, $E_{D}$ is the deuteron binding energy, $m_{\pi}$ is the neutral pion mass, and $f_{\pi}$ is the pion decay constant. The renormalisation scale $\mu$ is taken to be at the neutral pion mass $m_{\pi}$.

\subsection{Inclusion of sterile neutrinos}
\label{sec:SMNEFT}
We now briefly return to the question of matching the LEFT + $\nu_{R}$  ($\nu$LEFT) operators with a SM gauge-invariant EFT. As stated before, the commonly-studied SMEFT does not contain $\nu_{R}$ and so can only produce a subset of the $\nu$LEFT operators containing just $\nu_{L}$. It may not even be possible to match these operators if there are new particles with masses $\lesssim\mathcal{O}(100)$ GeV or if EW symmetry breaking is non-linear in the new physics sector~\cite{Falkowski:2019xoe}. Assuming that the matching is possible, in order to produce all low-energy operators, one can introduce $n$ number of sterile states $N_{R}$ to the SMEFT. To this end, a complete basis of lepton number and baryon number conserving and violating operators has been considered in the literature at dimension-five~\cite{Aparici:2009fh}, dimension-six~\cite{delAguila:2008ir} and dimension-seven~\cite{Bhattacharya:2015vja,Liao:2016qyd}. Refs.~\cite{Bischer:2019ttk,Terol-Calvo:2019vck,Chala:2020vqp} provide the matching conditions between the $\nu$LEFT (in the basis of Appendix \ref{sec:compare}) and the SMEFT + $N_{R}$.

At dimension-four in the SMEFT + $N_{R}$ there are the terms 
\begin{align}
\mathcal{L}_{N_{R}}=i \overline{N_{R}^{\prime}} \slashed \partial N_{R}^{\prime}-\left(\frac{1}{2} \overline{N_{R}^{\prime c}} M N_{R}^{\prime}+\mathrm{h.c.}\right)-\left(\bar{L} Y_{\nu} N_{R}^{\prime} \tilde{H}+\mathrm{h.c.}\right)\,,
\end{align}
where $M_{R}$ is the Majorana mass for the sterile states and we have the usual Yukawa term. At dimension-five, in addition to the Weinberg operator of Eq. \eqref{eq:weinberg} we have~\cite{Aparici:2009fh}
\begin{align}
\label{eq:Leff5}
\mathcal{L}_{\mathrm{eff}}^{(5)}=-\frac{1}{2}\overline{N^{\prime c}_{R}} \zeta \sigma^{\mu \nu} N_{R}^{\prime} B_{\mu \nu}-\frac{1}{\Lambda_{\text{NP}}}\left(H^{\dagger} H\right) \overline{N_{R}^{\prime c}} \xi N_{R}^{\prime}+\mathrm{h.c.}\,,
\end{align}
where the former is an EW coupling $\zeta $ to the $U(1)_{Y}$ field strength operator $B_{\mu\nu}$ and the latter is a Majorana mass-like coupling $\chi$ to Higgs doublets. After EW symmetry breaking we obtain the following mass terms
\begin{align}
\label{eq:Lmass}
\mathcal{L}_{m}=-\overline{\nu_{L}^{\prime}} M_{D} N_{R}^{\prime}-\frac{1}{2} \overline{\nu_{L}^{\prime c}} M_{L} \nu_{L}^{\prime}-\frac{1}{2} \overline{N_{R}^{\prime c}} M_{R} N_{R}^{\prime}+\mathrm{h.c.}\,,
\end{align}
where $M_{D}=Y_{\nu}v/\sqrt{2}$, $M_{L}= \frac{\chi}{\Lambda_{\text{NP}}} v^2$, $M_{R}= M+\frac{\xi}{\Lambda_{\text{NP}}} v^2$ and $v$ is the SM Higgs VEV. Various limiting cases can now be obtained depending on the matrices $M_{D}$, $M_{L}$ and $M_{R}$. If $M_{D}\ll M_{L}\ll M_{R}$ then we obtain a type-I seesaw-like scenario, whereas if $M_{D}\ll M_{L,R}$ we obtain quasi-Dirac neutrinos such as those studied in Refs.~\cite{Anamiati:2016uxp,Anamiati:2017rxw}. In the former case the mass matrices for the light and heavy neutrinos are approximately
\begin{align}
\begin{aligned}
&\mathcal{M}_{\nu} \approx M_{L}-M_{D}^{*} (M_{R}^{\dagger})^{-1} M_{D}^{\dagger}\,,\\
&\mathcal{M}_{N} \approx M_{R}\,,\\
\end{aligned}
\end{align}
respectively, which are then diagonalised to the mass basis via $M_{\nu}=U_{\nu}^{T}\cdot \mathcal{M}_{\nu}\cdot U_{\nu}$ and $M_{N}=U_{N}^{T}\cdot \mathcal{M}_{N}\cdot U_{N}$. In the mass basis we therefore obtain three light Majorana neutrinos $\nu = \nu^c$ and $n$ relatively heavier Majorana neutrinos $N = N^c$. The weak eigenstates are given in terms of the mass eigenstates by 
\begin{align}
\label{eq:seesawrotation}
\begin{aligned}
\nu_{L}^{\prime}&=P_{L}\left(U_{\nu} \nu+\varepsilon U_{N} N+...\right) \equiv P_{L} U n\,,\\
N_{R}^{\prime c}&=P_{L}\left(-\varepsilon^{\dagger} U^{*}_{\nu} \nu+U^{*}_{N} N+...\right)\equiv P_{L}\tilde{U}n\,,
\end{aligned}
\end{align}
where $n=(\nu_{1},\nu_2,\nu_3,N_{1},N_{2},...)$. As all the light neutrinos are Majorana in this case, the relevant operators with coefficients $c^{LL}$ and $c^{RL}$ (the $c^{LR}$ and $c^{RR}$ operators are equivalent by Eq. \eqref{eq:Majorana_relations}) are naively rotated by the $3\times (3+n)$ matrix $U$. However, if any of the $N$ are above the EW scale then they are integrated out of the EFT, leaving only those below. In the presence of light sterile states one can have terms such as
\begin{align}
\label{eq:GFNN}
\frac{4G_{F}}{\sqrt{2}}\left[\tilde{c}^{LR}_{\alpha\beta;ss'} (\bar f_{\alpha L} \gamma_\mu  f_{\beta L})+\tilde{c}^{RR}_{\alpha\beta;ss'} (\bar f_{\alpha R} \gamma_\mu  f_{\beta R})\right](\overline{N'_{s R}}
\gamma^\mu N'_{s' R})\,,
\end{align}
where $s, s'$ label the weak eigenstates of $N_{R}$, which are rotated to the mass basis by the $n\times (3+n)$ matrix $\tilde{U}$ and $n$ is the number of \textit{light} sterile states $N$.

To summarise, in the SMEFT + $N_{R}$, the fields $N'_{R}$ in the flavour basis are rotated to the mass basis by the matrix $\tilde{U}$. This corresponds to an extended block of the enlarged mixing matrix diagonalising the full neutrino mass matrix. For three light (mostly active) Majorana neutrinos, the operator with coefficient $c^{RR}$ appearing in Eq.~\eqref{eq:Leff1} is written in terms of the $\nu_L$ states, $\nu_R = \mathbb{C}\bar{\nu}_{L}^{T}$, and so can be related to $c^{RL}$ by the symmetry relations of Eq.~\eqref{eq:Majorana_relations}. $c^{RR}$ is thus instead rotated by the block $U$ as in Eq.~\eqref{eq:LHweaktomass}. Therefore Eq.~\eqref{eq:RHweaktomass} is only strictly true for the coefficients $\tilde{c}^{LR}$ and $\tilde{c}^{RR}$ appearing in Eq.~\eqref{eq:GFNN} for additional (mostly sterile) Majorana states. On the other hand, if neutrinos are Dirac and there are right-handed gauge interactions, $\tilde{U}$ corresponds to the right-handed analogue of the PMNS mixing matrix diagonalising the right-handed charged-current interactions~\cite{Cepedello:2018zvr,Bolton:2019bou}. However, we have already discussed how it multiplies similarly unknown Wilson coefficients and can thus be subsumed.

In this work, we are mainly concerned with the two extreme limits discussed below Eq.~\eqref{eq:Lmass}. The first is the scenario in which lepton number is conserved -- $M_{L} = M_{R} = 0$ and the right-handed states $N_{R}\equiv \nu_{R}$ form three light Dirac neutrinos with the $\nu_L$ states. The second is if lepton number is violated and any sterile mass eigenstate fields $N$ are integrated out, leaving three light Majorana neutrinos. However, we will see that light sterile fields $N$ (which may or may not be related to the seesaw mechanism generating light left-handed neutrino masses) can be of relevance in Sec.~\ref{sec:mag}.

\subsection{Bounds from other probes}
\label{sec:other-probes}
In the SM, the left-handed neutrinos are part of $SU(2)_L$ doublet with the charged leptons as their partners. Therefore, for a given new physics model at a scale higher than the EW symmetry breaking scale, the SM gauge-invariant
operators that mediate the long-range neutrino interactions can also mediate charged lepton flavour changing processes \cite{Deppisch:2012vj} which are constrained stringently from experimental non-observation of various cLFV observables~\cite{Antusch:2008tz, Gavela:2008ra, Davidson:2019iqh}. The cLFV radiative decays and $\mu\rightarrow e$ conversion are particularly relevant in this context as they are subject to intensive searches at various ongoing and upcoming experiments. The decays of tau into light mesons accompanied by a lepton are also relevant since the relevant bounds are expected to be improved significantly in Belle II~\cite{Kou:2018nap}. To illustrate the relevance of cLFV processes let us consider the contact interaction
of Eq.~\eqref{eq:Leff1}
\begin{equation}
\frac{4G_F}{\sqrt{2}}
c^{LL}_{\alpha\beta;\rho \sigma} (\bar f_{\alpha L} \gamma_\mu  f_{\beta L})(\bar \nu_{\rho L}
\gamma^\mu \nu_{\sigma L})\,.
\label{SMEFT1}
  \end{equation}
For $f=e$ this operator can be generated by
the dimension-six SMEFT~\cite{Buchmuller:1985jz,Grzadkowski:2010es,Jenkins:2013zja,Jenkins:2013wua,Alonso:2013hga} operator
\begin{equation}
\frac{4G_F}{\sqrt{2}}\varepsilon_e^{\rho\sigma}
(\overline{l}_e\gamma_\mu l_e)(\overline{l}_\rho\gamma^\mu l_\sigma)\,,
\label{SMEFT2}
  \end{equation}
where $\bar{l}$ is the SU(2) doublet $(\overline{\nu_L}, \overline{\ell_L})$. It is easy to see that such an operator can also induce the cLFV interaction
$(\overline{e }\gamma_\mu P_L e)(\overline{\ell}_\rho\gamma^\mu P_L  \ell_\sigma)$ (where $\rho\neq \sigma$) with the relevant Wilson coefficient constrained by the experimental limits on cLFV
decays, e.g., $\ell_\sigma \to \ell_\rho e^{-}e^{+}$.  However, such a scenario
can be avoided by instead constructing such operators at higher dimension, e.g., dimension-eight operator in SMEFT
\begin{equation}
\frac{
C_{1f,\rho\sigma}^{(8)}}{\Lambda_{\text{NP}}
^4}
(\overline{l}^p_\rho \epsilon_{pQ}H^{Q*}) \gamma^\mu (H^R \epsilon_{Rs} l^s_\sigma)(
\overline{f}\gamma_\mu f )
\label{SMEFT3}
  \end{equation}
where $\epsilon_{pQ}$ is the usual antisymmetric SU(2) contraction and $\Lambda_{\text{NP}}$ is the heavy new physics scale. When the neutral component of Higgs $H= (H^+,\,H_0)$ acquires a vacuum expectation
value $\langle H_0 \rangle = v$,  the dimension-eight operator leads to
the contact interaction in Eq.~\eqref{SMEFT1}.

Assuming a new physics model valid at a scale $\Lambda_{\text{NP}} > m_W$, which at tree level generates the relevant long-range interactions but avoids inducing cLFV at the tree level, it is important to note that such long-range interactions can still induce cLFV operators through Higgs or $W$ loops. The vector operators of dimension-six and eight which can be added to the SM  Lagrangian are of the form
\begin{equation}
\delta {\cal L} =  \sum_{X,\zeta} \frac{C^{(6)}_{X,\zeta}}{\Lambda_{\text{NP}}^{2}}\,{\cal O}^{(6)}_{X,\zeta} + \sum_{X',\zeta} \frac{C^{(8)}_{X',\zeta}}{\Lambda_{\text{NP}}^{4}}\,{\cal O}^{(8)}_{X',\zeta} +\mathrm{h.c.}\,,
\label{SMEFT4}
 \end{equation}
where $X$ ($X'$) labels operators with the Lorentz structure $\gamma^\mu
\times \gamma_\mu$ and  $\zeta$ denotes the flavour indices. A complete list of all the relevant dimension-six operators in the `Warsaw' basis can be found in~\cite{Grzadkowski:2010es}, while the relevant dimension-eight operators when the external fermion is a $SU(2)_L$ doublet can be found, for example, in~\cite{Berezhiani:2001rs}. A particularly interesting basis has also been proposed recently in~\cite{Davidson:2019iqh}; following this basis in the case where the external fermion is a $SU(2)_L$ doublet quark or lepton ($f=q, l_\alpha$ with $\alpha\neq \rho,\sigma$), the relevant dimension-six SMEFT operators are
\begin{eqnarray}
{\cal O}^{(6)}_{1f,\rho\sigma}\equiv
(\overline{l}_\rho \gamma^\mu   l_\sigma) (\overline{f}  \gamma_\mu  f)
\,&,&~~
{\cal O}^{(6)}_{2f,\rho\sigma} \equiv
(\overline{l}_\rho \gamma^\mu   f) (\overline{f}  \gamma_\mu  l_\sigma)\, ,
\label{SMEFT5}
\end{eqnarray}
where the $SU(2)_L$ contractions are understood to be inside the parentheses.
At dimension-eight, the relevant SMEFT operators are given by
\begin{align}
&{\cal O}^{(8)}_{1f,\rho\sigma} \equiv (\overline{l}_\rho \epsilon H^{*}) \gamma^\mu
(H \epsilon l_\sigma)(\overline{f}  \gamma_\mu  f) \,,~~
{\cal O}^{(8)}_{2f,\rho\sigma}  \equiv (\overline{l}_\rho H) \gamma^\mu  (H^\dagger  l_\sigma)
(\overline{f}  \gamma_\mu  f)\, , \nonumber\\
&{\cal O}^{(8)}_{3f,\rho\sigma} \equiv
(\overline{l}_\rho \gamma^\mu   f) (\overline{f} H) \gamma_\mu  (H^\dagger l_\sigma)
\,,~~~~\,
{\cal O}^{(8)\dagger}_{4f,\rho\sigma} \equiv
(\overline{l}_\rho H) \gamma^\mu (H^\dagger  f) (\overline{f} \gamma_\mu   l_\sigma)\, ,
\label{SMEFT6}\\
&{\cal O}^{(8)}_{5f+,\rho\sigma} \equiv ({\cal O}^{(8)}_{5f,\rho\sigma} +
{\cal O}^{(8)\dagger}_{5f,\rho\sigma})
\equiv
(\overline{l}_\rho \gamma^\mu   f) (\overline{f} \epsilon H^{*} ) \gamma_\mu
(H \epsilon l_\sigma)
+
(\overline{l}_\rho \epsilon H^{*})  \gamma^\mu  (H \epsilon  f)  (\overline{f}  \gamma_\mu
l_\sigma)\,.
\nonumber
\end{align}
In the case where the external fermion $f = l_\alpha$ with $\alpha = \rho$, the relevant operators reduce to
\begin{align}
\begin{aligned}
{\cal O}^{(6)}_{1l,\rho\sigma} &\equiv
(\overline{l}_\rho \gamma^\mu  l_\sigma) (\overline{l}_\rho  \gamma_\mu  l_\rho)\, ,\\
{\cal O}^{(8)}_{1l,\rho\sigma}&\equiv (\overline{l}_\rho \epsilon H^{*}) \gamma^\mu
(H \epsilon l_\sigma)(\overline{l}_\rho  \gamma_\mu  l_\rho)\, ,\\
{\cal O}^{(8)}_{2l,\rho\sigma} &\equiv (\overline{l}_\rho H) \gamma^\mu  (H^\dagger  l_\sigma)
(\overline{l}_\rho  \gamma_\mu  l_\rho)\, ,\\
{\cal O}^{(8)}_{5l+,\rho\sigma} 
 &\equiv
(\overline{l}_\rho \gamma^\mu   l_\rho) (\overline{l_\rho} \epsilon H^*) \gamma_\mu  (H \epsilon l_\sigma)+ (\overline{l_\rho} \epsilon H^*) \gamma_\mu
(H \epsilon l_\rho)(\overline{l}_\rho \gamma^\mu   l_\sigma)\, .
\label{SMEFT7}
\end{aligned}
\end{align}
In the case where the external fermion $f$ is an $SU(2)_L$ singlet quark or lepton, the relevant dimension-six and eight SMEFT operators are given by
\begin{align}
\begin{aligned}
&{\cal O}^{(6)}_{1sf,\rho\sigma} \equiv
(\overline{l}_\rho  \gamma^\mu   l_\sigma)(\overline{f}  \gamma_\mu  f) \,,\\
&{\cal O}^{(8)}_{1sf,\rho\sigma} \equiv (\overline{l}_\rho \epsilon H^{*}) \gamma^\mu
(H \epsilon l_\sigma)(\overline{f}  \gamma_\mu  f) \,,\\
&{\cal O}^{(8)}_{2sf,\rho\sigma}  \equiv
(\overline{l}_\rho H \gamma^\mu  H^\dagger  l_\sigma)(\overline{f}  \gamma_\mu  f)\,.
\label{SMEFT8}
\end{aligned}
\end{align}

\begin{table}[t!]
\setlength{\tabcolsep}{8pt}
\begin{center}
\begin{tabular}{ccc}
\hline\hline
$\nu$LEFT Wilson coefficient & Relevant cLFV process & Current cLFV sensitivity  \\
\hline
   $c^{LL}_{ee;\mu e}$ 	 &  \quad $\text{BR}(\mu \to 3 e)$             & \quad $7.8\times 10^{-7}$ \\
   $c^{LL}_{dd;\mu e}$ 	 &  \quad $\text{CR}(\mu- e, \text{Au})$       & \quad $5.3\times 10^{-8}$ \\
   $c^{LL}_{uu;\mu e}$ 	 &  \quad $\text{CR}(\mu- e, \text{Au})$       & \quad $6.0\times 10^{-8}$ \\
   $c^{RL}_{ee;\mu e}$ 	 &  \quad $\text{BR}(\mu \to 3 e)$             & \quad $9.3\times 10^{-7}$ \\
   $c^{RL}_{dd;\mu e}$ 	 &  \quad $\text{CR}(\mu- e, \text{Au})$       & \quad $5.4\times 10^{-8}$ \\
   $c^{RL}_{uu;\mu e}$ 	 &  \quad $\text{CR}(\mu- e, \text{Au})$       & \quad $6.3\times 10^{-8}$ \\
\hline

   $c^{LL}_{ee;\tau e}$   & \quad $\text{BR}(\tau \to 3 e)$	           & \quad $2.8\times 10^{-4}$ \\
   $c^{LL}_{ee;\tau \mu}$ & \quad $\text{BR}(\tau \to \mu e \bar{e})$  & \quad $3.2\times 10^{-4}$ \\
   $c^{RL}_{ee;\tau e}$   & \quad $\text{BR}(\tau \to 3 e)$	           & \quad $4.0\times 10^{-4}$ \\
   $c^{RL}_{ee;\tau \mu}$ & \quad $\text{BR}(\tau \to \mu e \bar{e})$  & \quad $3.2\times 10^{-4}$ \\
\hline
    $\left|c^{LL(RL)}_{dd(uu);\tau e}\right|$ 	 &\quad $\text{BR}(\tau \to e \rho;e \eta)$       & \quad $7.1\times 10^{-4}$ \\
    $\left|c^{LL(RL)}_{dd(uu);\tau \mu}\right|$  &\quad $\text{BR}(\tau \to \mu \rho; \mu\eta)$   & \quad $5.9\times 10^{-4}$ \\
\hline\hline
\end{tabular}
\caption{
Experimental sensitivities of various relevant $\nu$LEFT Wilson coefficients in Eq. \eqref{eq:Leff1}~\cite{Davidson:2019iqh} based on the current best limits from various cLFV experiments. To derive the bottom two constraints it has been assumed that LFV is induced either on left-handed or right-handed quarks, but not both simultaneously.
}
\label{tab:WETlim}
\end{center}
\end{table}
After the EW symmetry breaking, at the $Z$-pole ($\mu=m_Z$) the SM gauge group invariant operators are matched onto the $\nu$LEFT operators given in Eq. \eqref{eq:Leff1}. However, it is important to include the RG running induced mixings via $W$ and Higgs loops exchange between various SMEFT operators discussed above. This is particularly relevant because even if the relevant interactions in a given new physics model may not induce cLFV at the tree level, such operators get induced via the mixings at the one-loop level. A detailed discussion of the relevant matching and mixing effects is beyond the scope of the current work and can be found for example in~\cite{Davidson:2019iqh}. In Table \ref{tab:WETlim}, we summarise the phenomenological limits on the relevant $\nu$LEFT Wilson coefficients that can be derived from the negative search limits from various experimental cLFV searches~\cite{Davidson:2019iqh}. From an EFT point of view, it is important to notice that the relevant limits on the Wilson coefficients can be derived under varying assumptions about the cancellation among the SMEFT operators of different dimensions and with different powers of $\log (\Lambda_{\text{NP}}/m_Z)$~\cite{Antusch:2008tz, Gavela:2008ra, Davidson:2019iqh} and therefore such limits are to be interpreted with more care under the given assumptions. On the other hand, for a given new physics model one can numerically check for any such possible cancellation and the constraints can be interpreted unambiguously. One important point to note regarding the existing limits such as from cLFV processes here is that the corresponding processes occur at energy scales of the decaying muon or tau mass. Therefore, the analysis in the $\nu$LEFT framework is valid at those energy scales and the Wilson coefficients are sensitive to new physics scales heavier than these mass scales. On the other hand for two neutrino exchange, the scale of the process corresponds to the Bohr radius scale $a_{0}^{-1} = \alpha m_e \approx {\mathcal{O}}(10)$~keV in atomic systems and as small as the neutrino mass ${\mathcal{O}}(\mathrm{eV})$ for macroscopic-scale forces. The $\nu$LEFT framework, in this case, is therefore generally applicable for much lighter new physics scales. This opens up the possibility of exploring a lot of interesting light new physics scenarios with non-trivial couplings to neutrinos and other SM fermions.

Other than the cLFV processes discussed above, the Wilson coefficients relevant to first- and second-generation leptons are also subjected to direct bounds from the experimental data on various scattering processes such as $\nu_\mu e$ scattering in CHARM-II~\cite{Vilain:1993kd,Vilain:1994qy} (which is supposed to be improved by an order of magnitude at the DUNE near detector~\cite{Bischer:2018zcz}), neutrino-nucleon scattering data at CHARM and CDHS ~\cite{Vilain:1993kd,Vilain:1994qy,Farzan:2017xzy}. The Wilson coefficients relevant to tau are constrained from $e\bar{e}\rightarrow \nu\bar{\nu} \gamma$ data at LEP~\cite{Barranco:2007ej}. However, these bounds are orders of magnitude weaker as compared to the bounds from cLFV processes. Some relevant discussion about these bounds can be found for example in~\cite{Barranco:2007ej,Biggio:2009nt,Escrihuela:2011cf,Cirigliano:2013xha,Alcaide:2019pnf,Butterworth:2019iff}. In addition, the observation of coherent elastic neutrino-nucleus scattering at COHERENT~\cite{Akimov:2017ade,Freedman:1973yd} and beta decays~\cite{Gonzalez-Alonso:2018omy} are also relevant for deriving bounds on the relevant Wilson coefficients~\cite{Bischer:2019ttk}.

\section{Long-Range Potentials and Scattering Amplitudes}
\label{sec:derivepotential}

\begin{figure}[t!]
	\centering
	\includegraphics[width=0.32\textwidth]{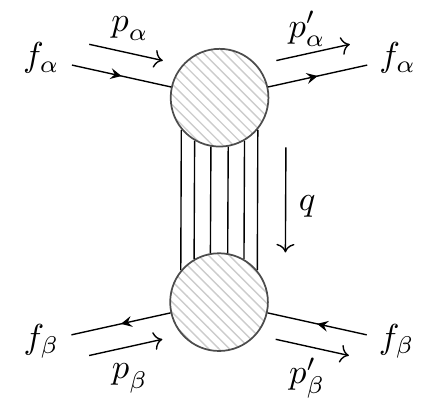}
	\hspace{2em}
	\includegraphics[width=0.32\textwidth]{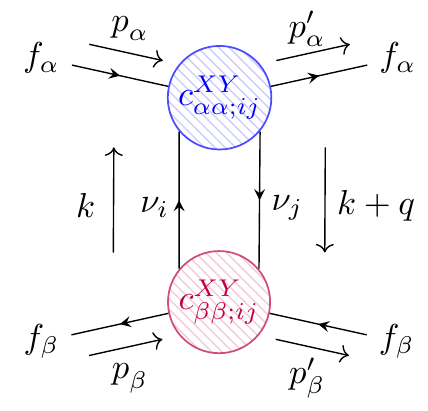}
	\caption{Left: Long-range force mediated between particles $f_{\alpha}$ and $f_{\beta}$ by virtual particles carrying the momentum exchange $q = p^{}_{\alpha}-p_{\alpha}'=p_{\beta}'-p^{}_{\beta}$. Right: Diagram depicting the exchange of two mass-eigenstate neutrinos between fermions $f_{\alpha}$ and $f_{\beta}$. The interaction vertices are four-fermion interactions with coefficients $c^{XY}_{\alpha\alpha;ij}$ and $c^{XY}_{\beta\beta;ij}$ respectively, where the superscripts $X, Y = L, R$ refer to the chirality of the fermion and neutrino currents.}
	\label{fig:effpotential}
\end{figure}

It has long been known that a force acting at a distance can be interpreted as the exchange of a virtual particle (or multiple particles) between external on-shell states. As depicted to the left of Fig.~\ref{fig:effpotential}, a mediator or mediators are necessary to exchange the momentum $q=p^{}_{\alpha}-p'_{\alpha}=p'_{\beta}-p^{}_{\beta}$ between the two interacting particles $f_{\alpha}$ and $f_{\beta}$ with initial momenta $p^{}_{\alpha}$ and $p^{}_{\beta}$ and final momenta $p'_{\alpha}$ and $p'_{\beta}$ respectively. 

In the Feynman diagrammatic approach it is possible to derive a long-range potential $V(\boldsymbol{r}, \boldsymbol{v})$ for an interaction -- most generally a function of the relative displacement between the particles $\boldsymbol{r}$ and the average velocity of the system,
\begin{align}
\label{eq:averagevelocity}
\boldsymbol{v}=\frac{1}{2}\left(\frac{\mathbf{p}_{\alpha}}{m_{\alpha}}+\frac{\mathbf{p}_{\beta}}{m_{\beta}}\right)\,,
\end{align} 
by taking the Fourier transform of the invariant amplitude of the scattering process \cite{sakurai_napolitano_2017}, i.e.,
\begin{align}
\label{eq:fouriertransform}
V(\boldsymbol{r}, \boldsymbol{v})=\int \frac{d^{3} \mathbf{q}}{(2 \pi)^{3}} \,e^{i \mathbf{q} \cdot \boldsymbol{r}} \,\mathcal{M}(s, t)\,,
\end{align}
where the invariant amplitude $\mathcal{M}(s,t)$ is an analytic function of the Mandelstam variables $s= P^2=(p^{}_{\alpha}+p^{}_{\beta})^2=(p_{\alpha}'+p_{\beta}')^2 $ and $t= q^2=(p^{}_{\alpha}-p'_{\alpha})^2=(p'_{\beta}-p^{}_{\beta})^2 $. 
The potential is time independent in the static limit of momentum transfer, $q\approx(0,\mathbf{q})$ and $t\approx -\mathbf{q}^2$, which is an accurate approximation for particles interacting at a distance. Furthermore, one can also exploit the analyticity properties of $\mathcal{M}(s,t)$ which enable the spectral decomposition \cite{Feinberg:1989ps}
\begin{align}
\label{eq:spectraldecomp}
\mathcal{M}(s,-\mathbf{q}^2) = -\int_{0}^{\infty} dt' ~	\frac{\rho^{}(s,t')}{t'+\mathbf{q}^2}\,,
\end{align}
where $\rho(s,t')$ is the so-called ``spectral function" of the process. The spectral function is related to the imaginary part of the discontinuity on the real $t$-axis of $\mathcal{M}(s,t)$,
\begin{align}
\label{eq:spectraldefinition}
\rho^{}(s,t)=\frac{1}{\pi}\,\mathrm{Im}[\mathcal{M}(s,t)]=\frac{1}{2\pi i}\,\mathrm{disc}[\mathcal{M}(s,t)]\,,
\end{align}
where
\begin{align}
\label{eq:discontinuitydef}
\mathrm{disc}[\mathcal{M}(s,t)] = \mathcal{M}(s,t+i\epsilon)- \mathcal{M}(s,t-i\epsilon)\,,
\end{align}
for $\epsilon\rightarrow 0$. One can now insert the decomposition of Eq. \eqref{eq:spectraldecomp} into Eq. \eqref{eq:fouriertransform} and evaluate the angular integral $d\Omega = d\theta \,d\phi \, \sin\theta$ contained in $d^{3}\mathbf{q}$. This integration is non-trivial if $\mathcal{M}(s,t)$ and therefore $\rho(s,t)$ depend on $\theta$ and $\phi$ -- for example if there are spin-dependent terms containing the dot product of $\mathbf{q}$ and a particle spin $\boldsymbol{\sigma}$. In fact, such terms arise naturally when taking the non-relativistic limit of the scattering amplitude. 

We follow the approach of Ref. \cite{Dobrescu:2006au} and divide the spectral function $\rho(t)$ (omitting the dependence on $s$) according to a basis of 16 spin operators,
\begin{align}
\label{eq:kexpansion}
\rho(t) &= \sum_{k=1}^{16}\,\rho_{k}(t)\,\mathcal{O}_{k}(\mathbf{q},\mathbf{P})\,f_{k}(\boldsymbol{v}^2)\,,
\end{align}
where $f_{k}(\boldsymbol{v}^2)$ are polynomials in powers of $\boldsymbol{v}^2$ corresponding to higher order terms in the non-relativistic expansion. The operators $\mathcal{O}_{k}$ form a complete basis constructed from the relevant three-momenta ($\mathbf{q}$ and $\mathbf{P}$) and the interacting particle spins ($\boldsymbol{s}_{\alpha}=\boldsymbol{\sigma}_{\alpha}/2$ and $\boldsymbol{s}_{\beta}=\boldsymbol{\sigma}_{\beta}/2$),
\begin{align}
\label{eq:spinops}
\begin{aligned}
\begin{split}
&\mathcal{O}_{1}=1\,,\\
&\mathcal{O}_{3}=(\boldsymbol{\sigma}_{\alpha}\cdot\mathbf{q})(\boldsymbol{\sigma}_{\beta}\cdot\mathbf{q})\,,\\
&\mathcal{O}_{6,7}=\frac{i}{2}\big[(\boldsymbol{\sigma}_{\alpha}\cdot\mathbf{P})(\boldsymbol{\sigma}_{\beta}\cdot\mathbf{q})\pm(\alpha,\beta)\big]\,,\\
&\mathcal{O}_{9,10}=\frac{i}{2}(\boldsymbol{\sigma}_{\alpha}\pm\boldsymbol{\sigma}_{\beta})\cdot\mathbf{q}\,,\\
&\mathcal{O}_{12,13}=\frac{1}{2}(\boldsymbol{\sigma}_{\alpha}\pm\boldsymbol{\sigma}_{\beta})\cdot\mathbf{P}\,,\\
&\mathcal{O}_{15}=\frac{1}{2}\big[\boldsymbol{\sigma}_{\alpha}\cdot(\mathbf{P}\times\mathbf{q})(\boldsymbol{\sigma}_{\beta}\cdot\mathbf{q})+(\alpha,\beta)\big]\,,
\end{split}
\hspace{-0.2em}
\begin{split}
&\mathcal{O}_{2}=\boldsymbol{\sigma}_{\alpha}\cdot\boldsymbol{\sigma}_{\beta}\,,\\
&\mathcal{O}_{4,5}=\frac{i}{2}(\boldsymbol{\sigma}_{\alpha}\pm\boldsymbol{\sigma}_{\beta})(\mathbf{P}\times\mathbf{q})\,,\\
&\mathcal{O}_{8}=(\boldsymbol{\sigma}_{\alpha}\cdot\mathbf{P})(\boldsymbol{\sigma}_{\beta}\cdot\mathbf{P})\,,\\
&\mathcal{O}_{11}=i(\boldsymbol{\sigma}_{\alpha}\times\boldsymbol{\sigma}_{\beta})\cdot\mathbf{q}\,,\\
&\mathcal{O}_{14}=(\boldsymbol{\sigma}_{\alpha}\times\boldsymbol{\sigma}_{\beta})\cdot\mathbf{P}\,,\\
&\mathcal{O}_{16}=\frac{i}{2}\big[\boldsymbol{\sigma}_{\alpha}\cdot(\mathbf{P}\times\mathbf{q})(\boldsymbol{\sigma}_{\beta}\cdot\mathbf{P})+(\alpha,\beta)\big]\,,
\end{split}
\end{aligned}
\end{align}
where $(\alpha,\beta)$ is a shorthand for $(\alpha\leftrightarrow\beta)$.

Combining Eqs.~\eqref{eq:fouriertransform}, \eqref{eq:spectraldecomp} and \eqref{eq:kexpansion}, the potential can also be split up as
\begin{align}
\label{eq:Vsum}
V(\boldsymbol{r}, \boldsymbol{v}) =  \sum_{k=1}^{16}\mathcal{V}_{k}(\boldsymbol{r}, \boldsymbol{v})f_{k}(\boldsymbol{v}^2)\,,
\end{align}
where
\begin{align}
\label{eq:Vk}
\mathcal{V}_{k}(\boldsymbol{r}, \boldsymbol{v})= -\int \frac{d^{3} \mathbf{q}}{(2 \pi)^{3}} \,e^{i \mathbf{q} \cdot \boldsymbol{r}}\int_{0}^{\infty} dt' \,	\frac{\rho_{k}(t')\,\mathcal{O}_{k}(\mathbf{q}',\mathbf{P})}{t'+\mathbf{q}^2}\,,
\end{align}
and the variable $t' = -(\mathbf{q}')^2$ is integrated over $dt'$ while $\mathbf{q}$ is integrated over $d^{3}\mathbf{q}$.

The functions $\mathcal{V}_{k}(\boldsymbol{r}, \boldsymbol{v})$ can be computed by first evaluating the integral without the factor $\mathcal{O}_{k}$ and multiplying by a single power of $r$,
\begin{align}
\label{eq:simplefourier}
\mathcal{V}'_{k}(r)&\equiv - r\int \frac{d^{3} \mathbf{q}}{(2 \pi)^{3}} \,e^{i \mathbf{q} \cdot \boldsymbol{r}}\int_{0}^{\infty} dt'\,	\frac{\rho_{k}(t')}{t'+\mathbf{q}^2}=\frac{1}{4\pi}\int_{0}^{\infty} dt \,\rho_{k}(t)\,e^{-r\sqrt{t}}\,,
\end{align}
where for the second equality we have integrated over $\mathbf{q}$, $\theta$ and $\phi$ and relabelled the dummy variable $t'$ as $t$. As outlined in Ref. \cite{Dobrescu:2006au}, the functions $\mathcal{V}_{k}(r)$ can be readily computed by applying derivatives to the $\mathcal{V}'_{k}(r)$ functions. We have for example the following operations for the operators $\mathcal{O}_1$, $\mathcal{O}_2$ and $\mathcal{O}_3$,
\begin{align}
\begin{aligned}
\label{eq:VprimetoV}
\mathcal{V}_{1}(r)&=\frac{1}{r}\,\mathcal{V}'_{1}(r)\,,\\
\mathcal{V}_{2}(r)&=\frac{1}{r}\,(\boldsymbol{\sigma}_{\alpha}\cdot\boldsymbol{\sigma}_{\beta})\,\mathcal{V}'_{2}(r)\,,\\
\mathcal{V}_{3}(r)&=\frac{1}{r^3}\bigg[(\boldsymbol{\sigma}_{\alpha}\cdot\boldsymbol{\sigma}_{\beta})\,\bigg(1-r\frac{d}{dr}\bigg)-3(\boldsymbol{\sigma}_{\alpha}\cdot\mathbf{q})(\boldsymbol{\sigma}_{\beta}\cdot\mathbf{q})\bigg(1-r\frac{d}{dr}+\frac{r^2}{3}\frac{d^2}{dr^2}\bigg)\bigg]\mathcal{V}'_{3}(r)\,.
\end{aligned}
\end{align}

\section{Long-Range Potentials From Two-Neutrino Exchange}
\label{sec:effectivepotential}

In this section we will derive, using Eq.~\eqref{eq:fouriertransform}, the potentials $V_{\alpha\beta}(r)$ induced by the exchange of two neutrinos between fermions $f_{\alpha}$ and $f_{\beta}$, depicted in Fig.~\ref{fig:effpotential} (right). We consider the neutrinos in the loop and the external fermions to interact via the four-fermion $\nu$LEFT Lagrangian of Eq.~\eqref{eq:Leff1}, which includes SM CC and NC interactions in the operators with coefficients $c^{LL}$ and $c^{RL}$ in addition to non-standard operators of vector ($c^{XY}$), scalar ($g^{XY}$) and tensor ($h^{XX}$) type. The external fermions may be charged leptons ($f=\ell$) or up-type and down-type quarks ($f=u,d$) within a nucleon or nucleus $\mathcal{N}$. The quark-level coefficients $c^{XY}$ must be matched to nucleon/nucleus-level coefficients using Eq.~\eqref{eq:nucleoncoefficients}.

In Sec.~\ref{sec:SMpotential} we derive the potential $V^{LL}_{\alpha\beta}(r)$ when only the SM CC and NC interactions are present. In Sec.~\ref{sec:RHpotential} we include right-handed vector-type neutrino currents and derive the potentials $V^{LR}_{\alpha\beta}(r)$ and $V^{RR}_{\alpha\beta}(r)$ when one or both of the neutrino currents are right-handed. In Sec.~\ref{sec:scalarpotential} we introduce scalar interactions and derive the vector-scalar and scalar-scalar potentials $V^{VS}_{\alpha\beta}(r)$ and $V^{SS}_{\alpha\beta}(r)$. In Sec.~\ref{sec:tensorpotential} we consider tensor interactions, determining the vector-tensor potential $V^{VT}_{\alpha\beta}(r)$. We derive each potential for Dirac and Majorana neutrinos, examining the dependence on the distance in the short- and long-range limits and on the spins of the external states. We finally plot and compare the potentials in Fig.~\ref{fig:potentialplot} of Sec.~\ref{sec:comparepotentials}.

\subsection{Standard Model charged and neutral currents}
\label{sec:SMpotential}

\begin{figure}[t!]
	\centering
	\includegraphics[width=0.24\textwidth]{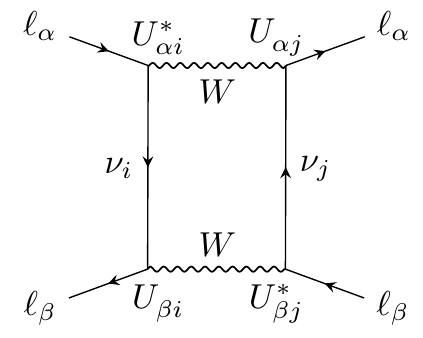}
	\includegraphics[width=0.24\textwidth]{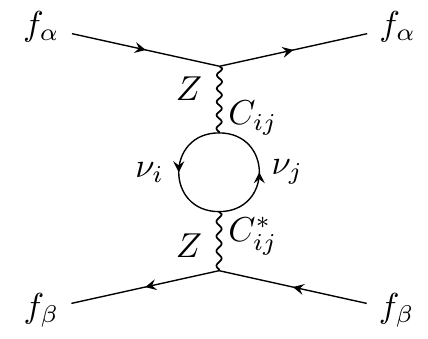}
	\includegraphics[width=0.24\textwidth]{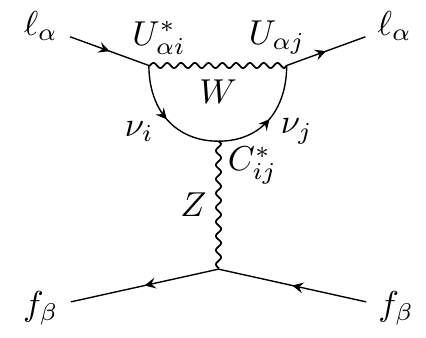}
	\includegraphics[width=0.24\textwidth]{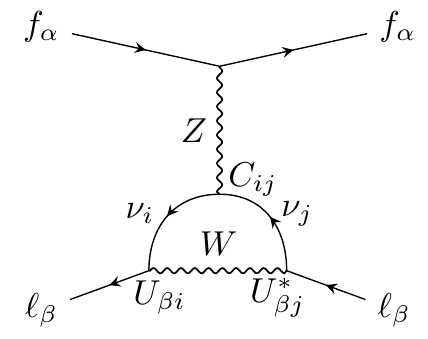}
	\caption{Diagrams depicting the exchange of two massive neutrinos between fermions $f_{\alpha}$ and $f_{\beta}$  with SM CC and NC interactions at each vertex. For the CC interactions $f=\ell$.}
	\label{fig:SMpotential}
\end{figure}

We begin by deriving the potential $V^{LL}_{\alpha\beta}(r)$ arising from the SM diagrams in Fig.~\ref{fig:SMpotential}. For simplicity we determine the amplitude $\mathcal{M}_{\alpha\beta}$ (and the corresponding spectral function $\rho_{\alpha\beta}$) by integrating out the $W^{\pm}$ and $Z$ boson propagators and using the $\nu$LEFT interaction Lagrangian of Eq.~\eqref{eq:Leff1}. We see that $W^{\pm}$ exchange can only occur for charged leptons while $Z$ exchange is possible for both leptons and quarks within a nucleon/nucleus $\mathcal{N}$. Both $W^{\pm}$ and $Z$ exchange contribute to the coefficient $c^{LL}_{\alpha\beta;ij}$ while only $Z$ exchange contributes to $c^{RL}_{\alpha\beta;ij}$ -- the values for these are given in Eq.~\eqref{eq:SMcLLcRL}. The external fermion currents are therefore either left- or right-handed while the neutrino currents are strictly left-handed.

Applying the appropriate Feynman rules from the interaction Lagrangian of Eq.~\eqref{eq:Leff1}, we can write the invariant amplitude of the scattering process in Fig.~\ref{fig:effpotential} (right) in the convenient form
\begin{align}
\label{eq:SMamplitude}
-i\mathcal{M}_{\alpha\beta}=\frac{1}{4m_{\alpha}m_{\beta}}\left(-i\frac{4G_{F}}{\sqrt{2}}\right)^2\sum_{i,j=1}^{N}\sum_{X,Y=L,R}c^{XL}_{\alpha;ij}\,c^{YL}_{\beta;ij}\,\mathcal{H}^{\alpha\beta}_{\mu\nu}\,\mathcal{N}_{ij}^{\mu\nu}\,,
\end{align}
where $1/(4m_{\alpha}m_{\beta})$ is a normalisation factor convenient in the non-relativistic limit \cite{Feinberg:1989ps}. The amplitude firstly contains the sum over the neutrino mass eigenstates, $i,j$, which we allow to run from 1 to $N=3+n$ to allow for the presence of $n$ additional Dirac or Majorana states. It also contains the sum over the possible chiralities ($X,Y = L,R$) of the external fermion currents. As we are focussing on scattering processes in which the flavours of the interacting fermions do not change, the coefficients $c^{XY}_{\alpha\beta;ij}$ will always be diagonal in the flavour of the external fermions ($\alpha=\beta$). We therefore relabel $c^{XY}_{\alpha\alpha;ij}\equiv c^{XY}_{\alpha;ij}$ in Eq.~\eqref{eq:SMamplitude} and the following discussion. 

The amplitude in Eq.~\eqref{eq:SMamplitude} is also split conveniently into two Lorentz tensor factors. The first is the product of external fermion bilinears
\begin{align}
\label{eq:Hmunu}
\mathcal{H}^{\alpha\beta}_{\mu\nu}=[\bar{u}_{s_{\alpha}^{\prime}}(\mathbf{p}_{\alpha}^{\prime}) \,\gamma_{\mu}\,\mathbb{P}_{X}\, u_{s_{\alpha}}(\mathbf{p}_{\alpha})][\bar{u}_{s_{\beta}^{\prime}}(\mathbf{p}_{\beta}^{\prime}) \,\gamma_{\nu}\,\mathbb{P}_{Y}\, u_{s_{\beta}}(\mathbf{p}_{\beta})]\equiv[\gamma_{\mu}\,\mathbb{P}_{X}]_{\alpha}[\gamma_{\nu}\,\mathbb{P}_{Y}]_{\beta}\,,
\end{align}
where $u_{s_{\alpha}}(\mathbf{p}_{\alpha})$ and $u_{s_{\beta}}(\mathbf{p}_{\beta})$ are four-component Dirac spinors for the fermions $f_{\alpha}$ and $f_{\beta}$ (or nucleon $\mathcal{N}$) and $\mathbb{P}_{X}$ and $\mathbb{P}_{Y}$ are the usual chirality projection operators $\mathbb{P}_{L}=(1-\gamma_{5})/2$ and $\mathbb{P}_{R}=(1+\gamma_{5})/2$ for $X,Y=L,R$. The second factor $\mathcal{N}_{ij}^{\mu\nu}$ integrates the product of massive neutrino propagators over the loop momentum $k$,
\begin{equation}
\mathcal{N}^{\mu \nu}_{ij} =  \int {d^4 k \over (2 \pi)^4} {{\rm Tr} [ \gamma^\mu \mathbb{P}_L
	(\slashed{q}+\slashed{k} +m_j)\gamma^\nu \mathbb{P}_L(\slashed{k} +m_i) ] \over {(k^2-m_i^2)\,((q+k)^2-m_j^2)}}\,.
\label{eq:Nmunu}
\end{equation}

We now use the method from Eq.~\eqref{eq:Vsum} onwards to calculate the potential. Using Eq.~\eqref{eq:spectraldefinition} we first determine the spectral function by taking the discontinuity the amplitude. The discontinuity only needs to be taken for the neutrino loop factor,
\begin{align}
\begin{aligned}
\mathrm{disc}(\mathcal{N}_{ij}^{\mu \nu}) =~&\frac{\Lambda^{1/2} (q^2,m_{i}^2,m_{j}^2)}{12\pi} \,
\Bigg\{- \left(1-\frac{\overline{m_{ij}^2}}{q^2}-\frac{(\Delta m_{ij}^2)^2}{2q^4}\right)g^{\mu \nu}\\
&+ \left(1+\frac{2\overline{m_{ij}^2}}{q^2}-\frac{2(\Delta m_{ij}^2)^2}{q^4}\right)\frac{q^{\mu} q^{\nu}}{q^2} \Bigg\}\,\Theta\big(q^2-(m_{i}+m_{j})^2\big)\, ,
\label{eq:discontinuity}
\end{aligned}
\end{align}
where $\Theta(x)$ is the Heaviside step function, $\overline{m_{ij}^2}=(m_{i}^2+m_{j}^2)/2$ is the average of the squares of the neutrino masses, $\Delta m_{ij}^2=m_{i}^2-m_{j}^2$ is the difference in the squares and $\Lambda(x,y,x)$ is the K\"all\'en function,
\begin{align}
\Lambda(x,y,z)=x^2+y^2+z^2-2xy-2yz-2zx\,.
\end{align}
To compute the spectral function we contract $\mathcal{H}^{\alpha\beta}_{\mu\nu}$ with $\mathrm{disc}(\mathcal{N}^{\mu \nu}_{ij})$. The Lorentz indices of $\mathcal{H}^{\alpha\beta}_{\mu\nu}$ either contract with $g^{\mu\nu}$ in $\mathrm{disc}(\mathcal{N}^{\mu \nu}_{ij})$ to give $[\gamma_{\mu}\,\mathbb{P}_{X}]_{\alpha}[\gamma^{\mu}\,\mathbb{P}_{Y}]_{\beta}$ or with $q^{\mu}q^{\nu}$ to give $[\slashed q \,\mathbb{P}_{X}]_{\alpha}[\slashed q\,\mathbb{P}_{Y}]_{\beta}$. 

An assumption we now make is that the external fermions are non-relativistic. In this limit it is possible to replace $[\gamma_{\mu}\,\mathbb{P}_{X}]_{\alpha}[\gamma^{\mu}\,\mathbb{P}_{Y}]_{\beta}$ and $[\slashed q \,\mathbb{P}_{X}]_{\alpha}[\slashed q\,\mathbb{P}_{Y}]_{\beta}$ with the lowest-order terms in the non-relativistic expansion -- Appendix \ref{sec:nonrel} lists the lowest-order terms for bilinear products such as $[\gamma_{\mu}]_{\alpha}[\gamma^{\mu}]_{\beta}$, $[\gamma_{\mu}]_{\alpha}[\gamma^{\mu}\gamma_{5}]_{\beta}$ and $[\slashed q \gamma_{5}]_{\alpha}[\slashed q \gamma_{5}]_{\beta}$. The terms that dominate are proportional to $4m_{\alpha}m_{\beta}$, cancelling the $1/(4m_{\alpha}m_{\beta})$ normalisation factor in the amplitude. 	Higher-order terms in the expansion are suppressed by powers of $\mathbf{q}/m_{\alpha}$ and can be neglected. 

The discussion has so far been valid for Dirac neutrinos. For Majorana neutrinos only the axial part contributes to the left-handed current and is a factor two larger than the Dirac axial vector current. The neutrino loop factor $\mathcal{N}_{\mu\nu}$ is instead given by
\begin{equation}
\mathcal{N}^{\mu \nu}_{ij,\,M} =  \frac{1}{2}\times 4\int {d^4 k \over (2 \pi)^4} {{\rm Tr} [ \gamma^\mu \gamma_{5}
	(\slashed{q}+\slashed{k} +m_j)\gamma^\nu \gamma_{5}(\slashed{k} +m_i) ] \over {(k^2-m_i^2)\,((q+k)^2-m_j^2)}}\,,
\label{eq:NmunuMaj}
\end{equation}
where an additional factor of 1/2 is required due to the permutation symmetry of the Majorana states in the loop.

Using Eq.~\eqref{eq:spectraldefinition} we can now write the spectral function as
\begin{align}
\begin{aligned}
\label{eq:rhoLLgeneral}
\rho_{\alpha\beta}^{LL}(t)&=-\frac{G_{F}^2}{\pi m_{\alpha}m_{\beta}}\sum_{i,j=1}^{N}\sum_{X,Y=L,R}c^{XL}_{\alpha;ij}\,c^{YL}_{\beta;ij}\,\mathcal{H}^{\alpha\beta}_{\mu\nu}\,\mathrm{disc}(\mathcal{N}_{ij}^{\mu\nu})\,,
\end{aligned}
\end{align}
where we choose the $LL$ superscript to indicate the presence of two left-handed neutrino currents. Now inserting Eqs.~\eqref{eq:discontinuity} and \eqref{eq:Hmunu} into Eq.~\eqref{eq:rhoLLgeneral} and taking the non-relativistic limit, we obtain to lowest-order
\begin{align}
\begin{aligned}
\label{eq:rhoLL}
&\rho_{\alpha\beta,D(M)}^{LL}(t)=\frac{G_{F}^2}{12\pi^2}\sum_{i,j=1}^{N}\Theta\big(t-(m_{i}+m_{j})^2\big)\,\Lambda^{1/2}(t,m_{i}^2,m_{j}^2)\\
&\hspace{4.4em}\times \bigg\{\Big[X^{LL}_{\alpha\beta;ij}-Y^{LL}_{\alpha\beta;ij}\,(\boldsymbol{\sigma}_{\alpha}\cdot\boldsymbol{\sigma}_{\beta})\Big]F^{D(M)}_{ij}(t)-Y^{LL}_{\alpha\beta;ij}\,(\boldsymbol{\sigma}_{\alpha}\cdot\mathbf{q})(\boldsymbol{\sigma}_{\beta}\cdot\mathbf{q})\,F^{V}_{ij}(t)\bigg\}\,,
\end{aligned}
\end{align}
which retains one spin-independent and two spin-dependent terms. The factors $X^{LL}_{\alpha\beta;ij}$ and $Y^{LL}_{\alpha\beta;ij}$ are given by the following combinations of the $\nu$LEFT coefficients,
\begin{align}
X^{LL}_{\alpha\beta;ij} &= (c^{LL}+c^{RL})_{\alpha;ij}(c^{LL}+c^{RL})_{\beta;ij}^*\,,\\
Y^{LL}_{\alpha\beta;ij} &= (c^{LL}-c^{RL})_{\alpha;ij}(c^{LL}-c^{RL})_{\beta;ij}^*\,,
\end{align} 
where we have used the shorthand notation $(c^{LL}\pm c^{RL})_{\alpha;ij}\equiv c_{\alpha;ij}^{LL}\pm c_{\alpha;ij}^{RL}$. Taking the SM values of the coefficients in Eq.~\eqref{eq:SMcLLcRL} and assuming a unitary light neutrino mixing matrix $U_{\alpha i}$ such that $C_{ij}=\delta_{ij}$, these factors for leptons ($\alpha,\beta = e,\mu,\tau$) are for example
\begin{align}
\label{eq:XYSM}
X^{LL}_{\alpha\beta;ij} &= (U_{\alpha i}^*U^{}_{\alpha j}+g^{\ell}_{V}\delta_{ij})(U_{\beta i}^*U^{}_{\beta j}+g^{\ell}_{V}\delta_{ij})^*\,,\\
Y^{LL}_{\alpha\beta;ij} &= (U_{\alpha i}^*U^{}_{\alpha j}+g^{\ell}_{A}\delta_{ij})(U_{\beta i}^*U^{}_{\beta j}+g^{\ell}_{A}\delta_{ij})^*\,.
\end{align} 
The functions $F_{ij}^{D}$, $F_{ij}^{M}$ and $F_{ij}^{V}$ in Eq. \eqref{eq:rhoLL} are given by
\begin{align}
\begin{aligned}
F_{ij}^{D}(t)&=1-\frac{\overline{m_{ij}^2}}{t}-\frac{(\Delta m_{ij}^2)^2}{2t^{2}}\,,\\
F_{ij}^{M}(t)&=1-\frac{\overline{m_{ij}^2}+3m_{i}m_{j}}{t}-\frac{(\Delta m_{ij}^2)^2}{2t^{2}}\,,\\
F^{V}_{ij}(t)&=\frac{1}{t}\left(1+\frac{2\overline{m_{ij}^2}}{t}-\frac{2(\Delta m_{ij}^2)^2}{t^{2}}\right)\,.
\end{aligned}
\end{align}
The difference between the Dirac ($D$) and Majorana ($M$) cases is reflected in the function $F_{ij}^{D(M)}$ multiplying the term in square brackets in Eq.~\eqref{eq:rhoLL}. In the Majorana case there is an additional term equal to $-3m_{i}m_{j}/t$ corresponding to the helicity-suppressed process of two left-handed neutrinos being created and two right-handed \textquoteleft anti-neutrinos' being annihilated. This process is not possible for left-handed Dirac neutrinos without introducing a right-handed current.

The spectral function of Eq.~\eqref{eq:rhoLL} contains terms proportional to the spin operators $\mathcal{O}_{1}=1$, $\mathcal{O}_{2}=\boldsymbol{\sigma}_{\alpha}\cdot\boldsymbol{\sigma}_{\beta}$ and $\mathcal{O}_{3}=(\boldsymbol{\sigma}_{\alpha}\cdot\mathbf{q})(\boldsymbol{\sigma}_{\beta}\cdot\mathbf{q})$. To determine the overall potential $V^{LL}_{\alpha\beta}(r)$ we evaluate the integral in Eq.~\eqref{eq:simplefourier} for each of the three parts of the spectral function multiplying these operators. We then take the appropriate derivatives in Eq.~\eqref{eq:VprimetoV} to derive the three components of the potential $\mathcal{V}^{LL}_{k}(r)$ ($k=1,2,3$) and add these to obtain
\begin{align}
\begin{aligned}
\label{eq:VLL}
V_{\alpha\beta,D(M)}^{LL}(r) = \frac{G_{F}^2}{4\pi^3r^{5}}&\sum_{i,j=1}^{N}\bigg\{X^{LL}_{\alpha\beta;ij}\,I_{ij}^{D(M)}(r)\\
&~~~~~~~-Y^{LL}_{\alpha\beta;ij}\left[ (\boldsymbol{\sigma}_{\alpha}\cdot\boldsymbol{\sigma}_{\beta})\,J_{ij}^{D(M)}(r)-(\boldsymbol{\sigma}_{\alpha}\cdot\hat{\boldsymbol{r}})(\boldsymbol{\sigma}_{\beta}\cdot\hat{\boldsymbol{r}}) \,J_{ij}^{V}(r)\right]\bigg\}\,,
\end{aligned}
\end{align}
where $\hat{\boldsymbol{r}}=\boldsymbol{r}/|\boldsymbol{r}|$ is the unit displacement between the interacting states and the integral functions $I^{D}_{ij}(r)$, $I^{M}_{ij}(r)$, $J^{D}_{ij}(r)$, $J^{M}_{ij}(r)$ and $J^{V}_{ij}(r)$ are given in Appendix~\ref{sec:integrals}. We define these functions to be dimensionless in order to take the dimensionful factor $G_{F}^2/(4\pi^3r^5)$ out of the sum. The potential therefore scales naively as $1/r^{5}$ though we will see that this behaviour changes in the long-range limit. The difference between the Dirac and Majorana cases is now a difference in the functions $I^{D(M)}_{ij}(r)$ and $J^{D(M)}_{ij}(r)$ in Eq.~\eqref{eq:VLL}.

The neutrino-mediated potential in Eq.~\eqref{eq:VLL} simplifies when only a single massive neutrino is considered. Firstly, the mixing factors in $X^{LL}_{\alpha\beta;ij}$ and $Y^{LL}_{\alpha\beta;ij}$ are replaced as $U^{*}_{\alpha i}U^{}_{\alpha j}\rightarrow 1$ and $\delta_{ij}\rightarrow 1$ and the summation is now over a single state $i=j\equiv\nu$. For the potential between two charged leptons we have for example $X^{LL}_{\alpha\beta;\nu\nu}=(1+g^{\ell}_{V})^2$ and $Y^{LL}_{\alpha\beta;\nu\nu}=(1+g^{\ell}_{A})^2$. Secondly, the functions $I_{ij}^{D}(r)$, $I_{ij}^{M}(r)$ and $I_{ij}^{V}(r)$ take the closed-forms
\begin{align}
\label{eq:closedform}
I^{D}_{\nu\nu}(r) &=m_{\nu}^3r^3K_{3}(2m_{\nu}r)\,,\nonumber\\
I^{M}_{\nu\nu}(r)&=2m_{\nu}^2r^2K_{2}(2m_{\nu}r)\,,\\
I^{V}_{\nu\nu}(r) &= 2m_{\nu}rK_{1}(2m_{\nu}r) +\frac{\pi^2m_{\nu}^2r^2}{2}\,G^{2,0}_{2,4}\left(m_{\nu}^2r^2\,\bigg|\begin{smallmatrix}\frac{1}{2},\frac{3}{2}\\0,0,\frac{1}{2},\frac{1}{2}\end{smallmatrix}\right)+2\pi m_{\nu}^3 r^3\,,\nonumber
\end{align}
where $K_{n}(x)$ are modified Bessel functions of the second kind and $G^{m,n}_{p,q}$ is the Meijer G-function. Using the relations in Appendix \ref{sec:integrals} we can also determine the functions $J^{D}_{\nu\nu}(r)$, $J^{M}_{\nu\nu}(r)$ and $J^{V}_{\nu\nu}(r)$. For interacting leptons the spin-independent parts of the Dirac and Majorana potentials become
\begin{align}
\label{eq:DandM}
V^{LL}_{\alpha\beta,D}(r)=\frac{G_{F}^2m_{\nu}^3(1+g_{V}^\ell)^2}{4\pi^3r^2}K_{3}(2m_{\nu}r)\,,~~V^{LL}_{\alpha\beta,M}(r)=\frac{G_{F}^2m_{\nu}^2(1+g_{V}^\ell)^2}{2\pi^3r^3}K_{2}(2m_{\nu}r)\,,
\end{align}
respectively, in agreement with previous results \cite{Thien:2019ayp}. 

The functions in Eq.~\eqref{eq:closedform} depend on the product $2m_{\nu}r$ -- given the behaviour of the modified Bessel functions $K_{n}(x)$ in the limits $x\ll\mathcal{O}(1)$ and $x\gg \mathcal{O}(1)$, the potential displays contrasting behaviour in the limits $r\ll r_{\nu}$ and $r\gg r_{\nu}$ where $r_{\nu}=1/(2m_{\nu})$ is half the Compton wavelength of the neutrino. In the `short-range' limit ($r\ll r_{\nu}$) the exchanged neutrinos are relativistic and their masses can be neglected. The Dirac or Majorana nature of neutrinos cannot be probed due to the suppression of the term $-3m_{i}m_{j}/t$ in $F_{ij}^{M}(t)$ and the converging of the Dirac and Majorana potentials. In the `long-range' limit ($r\gg r_{\nu}$) they become non-relativistic -- the neutrino masses are important and the potential is exponentially suppressed as $V\propto e^{-2m_{\nu}r}$. A priori the Dirac or Majorana nature can now be probed given the small difference in behaviour of Dirac and Majorana neutrinos.

To verify this quantitatively we expand the functions in Eq.~\eqref{eq:closedform} and therefore the single-neutrino potential in the opposing limits. For $r\ll r_{\nu}$ we find to lowest-order
\begin{align}
\label{eq:shortrange}
V_{\alpha\beta,D(M)}^{LL}(r) &\approx \frac{G_{F}^2}{4\pi^3r^{5}}\left\{X_{\alpha\beta;\nu\nu}^{LL}-Y_{\alpha\beta;\nu\nu}^{LL}\left[\frac{3}{2}(\boldsymbol{\sigma}_{\alpha}\cdot\boldsymbol{\sigma}_{\beta}) -\frac{5}{2}(\boldsymbol{\sigma}_{\alpha}\cdot\hat{\boldsymbol{r}})(\boldsymbol{\sigma}_{\beta}\cdot\hat{\boldsymbol{r}})\right]\right\}\,,
\end{align}
in both the Dirac and Majorana cases, as expected. The potentials therefore decrease with the distance as $1/r^{5}$ up to half the neutrino Compton wavelength. Eq.~\eqref{eq:shortrange} is not just valid for a single neutrino -- it can be obtained for three (or in general, $N$) neutrinos by neglecting the neutrino masses $m_{i}$ and $m_{j}$ appearing in the functions $I^{X}_{ij}(r)$ and $J^{X}_{ij}(r)$ in Eq.~\eqref{eq:VLL}. In this limit the functions tend to the constant values
\begin{align}
\label{eq:neglectij}
I^{D(M)}_{ij}(r) \approx 1\,,~~J^{D(M)}_{ij}(r)\approx\frac{3}{2}\,,~~J^{V}_{ij}(r) \approx\frac{5}{2}\,,
\end{align}
as outlined in Appendix~\ref{sec:integrals}. It is now possible to pull these constants out of the sum in Eq.~\eqref{eq:VLL} and identify $X^{LL}_{\alpha\beta;\nu\nu} = \sum\limits_{i,j}^{N}X^{LL}_{\alpha\beta;ij}$ and $Y^{LL}_{\alpha\beta;\nu\nu} = \sum\limits_{i,j}^{N}Y^{LL}_{\alpha\beta;ij}$. 

Expanding in the opposite limit $r\gg r_{\nu}$ gives in the Dirac case
\begin{align} \label{eq:diraclong}
V_{\alpha\beta,D}^{LL}(r) &\approx \frac{G_{F}^2m_{\nu}^{5/2}e^{-2m_{\nu}r}}{8\pi^{5/2}r^{5/2}}\bigg\{X_{\alpha\beta;\nu\nu}^{LL}-Y_{\alpha\beta;\nu\nu}^{LL}\Big[(\boldsymbol{\sigma}_{\alpha}\cdot\boldsymbol{\sigma}_{\beta}) -2(\boldsymbol{\sigma}_{\alpha}\cdot\hat{\boldsymbol{r}})(\boldsymbol{\sigma}_{\beta}\cdot\hat{\boldsymbol{r}})\Big]\bigg\}\,,
\end{align}
while in the Majorana case
\begin{align} \label{eq:majlong}
V_{\alpha\beta,M}^{LL}(r) &\approx \frac{G_{F}^2m_{\nu}^{3/2}e^{-2m_{\nu}r}}{4\pi^{5/2}r^{7/2}}\bigg\{X_{\alpha\beta;\nu\nu}^{LL}-Y_{\alpha\beta;\nu\nu}^{LL}\bigg[\frac{3}{2}(\boldsymbol{\sigma}_{\alpha}\cdot\boldsymbol{\sigma}_{\beta}) -m_{\nu}r(\boldsymbol{\sigma}_{\alpha}\cdot\hat{\boldsymbol{r}})(\boldsymbol{\sigma}_{\beta}\cdot\hat{\boldsymbol{r}})\bigg]\bigg\}\,.
\end{align}
In the Dirac case both spin-independent and spin-dependent terms scale as $e^{-2m_{\nu}r}/r^{5/2}$, while in the Majorana case the spin-independent and $\boldsymbol{\sigma}_{\alpha}\cdot\boldsymbol{\sigma}_{\beta}$ terms scale as $e^{-2m_{\nu}r}/r^{7/2}$. The term containing $(\boldsymbol{\sigma}_{\alpha}\cdot\hat{\boldsymbol{r}})(\boldsymbol{\sigma}_{\beta}\cdot\hat{\boldsymbol{r}})$ however also scales as $e^{-2m_{\nu}r}/r^{5/2}$ in the Majorana case.

Finally, we note that the operators with coefficients $c^{LL}_{\alpha;ij}$ and $c^{RL}_{\alpha;ij}$ may include the effects of new physics, which can be parametrised as small corrections $\delta c^{LL}_{\alpha;ij}$ and $\delta c^{RL}_{\alpha;ij}$ to the SM values of $c^{LL}_{\alpha;ij}$ and $c^{RL}_{\alpha;ij}$. Deviations from the SM potential $V^{LL}_{\alpha\beta}(r)$ therefore arise as corrections to the factors $X^{LL}_{\alpha\beta;ij}$ and $Y^{LL}_{\alpha\beta;ij}$, 
\begin{align}
\label{eq:SMcorrection}
\delta X^{LL}_{\alpha\beta;ij} &= (c^{LL}+c^{RL})_{\alpha;ij}(\delta c^{LL}+\delta c^{RL})_{\beta;ij}^*~+~(\alpha,\beta)\,,\\
\delta Y^{LL}_{\alpha\beta;ij} &= (c^{LL}-c^{RL})_{\alpha;ij}(\delta c^{LL}-\delta c^{RL})_{\beta;ij}^*~+~(\alpha,\beta)\,,
\end{align} 
where the correction can either be at the vertex with fermion $f_{\alpha}$ or $f_{\beta}$.

\subsection{Right-handed vector non-standard interactions}
\label{sec:RHpotential}

\begin{figure}[t!]
	\centering
	\includegraphics[width=0.32\textwidth]{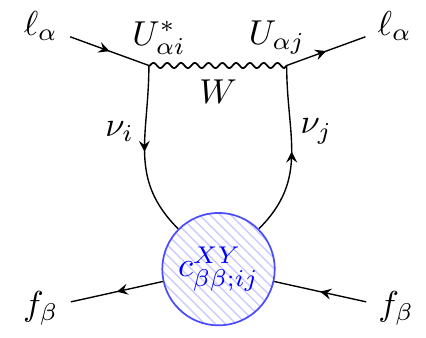}
	\hspace{2em}
	\includegraphics[width=0.32\textwidth]{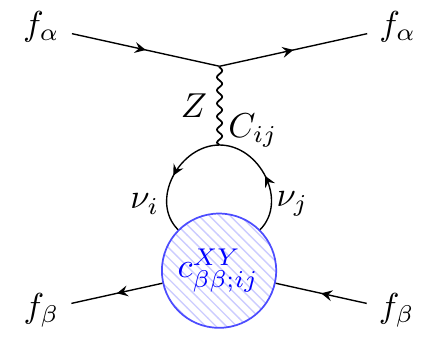}
	\caption{Diagrams depicting the exchange of two massive neutrinos between fermions $f_{\alpha}$ and $f_{\beta}$ with SM CC and NC interactions at one vertex and an effective four-fermion interaction at the other. In our framework the effective interaction may be of vector ($c^{XY}_{\alpha\beta;ij}$), scalar ($g^{XY}_{\alpha\beta;ij}$) or tensor ($h^{XY}_{\alpha\beta;ij}$) type.}
	\label{fig:RHpotential}
\end{figure}

Motivated by theories such as the Left-Right Symmetric Model (LRSM) we now introduce a right-handed neutrino current. We will first derive the neutrino-mediated potential $V^{LR}_{\alpha\beta}(r)$ induced when there is a SM CC or NC interaction at one vertex and a right-handed neutrino current at the other, depicted in Fig.~\ref{fig:RHpotential}. In the $\nu$LEFT interaction Lagrangian of Eq.~\eqref{eq:Leff1} we now allow the coefficients $c_{\alpha\beta;ij}^{LR}$ and $c_{\alpha\beta;ij}^{RR}$ to be non-zero along with $c_{\alpha\beta;ij}^{LL}$ and $c_{\alpha\beta;ij}^{RL}$.

The spectral function $\rho_{\alpha\beta}^{LR}(t)$ in this scenario is the same as Eq. \eqref{eq:rhoLL} but with one coefficient replaced as $c^{XL}_{\alpha;ij}\rightarrow c^{XR}_{\alpha;ij}$ and one chirality projection operator replaced as $\mathbb{P}_{L}\rightarrow\mathbb{P}_{R}$ in the neutrino loop factor $\mathcal{N}^{\mu \nu}_{ij}$. We also add an identical contribution with $(\alpha\leftrightarrow\beta)$ to account for the right-handed current being either at the vertex with the external fermion $f_{\alpha}$ or $f_{\beta}$. If the external fermions are identical ($\alpha=\beta$) we must multiply the spectral function by an additional factor of $1/2$ to avoid double counting -- this gives a factor $1/(1+\delta_{\alpha\beta})$.

The discontinuity of the neutrino loop factor $\mathcal{N}_{ij}^{\mu \nu}$ for Dirac neutrinos is
\begin{eqnarray}
\label{eq:NmunuLR}
\mathrm{disc}(\mathcal{N}_{ij}^{\mu \nu}) &=&-\frac{\Lambda^{1/2} (t,m_{i}^2,m_{j}^2)}{4\pi}\,\frac{m_{i}m_{j}}{q^2}\,g^{\mu \nu}\,\Theta\big(q^2-(m_{i}+m_{j})^2\big)\, ,
\label{eq:discontinuityLR}
\end{eqnarray}
which is suppressed by the factor $m_{i}m_{j}/q^2$ because the process is helicity-suppressed. A negative helicity neutrino $\nu_{i}$ created by the left-handed current will be annihilated by the right-handed current with an associated factor $m_{i}/q$ -- for both neutrinos this results in the $m_{i}m_{j}/q^2$ factor. The physics is identical to the helicity-suppressed contribution to Majorana neutrino exchange in the previous subsection.

Contracting the $g^{\mu\nu}$ factor in Eq.~\eqref{eq:NmunuLR} with the product of external fermion bilinears $\mathcal{H}^{\alpha\beta}_{\mu\nu}$ we obtain $[\gamma_{\mu}\,\mathbb{P}_{X}]_{\alpha}[\gamma^{\mu}\,\mathbb{P}_{Y}]_{\beta}$. The non-relativistic limit can now be taken to obtain the spectral function -- in the Dirac case
\begin{align}
\begin{aligned}
\rho^{LR}_{\alpha\beta,D}(t)=\frac{1}{1+\delta_{\alpha\beta}}\frac{G_{F}^2}{4\pi^2 }\sum_{i,j=1}^{3}&\Theta\big(t-(m_{i}+m_{j})^2\big)\,\Lambda^{1/2}(t,m_{i}^2,m_{j}^2)\,\\
&\times \frac{m_{i}m_{j}}{t}\Big\{X^{LR}_{\alpha\beta;ij} - Y^{LR}_{\alpha\beta;ij}\,(\boldsymbol{\sigma}_{\alpha}\cdot\boldsymbol{\sigma}_{\beta})\Big\}\,,
\end{aligned}
\end{align}
where the pre-factors $X^{LR}_{\alpha\beta;ij}$ and $Y^{LR}_{\alpha\beta;ij}$ are
\begin{align}
\label{eq:rhoLR}
X^{LR}_{\alpha\beta;ij} &= (c^{LL}+c^{RL})_{\alpha;ij}(c^{LR}+c^{RR})_{\beta;ij}^*~+~(\alpha, \beta)\,,\\
Y^{LR}_{\alpha\beta;ij} &= (c^{LL}-c^{RL})_{\alpha;ij}(c^{LR}-c^{RR})_{\beta;ij}^*~+~(\alpha, \beta)\,.
\end{align}
Using the same method as the previous section to derive the potential from the spectral function, we find in the Dirac case
\begin{align}
\label{eq:VLRdirac}
V^{LR}_{\alpha\beta,D}(r) &= \frac{1}{1+\delta_{\alpha\beta}}{G^2_{F} \over {8\pi^3r^3}}\sum^{3}_{i,j=1} m_i m_j\Big\{X^{LR}_{\alpha\beta;ij} - Y^{LR}_{\alpha\beta;ij}\,(\boldsymbol{\sigma}_{\alpha}\cdot\boldsymbol{\sigma}_{\beta})\Big\}\,I^{LR}_{ij}(r)\,,
\end{align} 
where the dimensionless function $I_{ij}^{LR}(r)$ is given in Appendix \ref{sec:integrals}

The Majorana case is different due to the symmetry relations of Eq. \eqref{eq:Majorana_relations} -- the right-handed current operator with coefficient $c^{LR}_{\alpha\beta;ij}$ is equivalent to the left-handed current operator with coefficient $c^{LL}_{\alpha\beta;ij}$ and thus the coefficients are related by $c^{LL}_{\alpha\beta;ij}=-c^{LR}_{\alpha\beta;ji}$. This is equivalent to the vector current vanishing for Majorana neutrinos. The potential we derive from the right-handed current operator is therefore identical to Eq. \eqref{eq:VLL} and the coefficient $c^{LR}_{\alpha\beta;ij}$ gets the same contributions from the SM CC and NC interactions as $c^{LL}_{\alpha\beta;ij}$. If on the other hand we were to introduce additional light sterile Majorana states $N_{R}$ with right-handed interactions as in Eq. \eqref{eq:GFNN}, the coefficients $\tilde{c}^{LR}_{\alpha\beta;ij}=-\tilde{c}^{LL}_{\alpha\beta;ji}$ get no SM contribution.

For the three light active neutrinos it therefore makes more sense to consider the corrections $\delta c^{LL}_{\alpha;ij} = -\delta c^{LR}_{\alpha;ji}$ to the SM-valued coefficients $c^{LL}_{\alpha;ij} = -c^{LR}_{\alpha;ji}$ from new physics. The correction to the spectral function is
\begin{align}
\begin{aligned}
\label{eq:rhoLRmaj}
&\rho^{LR}_{\alpha\beta,M}(t)=-\frac{1}{1+\delta_{\alpha\beta}}\frac{G_{F}^2}{12\pi^2}\sum_{i,j=1}^{3}\Theta\big(t-(m_{i}+m_{j})^2\big)\,\Lambda^{1/2}(t,m_{i}^2,m_{j}^2)\\
&~~~~~\times \bigg\{\Big[\delta X^{LR}_{\alpha\beta;ij}-\delta Y^{LR}_{\alpha\beta;ij}\,(\boldsymbol{\sigma}_{\alpha}\cdot\boldsymbol{\sigma}_{\beta})\Big]F^{M}_{ij}(t)-\delta Y^{LR}_{\alpha\beta;ij}\,(\boldsymbol{\sigma}_{\alpha}\cdot\mathbf{q})(\boldsymbol{\sigma}_{\beta}\cdot\mathbf{q})\,F^{V}_{ij}(t)\bigg\}\,,
\end{aligned}
\end{align}
where $\delta X^{LR}_{\alpha\beta;ij} = -\delta X^{LL}_{\alpha\beta;ji}$ and $\delta Y^{LR}_{\alpha\beta;ij} = -\delta Y^{LL}_{\alpha\beta;ji}$ are given in Eq. \eqref{eq:SMcorrection}. This gives the correction to the SM potential
\begin{align}
\begin{aligned}
\label{eq:VLR}
V^{LR}_{\alpha\beta,M}(r) = -\frac{1}{1+\delta_{\alpha\beta}}\frac{G_{F}^2}{4\pi^3r^{5}}\sum_{i,j=1}^{3}&\bigg\{\delta X^{LR}_{\alpha\beta;ij}\,I_{ij}^{M}(r)\\
&-\delta Y^{LR}_{\alpha\beta;ij}\Big[ (\boldsymbol{\sigma}_{\alpha}\cdot\boldsymbol{\sigma}_{\beta})\,J_{ij}^{M}(r)-(\boldsymbol{\sigma}_{\alpha}\cdot\hat{\boldsymbol{r}})(\boldsymbol{\sigma}_{\beta}\cdot\hat{\boldsymbol{r}}) \,J_{ij}^{V}(r)\Big]\bigg\}\,.
\end{aligned}
\end{align}  

We again us the single neutrino simplification to study the properties of the potentials in Eqs. \eqref{eq:VLRdirac} and \eqref{eq:VLR} -- the function $I^{LR}_{ij}(r)$ takes the closed form
\begin{align}
I_{\nu\nu}^{LR}(r)=2m_{\nu}r \,K_{1}(2m_{\nu}r)\,.
\end{align}  
In the short-range limit, or $r\ll r_{\nu}$, we expand the Dirac potential Eq.~\eqref{eq:VLRdirac} as
\begin{align}
\label{eq:VLRdiracshort}
V^{LR}_{\alpha\beta,D}(r) &= \frac{1}{1+\delta_{\alpha\beta}}{G^2_{F} m_{\nu}^2\over {8\pi^3r^3}}\Big\{X^{LR}_{\alpha\beta;\nu\nu} - Y^{LR}_{\alpha\beta;\nu\nu}\,(\boldsymbol{\sigma}_{\alpha}\cdot\boldsymbol{\sigma}_{\beta})\Big\}\,.
\end{align} 
This potential is also valid in the three (or $N$) neutrino picture by using that $I^{LR}_{ij}(r)\approx 1$ in the limit $m_{i} \approx 0$ and identifying $X^{LR}_{\alpha\beta;\nu\nu}=\sum\limits^{3}_{i,j}X^{LR}_{\alpha\beta;ij}$ and $Y^{LR}_{\alpha\beta;\nu\nu}=\sum\limits^{3}_{i,j}Y^{LR}_{\alpha\beta;ij}$. In the long-range limit, $r\gg r_{\nu}$, we obtain
\begin{align}
\label{eq:VLRdiraclong}
V^{LR}_{\alpha\beta,D}(r) &= \frac{1}{1+\delta_{\alpha\beta}}{G^2_{F} m_{\nu}^{5/2}e^{-2m_{\nu}r}\over {8\pi^{5/2}r^{5/2}}}\Big\{X^{LR}_{\alpha\beta;\nu\nu} - Y^{LR}_{\alpha\beta;\nu\nu}\,(\boldsymbol{\sigma}_{\alpha}\cdot\boldsymbol{\sigma}_{\beta})\Big\}\,.
\end{align} 
In the single neutrino simplification for the correction in Eq. \eqref{eq:VLR} to the SM Majorana potential takes the same form as Eqs. \eqref{eq:shortrange} and \eqref{eq:majlong} in the short- and long-range limits respectively.

We finish this subsection by considering the case where there are two right-handed neutrino currents at the interaction vertices. Now the potential takes the same form as Eq.~\eqref{eq:VLL}, 
\begin{align}
\begin{aligned}
\label{eq:VRR}
V_{\alpha\beta}^{RR}(r) = \frac{G_{F}^2}{4\pi^3r^{5}}\sum_{i,j=1}^{N}\bigg\{&X^{RR}_{\alpha\beta;ij}\,I_{ij}^{D(M)}(r)\\
&-Y^{RR}_{\alpha\beta;ij}\left[ (\boldsymbol{\sigma}_{\alpha}\cdot\boldsymbol{\sigma}_{\beta})\,J_{ij}^{D(M)}(r)-(\boldsymbol{\sigma}_{\alpha}\cdot\hat{\boldsymbol{r}})(\boldsymbol{\sigma}_{\beta}\cdot\hat{\boldsymbol{r}}) \,J_{ij}^{V}(r)\right]\bigg\}\,,
\end{aligned}
\end{align}
where
\begin{align}
\label{eq:XRR}
X^{RR}_{\alpha\beta;ij} &= (c^{LR}+c^{RR})_{\alpha;ij}(c^{LR}+c^{RR})^*_{\beta;ij}\,,\\
Y^{RR}_{\alpha\beta;ij} &= (c^{LR}-c^{RR})_{\alpha;ij}(c^{LR}-c^{RR})^*_{\beta;ij}\,.
\end{align}
Consequently, the short- and long-range potentials are given by Eqs. \eqref{eq:shortrange} and \eqref{eq:majlong} respectively with the replacements $X^{LL}_{\alpha\beta;ij}\rightarrow X^{RR}_{\alpha\beta;ij}$ and $Y^{LL}_{\alpha\beta;ij}\rightarrow Y^{RR}_{\alpha\beta;ij}$.

\subsection{Scalar non-standard interactions}
\label{sec:scalarpotential}

We now derive the neutrino-mediated potential in the presence of a \textit{scalar} non-standard interaction. In our framework these are the operators in Eq.~\eqref{eq:Leff1} with the coefficients $g^{LL}$, $g^{RL}$, $g^{LR}$ and $g^{RR}$ normalised to the Fermi constant $G_{F}$. We first focus on the case of a scalar interaction at one vertex and a SM CC or NC interaction at the other, as shown Fig. \ref{fig:RHpotential}.

The spectral function can be determined in this scenario according to
\begin{align}
\label{eq:rhoVSgeneral}
\rho_{\alpha\beta}^{VS}(t)=-\frac{G_{F}^2}{\pi m_{\alpha}m_{\beta}}\sum_{i,j=1}^{N}\sum_{X,Y,Z=L,R}\bigg\{c^{XL}_{\alpha\alpha;ij}\,g^{YZ}_{\beta\beta;ij}&\,\mathcal{H}^{\alpha\beta}_{\mu}\,\mathrm{disc}(\mathcal{N}_{ij}^{\mu}) ~+~ (\alpha,\beta)\bigg\}\,,
\end{align}
where the sum is over the possible chiralities of the external fermion currents ($X$ and $Y$) and the neutrino current of the scalar operator ($Z$). We have taken into account that the scalar interaction may be at either vertex by adding an identical contribution with ($\alpha\leftrightarrow\beta$). The Majorana case is treated in the same way by retaining only twice the axial vector current and dividing by a factor of two due to the permutation symmetry of the neutrinos in the loop. 

The discontinuity of $\mathcal{N}_{ij}^{\mu}$ in the Dirac case for example is
\begin{align}
\mathrm{disc}(\mathcal{N}_{ij}^{\mu}) =\mp\,\frac{\Lambda^{1/2} (q^2,m_{i}^2,m_{j}^2)}{8\pi} \, \frac{m_{i}q^{\mu}}{q^2}\,\bigg(1-\frac{m_{i}^2-m_{j}^2}{q^2}\bigg)\,\Theta\big(q^2-(m_{i}+m_{j})^2\big)\,,
\label{eq:discontinuityVS}
\end{align}
where the minus (positive) sign is for a left-handed (right-handed) neutrino current at the scalar interaction and the product of external fermion bilinears is
\begin{align}
\label{eq:HmunuVS}
\mathcal{H}^{\alpha\beta}_{\mu}=\,[\gamma_{\mu}\,\mathbb{P}_{X}]_{\alpha}[\,\mathbb{P}_{Y}]_{\beta}\,.
\end{align}
Contracting $\mathcal{H}^{\alpha\beta}_{\mu}$ with $\mathrm{disc}(\mathcal{N}_{ij}^{\mu})$ and making use of the non-relativistic limits of the fermion bilinear products given in Appendix \ref{sec:nonrel}, we obtain in the Dirac case
\begin{align}
\begin{aligned}
\label{eq:VSdirac}
\rho^{VS}_{\alpha\beta,D}(t)=\frac{1}{1+\delta_{\alpha\beta}}\frac{G_{F}^2}{8\pi^2}\sum_{i,j=1}^{3}&\Theta\big(t-(m_{i}+m_{j})^2\big)\,m_{i}\,\Lambda^{1/2}(t,m_{i}^2,m_{j}^2)\\
&\times \Big\{X_{\alpha\beta;ij}^{VS}\,(\boldsymbol{\sigma}_{\alpha}\cdot\mathbf{q})+(\alpha,\beta)\Big\}\,F^{\Delta}_{ij}(t)\,,
\end{aligned}
\end{align}
while in the Majorana case we obtain
\begin{align}
\begin{aligned}
\label{eq:VSmajorana}
\rho^{VS}_{\alpha\beta,M}(t)=\frac{1}{1+\delta_{\alpha\beta}}\frac{G_{F}^2}{8\pi^2}\sum_{i,j=1}^{N}&\Theta\big(t-(m_{i}+m_{j})^2\big)\,(m_{i}+m_{j})\,\Lambda^{1/2}(t,m_{i}^2,m_{j}^2)\\
&\times \Big\{X_{\alpha\beta;ij}^{VS}\,(\boldsymbol{\sigma}_{\alpha}\cdot\mathbf{q})+(\alpha,\beta)\Big\}\,F^{S}_{ij}(t)\,.
\end{aligned}
\end{align}
The factor $X^{VS}_{\alpha\beta;ij}$ containing the scalar coefficients is
\begin{align}
X_{\alpha\beta;ij}^{VS} &= (c^{LL}-c^{RL})_{\alpha;ij}(g^{LL}+g^{RL}-g^{LR}-g^{RR})_{\beta;ij}^*\,,
\end{align}  
and the functions $F_{ij}^{\Delta}$ and $F_{ij}^{S}$ are given by
\begin{align}
\begin{aligned}
\label{eq:FDelta}
F_{ij}^{\Delta}(t)&=\frac{1}{t}\left(1-\frac{\Delta m_{ij}^2}{t}\right),\\
F_{ij}^{S}(t)&=\frac{1}{t}\left(1-\frac{(m_{i}-m_{j})^2}{t}\right)\,.
\end{aligned}
\end{align}
 
These spectral functions only contain terms proportional to the parity-violating spin operators $\mathcal{O}'_{9}=\boldsymbol{\sigma}_{\alpha}\cdot\mathbf{q}$ and $\mathcal{O}'_{10}=\boldsymbol{\sigma}_{\beta}\cdot\mathbf{q}$ proportional to linear combinations of the operators $\mathcal{O}_{9}$ and $\mathcal{O}_{10}$ in Eq. \eqref{eq:spinops}. Taking the components of the spectral function multiplying these operators we compute the functions $\mathcal{V}'_{9}(r)$ and $\mathcal{V}'_{10}(r)$  in Eq. \eqref{eq:simplefourier} and from these the components of the overall potential using
\begin{align}
\begin{aligned}
\label{eq:VS9and10}
\mathcal{V}_{9}(r)&=\frac{i}{r^2}(\boldsymbol{\sigma}_{\alpha}\cdot\hat{\boldsymbol{r}})\bigg(1-r\frac{d}{dr}\bigg)\mathcal{V}^{'}_{9}(r)\,,\\
\mathcal{V}^{}_{10}(r)&=\frac{i}{r^2}(\boldsymbol{\sigma}_{\beta}\cdot\hat{\boldsymbol{r}})\bigg(1-r\frac{d}{dr}\bigg)\mathcal{V}^{'}_{10}(r)\,,
\end{aligned}
\end{align}
from Ref. \cite{Dobrescu:2006au}.

Thus we derive the following vector-scalar potentials for the Dirac and Majorana cases
\begin{align}
\label{eq:VVSdirac}
V^{VS}_{\alpha\beta,D}(r) &= \frac{1}{1+\delta_{\alpha\beta}}{iG^2_{F} \over {16\pi^3r^4}}\sum^3_{i,j=1} m_i\,\big\{X_{\alpha\beta;ij}^{VS}\,(\boldsymbol{\sigma}_{\alpha}\cdot \hat{\boldsymbol{r}})+\mathrm{(\alpha,\beta)}\big\} \,J^{\Delta}_{ij}(r)\,,\\
\label{eq:VVSmaj}
V^{VS}_{\alpha\beta,M}(r) &= \frac{1}{1+\delta_{\alpha\beta}}{iG^2_{F} \over {16\pi^3r^4}}\sum^3_{i,j=1} (m_i+m_j)\,\big\{X_{\alpha\beta;ij}^{VS}\,(\boldsymbol{\sigma}_{\alpha}\cdot \hat{\boldsymbol{r}})+\mathrm{(\alpha,\beta)}\big\} \,J^{S}_{ij}(r)\,\,,
\end{align}  
respectively, where the dimensionless functions $J^{\Delta}_{ij}(r)$ and $J^{S}_{ij}(r)$ are given in Appendix \ref{sec:integrals}. 

The first thing to note about these potentials is that they depend on the distance as $1/r^{4}$ and contain a single power of the neutrino mass $m_{\nu}$ in the numerator. This is more suppressed than the SM-SM potential in Eq. \eqref{eq:VLL} which scales as $1/r^{5}$ but less suppressed than the right-handed current potential for Dirac neutrinos in Eq. \eqref{eq:VLRdirac} which scales as $1/r^{3}$ but is suppressed by $m_{\nu}^2$. The second point to note is that the potentials written in Eqs. \eqref{eq:VVSdirac} and \eqref{eq:VVSmaj} retain a factor of $i$ -- this can simply be absorbed into the factor $X^{VS}_{\alpha\beta;ij}$ after a suitable redefinition of the scalar coefficients $g^{XY}_{\alpha;ij}$.

We now consider the case where both interactions are scalar. We now obtain the potential via the spectral function
\begin{align}
\label{eq:rhoSSgeneral}
\rho_{\alpha\beta}^{SS}(t)=-\frac{G_{F}^2}{\pi m_{\alpha}m_{\beta}}\sum_{i,j=1}^{N}\sum_{W,X,Y,Z=L,R}\bigg\{g^{WX}_{\alpha;ij}\,g^{YZ}_{\beta;ij}&\,\mathcal{H}^{\alpha\beta}\,\mathrm{disc}(\mathcal{N}_{ij}) \,+\, (\alpha,\beta)\bigg\}\,,
\end{align}
where the discontinuity of the neutrino loop factor $\mathcal{N}_{ij}$ is given for example in the Dirac case by
\begin{align}
\mathrm{disc}(\mathcal{N}_{ij}) =-\,\frac{\Lambda^{1/2} (q^2,m_{i}^2,m_{j}^2)}{4\pi} \,\frac{
	m_{i}m_{j}}{q^2}\,\Theta\big(q^2-(m_{i}+m_{j})^2\big)\,,
\label{eq:discontinuitySS}
\end{align}
if the chirality of the neutrino currents are the same ($X=Z=L,R$) and
\begin{align}
\mathrm{disc}(\mathcal{N}_{ij}) =\frac{\Lambda^{1/2} (q^2,m_{i}^2,m_{j}^2)}{8\pi} \,
\,\bigg(1-\frac{2\overline{m_{ij}^2}}{q^2}\bigg)\,\Theta\big(q^2-(m_{i}+m_{j})^2\big)\,,
\label{eq:discontinuitySS2}
\end{align}
if the chiralities of the neutrino currents are opposite ($X\neq Z$). The external fermion bilinear product is now $\mathcal{H}^{\alpha\beta}=[\mathbb{P}_{W}]_{\alpha}[\mathbb{P}_{Y}]_{\beta}$ and we obtain the following potential in the Dirac case
\begin{align}
\label{eq:SSdirac}
V^{SS}_{\alpha\beta,D}(r) &= \frac{G_{F}^2}{8\pi^3r^3}\sum_{i,j}^{N}m_{i}m_{j}X^{SS}_{\alpha\beta;ij}I^{LR}_{ij}(r)-\frac{3G_{F}^2}{8\pi^3r^5}\sum_{i,j}^{N}Y^{SS}_{\alpha\beta;ij}I^{SD}_{ij}(r)\,,
\end{align}  
where the combination of scalar coefficients are given by
\begin{align}
X^{SS}_{\alpha\beta;ij} = (g^{LL}+g^{RL})_{\alpha;ij}(g^{LL}+g^{RL})_{\beta;ij}+(g^{LR}+g^{RR})_{\alpha;ij}(g^{LR}+g^{RR})_{\beta;ij}\,,\\
Y^{SS}_{\alpha\beta;ij} = (g^{LL}+g^{RL})_{\alpha;ij}(g^{LR}+g^{RR})_{\beta;ij}+(g^{LR}+g^{RR})_{\alpha;ij}(g^{LL}+g^{RL})_{\beta;ij}\,,
\end{align}  
and the dimensionless functions $I^{SD}_{ij}(r)$ and $I^{LR}_{ij}(r)$ are given in Appendix \ref{sec:integrals}.
For Majorana neutrinos we instead obtain
\begin{align}
\label{eq:SSmaj}
V^{SS}_{\alpha\beta,M}(r) &= \frac{3G_{F}^2}{8\pi^3r^5}\sum_{i,j}^{N}Z^{SS}_{\alpha\beta;ij}I^{SM}_{ij}(r)\,,
\end{align}
where the combination of scalar coefficients is
\begin{align}
Z^{SS}_{\alpha\beta;ij} = (g^{LL}+g^{RL}-g^{LR}-g^{RR})_{\alpha;ij}(g^{LL}+g^{RL}-g^{LR}-g^{RR})_{\beta;ij}\,,
\end{align}
and the function $I^{SM}_{ij}(r)$ is also given in Appendix \ref{sec:integrals}. 

We see that the Dirac potential depends on the distance as $1/r^{5}$ only when the neutrino currents are of opposite chirality -- when they are the both left- or right-handed the potential becomes suppressed as $m_{\nu}^2/r^{3}$. This suppression does not occur for Majorana neutrinos --  the potential scales as $1/r^{5}$ for any combination of the coefficients $g^{XY}_{\alpha;ij}$.

\subsection{Tensor non-standard interactions}
\label{sec:tensorpotential}

We now derive the neutrino-mediated potential in the presence of a \textit{tensor} non-standard interaction. In our framework these are the operators in Eq.~\eqref{eq:Leff1} with the coefficients $h^{LL}$ and $h^{RR}$ normalised to the Fermi constant $G_{F}$. We first focus on the case of a tensor interaction at one vertex and a SM CC or NC interaction at the other, as shown Fig. \ref{fig:RHpotential}.

The spectral function can be determined in this scenario according to
\begin{align}
\label{eq:rhoVTgeneral}
\rho_{\alpha\beta}^{VT}(t)=-\frac{G_{F}^2}{\pi m_{\alpha}m_{\beta}}\sum_{i,j=1}^{N}\sum_{X,Y,Z=L,R}\bigg\{c^{XL}_{\alpha;ij}\,h^{YZ}_{\beta;ij}&\,\mathcal{H}^{\alpha\beta}_{\mu\nu\rho}\,\mathrm{disc}(\mathcal{N}_{ij}^{\mu\nu\rho}) + (\alpha,\beta)\bigg\}\,,
\end{align}
where the sum is over the possible chiralities of the external fermion currents ($X$ and $Y$) and the neutrino current of the tensor operator ($Z$). We have again taken into account that the tensor interaction may be at either vertex by adding an identical contribution with ($\alpha\leftrightarrow\beta$). The Majorana case is treated in the same way as previous subsections.

The discontinuity of $\mathcal{N}_{ij}^{\mu\nu\rho}$ in the Dirac case is
\begin{align}
\begin{aligned}
\mathrm{disc}(\mathcal{N}_{ij}^{\mu\nu\rho}) =\frac{\Lambda^{1/2} (q^2,m_{i}^2,m_{j}^2)}{8\pi}
\,&\frac{i m_{i}}{q^2}\,\big(g^{\mu\nu}q_{\rho}-g^{\mu\rho}q_{\nu}\mp i\varepsilon^{\mu\nu\rho\sigma}q_{\sigma}\big)\\
&\times\,\bigg(1-\frac{m_{i}^2-m_{j}^2}{q^2}\bigg)\,\Theta\big(q^2-(m_{i}+m_{j})^2\big)\,,
\label{eq:discontinuityVT}
\end{aligned}
\end{align}
and the external fermion bilinear product is
$\mathcal{H}^{\alpha\beta}_{\mu\nu\rho}=\,[\gamma_{\mu}\,\mathbb{P}_{X}]_{\alpha}[\,\sigma_{\nu\rho}\,\mathbb{P}_{Y}]_{\beta}$.

Contracting these factors and using the non-relativistic limits in Appendix \ref{sec:nonrel}, we obtain the spectral function in the Dirac case
\begin{align}
\begin{aligned}
\rho^{VT}_{\alpha\beta,D}(t)=-\frac{1}{1+\delta_{\alpha\beta}}\frac{G_{F}^2}{2\pi^2}&\sum_{i,j=1}^{3}\Theta\big(t-(m_{i}+m_{j})^2\big)~m_{i}\,\Lambda^{1/2}(t,m_{i}^2,m_{j}^2)\\
&\times \bigg\{X_{\alpha\beta;ij}^{VT}\,(\boldsymbol{\sigma}_{\beta}\cdot\mathbf{q})+iY_{\alpha\beta;ij}^{VT}(\boldsymbol{\sigma}_{\alpha}\times\boldsymbol{\sigma}_{\beta})\cdot\mathbf{q}+(\alpha,\beta)\bigg\}F_{ij}^{\Delta}(t)\,,
\end{aligned}
\end{align}
while in the Majorana case we obtain
\begin{align}
\begin{aligned}
\rho^{VT}_{\alpha\beta,M}(t)=-\frac{1}{1+\delta_{\alpha\beta}}\frac{G_{F}^2}{2\pi^2}&\sum_{i,j=1}^{N}\Theta\big(t-(m_{i}+m_{j})^2\big)~(m_{i}-m_{j})\,\Lambda^{1/2}(t,m_{i}^2,m_{j}^2)\\
&\times \bigg\{X_{\alpha\beta;ij}^{VT}\,(\boldsymbol{\sigma}_{\beta}\cdot\mathbf{q})+iY_{\alpha\beta;ij}^{VT}\,(\boldsymbol{\sigma}_{\alpha}\times\boldsymbol{\sigma}_{\beta})\cdot\mathbf{q}+(\alpha,\beta)\bigg\}F_{ij}^{T}(t)\,.
\end{aligned}
\end{align}
The coefficients $X_{\alpha\beta;ij}^{VT}$ and $Y_{\alpha\beta;ij}^{VT}$ containing the dependence on the tensor coefficients are
\begin{align}
X_{\alpha\beta;ij}^{VT} &= (c^{LL}+c^{RL})_{\alpha;ij}(h^{LL}-h^{RR})_{\beta;ij}^*\,,\\
Y_{\alpha\beta;ij}^{VT} &= (c^{LL}-c^{RL})_{\alpha;ij}(h^{LL}+h^{RR})_{\beta;ij}^*\,.
\end{align} 
The function $F^{\Delta}_{ij}$ is the same as Eq.~\eqref{eq:FDelta} and $F^{T}_{ij}$ is given by
\begin{align}
F_{ij}^{T}(t)=\frac{1}{t}\left(1-\frac{(m_{i}+m_{j})^2}{t}\right)\,.
\end{align}

The spectral functions above contain terms proportional to the parity-violating spin operators $\mathcal{O}'_{9}=\boldsymbol{\sigma}_{\alpha}\cdot\mathbf{q}$, $\mathcal{O}'_{10}=\boldsymbol{\sigma}_{\beta}\cdot\mathbf{q}$ and $\mathcal{O}_{11}=(\boldsymbol{\sigma}_{\alpha}\times\boldsymbol{\sigma}_{\beta})\cdot\mathbf{q}$. We can again take the components of the spectral functions multiplying these operators and evaluate the functions $\mathcal{V}'_{9}(r)$, $\mathcal{V}'_{10}(r)$ and $\mathcal{V}'_{11}(r)$ of Eq. \eqref{eq:simplefourier}. From these we use Eq. \eqref{eq:VS9and10} and
\begin{align}
\label{eq:VS11}
\mathcal{V}_{11}(r)&=\frac{i}{r^2}\,(\boldsymbol{\sigma}_{\alpha}\times\boldsymbol{\sigma}_{\beta})\cdot\hat{\boldsymbol{r}}\,\bigg(1-r\frac{d}{dr}\bigg)\mathcal{V}^{'}_{11}(r)\,,
\end{align}
to derive the full vector-tensor potential in the Dirac case
\begin{align}
\begin{aligned}
V^{VT}_{\alpha\beta,D}(r) = -\frac{1}{1+\delta_{\alpha\beta}}{G^2_{F} \over {4\pi^3r^4}}\sum^3_{i,j=1} m_i\,\bigg\{&iX_{\alpha\beta;ij}^{VT}\,(\boldsymbol{\sigma}_{\beta}\cdot \hat{\boldsymbol{r}})\\
&-Y_{\alpha\beta;ij}^{VT}\,(\boldsymbol{\sigma}_{\alpha}\times\boldsymbol{\sigma}_{\beta})\cdot\hat{\boldsymbol{r}}+(\alpha,\beta)\bigg\}\,J^{\Delta}_{ij}(r)\,,
\end{aligned}
\end{align}  
and in the Majorana case
\begin{align}
\begin{aligned}
V^{VT}_{\alpha\beta,M}(r)  = -\frac{1}{1+\delta_{\alpha\beta}}{G^2_{F} \over {4\pi^3r^4}}\sum^N_{i,j=1} (m_{i}-m_{j})\,\bigg\{&iX_{\alpha\beta;ij}^{VT}\,(\boldsymbol{\sigma}_{\beta}\cdot \hat{\boldsymbol{r}})\\
&-Y_{\alpha\beta;ij}^{VT}\,(\boldsymbol{\sigma}_{\alpha}\times\boldsymbol{\sigma}_{\beta})\cdot\hat{\boldsymbol{r}}+(\alpha,\beta)\bigg\}\,J^{T}_{ij}(r)\,,
\end{aligned}
\end{align}  
where the dimensionless functions $J^{\Delta}_{ij}(r)$ and $J^{T}_{ij}(r)$ are given in Appendix \ref{sec:integrals}.

We note that these potentials, like the vector-scalar potentials of the previous section, scale as $m_{\nu}/r^{4}$. They are similarly contain only parity-violating spin operators. The difference between the potentials for Dirac and Majorana neutrinos arises from the different $r$-dependence of the functions $J^{\Delta}_{ij}(r)$ and $J^{T}_{ij}(r)$. Finally, we see that the diagonal elements in the $i,j$ sum vanish for Majorana neutrinos.

\subsection{Comparison of potentials}
\label{sec:comparepotentials}

\begin{figure}[t!]
	\centering
	\includegraphics[width=0.49\textwidth]{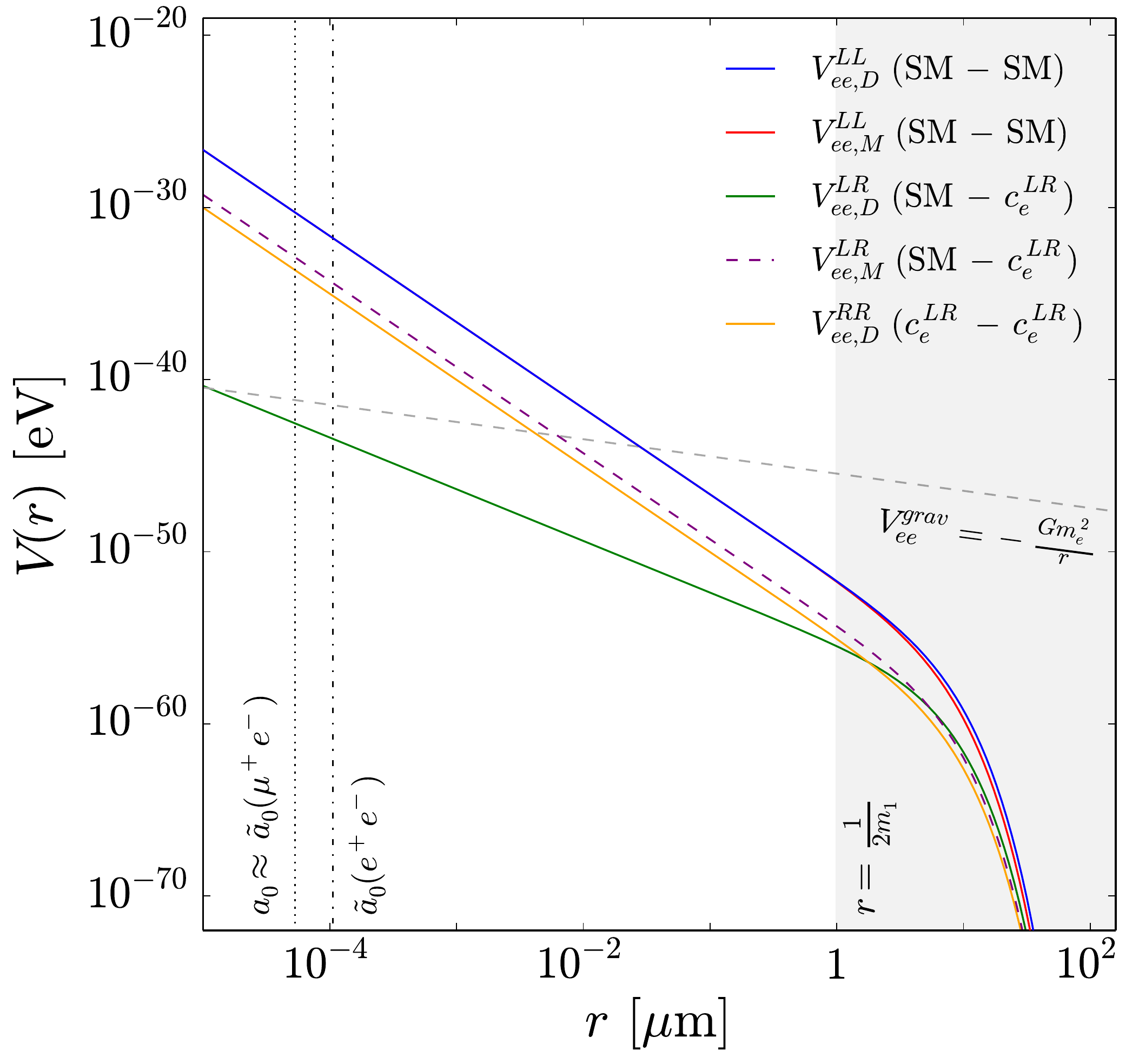}
	\includegraphics[width=0.49\textwidth]{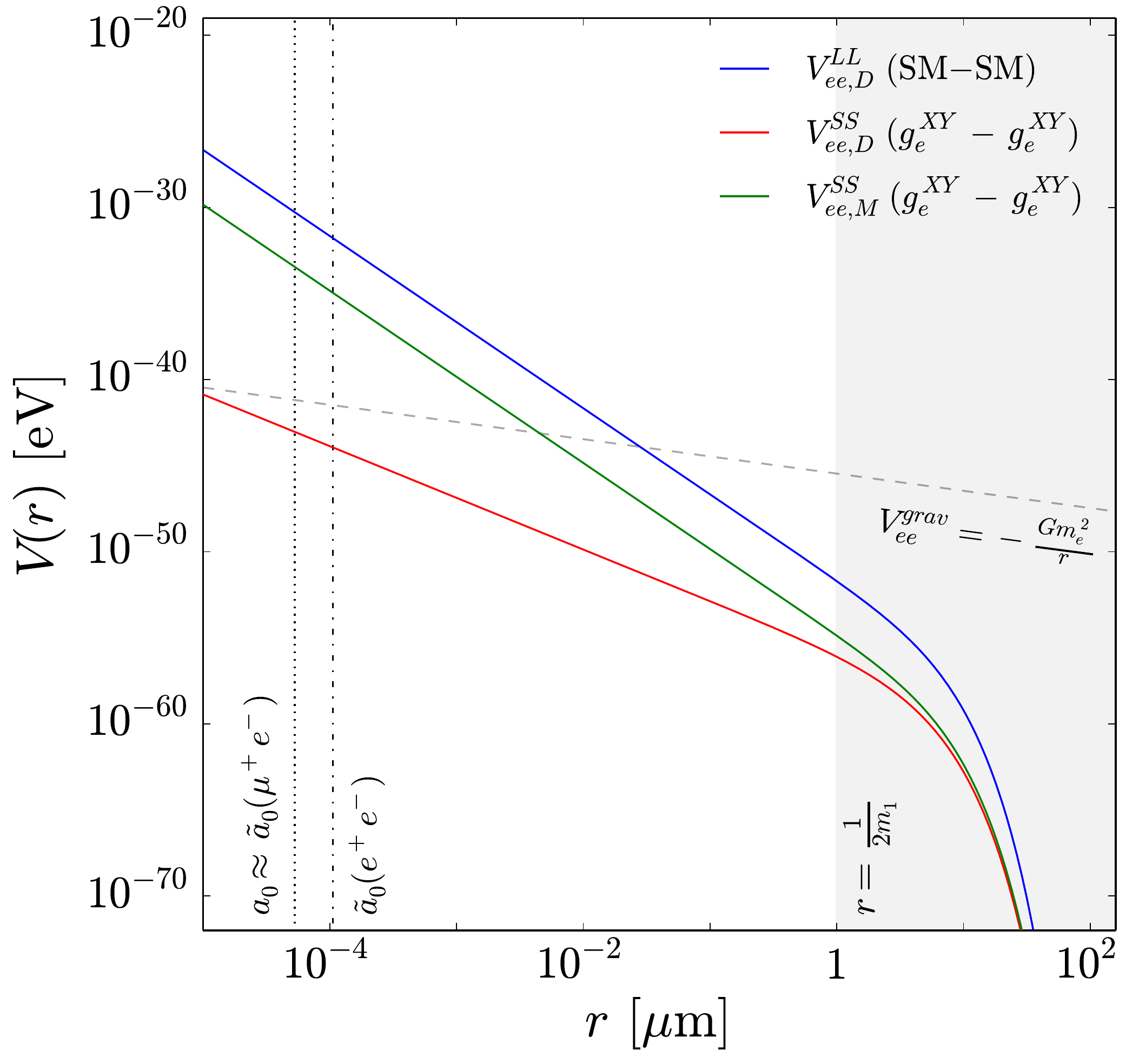}
	\caption{\textit{Left}: Spin-independent parts of the potentials $V^{LL}_{\alpha\beta}(r)$, $V^{LR}_{\alpha\beta}(r)$ and $V^{RR}_{\alpha\beta}(r)$ plotted for positronium ($e^{-}e^{+}$) and for the exchange of either three light active Dirac ($D$) or Majorana ($M$) neutrinos with $m_{1} = 0.1$ eV and NO mixing parameters. $V^{LL}_{ee}(r)$ is calculated with SM CC and NC interactions at each vertex, while $V^{LR}_{ee}(r)$ and $V^{RR}_{ee}(r)$ assume non-zero non-standard coefficients $c_{ee;ij}^{LR} \equiv c_{e}^{LR}\delta_{ij}= 10^{-2}\delta_{ij}$. The solid (dashed) lines indicate a positive (negative) potentials. \textit{Right}: Spin-independent parts of the potentials $V^{SS}_{ee}(r)$ in the Dirac and Majorana cases compared to $V^{LL}_{ee}(r)$, using $g_{ee;ij}^{XY}\equiv g_{e}^{XY}\delta_{ij}=10^{-2}\delta_{ij}$ and for a single combination of $X,Y = L,R$. In both plots the neutrino-mediated potentials are compared with the gravitational potential $V^{grav}_{ee}(r)$ between the electron and positron.}
	\label{fig:potentialplot}
\end{figure}

In Fig.~\ref{fig:potentialplot} we compare a selection the potentials derived in the previous subsections. To the left of Fig.~\ref{fig:potentialplot} we plot the spin-independent parts of the vector-vector potentials $V^{LL}_{\alpha\beta}(r)$, $V^{LR}_{\alpha\beta}(r)$ and $V^{RR}_{\alpha\beta}(r)$ for positronium ($e^{-}e^{+}$) and for either three Dirac or Majorana neutrinos. The potential $V^{LL}_{ee}(r)$ is calculated using SM values for the factors $X^{LL}_{ee;ij}$ and $Y^{LL}_{ee;ij}$ in Eq.~\eqref{eq:XYSM}. The potential $V^{LR}_{ee}(r)$ has a single SM vertex and is interpreted as a correction to $V^{LL}_{ee}(r)$ in the Majorana case, though we plot it separately. The potential $V^{RR}_{ee}(r)$ is derived from two non-standard right-handed neutrino currents. We set $m_{1}=0.1$ eV and take normal-ordered (NO) values of the mixing angles $\theta_{12}= 33.8\degree$, $\theta_{23}= 48.6\degree$ and $\theta_{13}= 8.6\degree$, the CP phase $\delta = 221\degree$ and the mass splittings $\Delta m_{21}^2=7.55\cdot10^{-5}~\mathrm{eV}^2$ and $\Delta m_{31}^2=2.5\cdot10^{-3}~\mathrm{eV}^2$. We set the non-standard coefficients to be $c^{LR}_{ee;ij} \equiv c^{LR}_{e}\delta_{ij}=10^{-2}\delta_{ij}$, i.e. only non-zero for diagonal $i,j$.

We first note the small difference between Dirac and Majorana potentials $V^{LL}_{ee,D}(r)$ and $V^{LL}_{ee,M}(r)$ . In the short-range limit $r\ll 1/2m_{1}$ the potentials are identical while in the long-range limit $r\gg 1/2m_{1}$ the Majorana potential is slightly smaller the Dirac potential, in agreement with the results of Ref.~\cite{Segarra:2020rah}. The potentials are generally seen to fall off as $1/r^5$ until the neutrinos become non-relativistic around $r\sim 1/2m_{1}$ and the potentials are exponentially suppressed. We see that the potentials are many orders of magnitude smaller than the gravitational potential $V_{ee}^{grav}(r)$ between the electron and positron. We also note the large difference between the Dirac and Majorana potentials $V^{LR}_{ee,D}(r)$ and $V^{LR}_{ee,M}(r)$ -- while the Dirac potential is slightly larger than the Majorana potential in the long-range limit, in the short-range limit the former scales as $1/r^{3}$ and is suppressed by two powers of the neutrino masses while the latter scales as $1/r^{5}$ and is unsuppressed. This is because the Majorana potential is interpreted as a correction to the SM potential $V^{LL}_{ee,M}(r)$ and thus scales in the same way. $V^{LR}_{ee,M}(r)$ is around two orders of magnitude smaller than $V^{LL}_{ee,D}(r)\approx V^{LL}_{ee,M}(r)$ due to the suppression from $c^{LR}_{e}=10^{-2}$. The potential $V^{RR}_{ee}(r)$ is shown just for the Dirac case -- because it contains two factors of $c^{RL}_{e} = 10^{-2}$ it is seen to be below $V^{LR}_{ee}(r)$. To the right  of Fig.  \ref{fig:potentialplot} we plot the scalar-scalar potentials for Dirac and Majorana neutrinos $V^{SS}_{ee,D}(r)$ and $V^{SS}_{ee,M}(r)$ and compare them to the spin-independent part of $V^{LL}_{ee}(r)$ and the gravitational potential $V_{ee}^{grav}(r)$. We choose a scalar coefficient $g_{ee;ij}^{XY}\equiv g_{e}^{XY}\delta_{ij}=10^{-2}\delta_{ij}$ to be non-zero for a single choice of the chiralities $X,Y$ -- looking at Eqs. \eqref{eq:SSdirac} and \eqref{eq:SSmaj} we see that the surviving terms of the Dirac potential scale in the short-range limit as $1/r^{3}$ while for the Majorana potential as $1/r^{5}$, as can be seen in the diagram.

\section{Atomic Spectroscopy}
\label{sec:spec}

There are a number of ways to probe exotic long-range forces over a range of distances. Starting at the macroscopic scale, precision torsion balance experiments adopt the method originally used by Cavendish to measure the gravitational constant $G$. Theories looking to resolve the discrepancy between the observed dark energy density $\rho_{\mathrm{d}}\approx 3.8~\mathrm{keV}/\mathrm{cm}^3$ and the theoretical prediction from quantum field theory (a factor of $\sim10^{120}$ larger) predict Yukawa violations or power-law modifications of the gravitational force at length-scales of $r\sim1~\mu \mathrm{m} - 1~ \mathrm{mm}$ \cite{Adelberger:2003zx}. These and other theories involve extra time \cite{Dvali:1999hn} and space \cite{Antoniadis:1998ig} dimensions and new scalar and vector mediators such the axion \cite{Ferrer:1998}, dilaton \cite{Kaplan:2000hh}, dark photon and $Z'$ \cite{Heeck:2014zfa}, all of which can alter the typical $1/r$ scaling of the gravitational potential and break the weak equivalence principle. Torsion balance experiments have excluded a region in the $|\alpha|-\lambda$ parameter space of the Yukawa-type parametrisation of deviations from the $1/r$ potential \cite{Kapner:2006si,Hoyle:2004,Spero:1980,Hoskins:1985,Long:2003,Chiaverini:2003,Smullin:2005,Adelberger:2007,Lee:2020}. Other experiments probing macroscopic distances have used optical levitation \cite{Rider:2016xaq,Jaffe:2016fsh} and atom interferometry \cite{Sabulsky:2018jma}. Finally, experiments using polarised electrons have been able to constrain macroscopic spin-dependent potentials \cite{Heckel:2013,Terrano:2015sna}.

As can be seen in Fig. \ref{fig:SMpotential}, the neutrino-mediated potentials fall off exponentially for $r\gtrsim 1~\mu\mathrm{m}$, roughly corresponding to the Compton wavelength of the lightest neutrino with $m_{1}=0.1$ eV. For point sources such as an electron and positron the associated forces are many orders of magnitude smaller than their gravitational attraction. In theory this can be overcome by using neutral aggregate matter with a coherent weak charge, boosting the strength of the neutrino-mediated force with respect to the gravitational force \cite{Segarra:2020rah}. It remains to be seen if torsion balance experiments can overcome the strong effect of the Earth's gravity to measure this. Another method is to measure the pressure exerted on two parallel plates by the Casimir-like force induced by the neutrino potential \cite{Costantino:2020bei}. Current experiments are however $\sim20$ orders of magnitude below the required sensitivity to measure the neutrino contribution.

To attain a greater sensitivity to the neutrino-mediated potentials one must therefore go to smaller distances where the potentials can be seen to exceed the gravitational potential in Fig.~\ref{fig:potentialplot}. The most stringent measurements come from nuclear and atomic spectroscopy probing $r\sim1~\mathrm{fm}$ and $r\sim 1~\mathrm{\AA}$, respectively. We outline some of the methods explored in the literature.

Atomic spectroscopy of heavy atomic species ($Z \gg 1$) might appear to be the most suitable method for probing the spin-independent part of the neutrino-mediated potential thanks to the coherent scaling of the nucleus -- going up roughly with the number of neutrons $N \gg 1$. The spin-dependent part on the other hand acts incoherently because nuclear pairing interactions leave the ground state nucleus with at most two unpaired nucleon spins. However, the complexity of many-electron interactions in heavy atoms makes the theoretical predictions for transitional frequencies inadequate for the current experimental precision. One can instead measure the isotope shift -- the difference in atomic splittings for different isotopes -- in systems such as $\mathrm{Ca}^{+}$ by observing a non-linearity in the King plot \cite{King:63}. This has been used to constrain models with $Z'$ bosons, exotic Higgs bosons and chameleon particles \cite{Delaunay:2017,Frugiuele:2017,Berengut:2017zuo,Flambaum:2017onb} and more recently the neutrino-mediated potential \cite{Stadnik:2017yge}.

A relevant probe at nuclear length scales is the binding energy of the deuteron $D^{+}$, a bound state of a proton and a neutron. One can model the binding energy with a spherical potential well with an infinitely repulsive inner hard core in order to find the radial wave-function of the system. This in turn can be used to calculate the expectation value of the neutrino-mediated potential and the shift to the binding energy. The difference in the measured  \cite{Kessler:1999zz} and predicted \cite{Entem:2003ft, Ekstrom:2015rta} binding energies has been used to constrain the neutrino-mediated potential \cite{Stadnik:2017yge}.

The sensitivity of simple atomic-like systems such as positronium ($e^{-}e^{+}$) and muonium ($e^{-}\mu^{+}$) to the neutrino-mediated potential may be more promising than the deuteron and other nuclear probes. As we will see, for these leptonic systems the characteristic cut-off scale (below which the $r$-dependence of the potential no longer holds) is provided by the cut-off of validity of the EFT, not the charge-radius of the nucleon or nucleus in semi-leptonic systems like hydrogen ($e^{-}p$), deuterium ($e^{-}\mathrm{D}^{+}$) or their muonic counterparts ($\mu^{-}p$ and $\mu^{-}\mathrm{D}^{+}$). At present the best measured splittings of these systems are the $1S-2S$ and ground state hyperfine splittings. These splittings have also been predicted to high accuracy and used as precision tests of QED. For example, the dominating Dirac, radiative, recoil and radiative-recoil QED corrections to the Fermi expression of the ground state hyperfine splitting $E_{F}$ have been calculated up to orders $\alpha^{2}(Z\alpha)^{2}E_{F}$ \cite{Eides:2018rph,Eides:2000xc,Mohr:2015ccw}. Smaller weak \cite{Eides:1996,Asaka:2018qfg} and hadronic corrections \cite{Nomura:2012sb} have also been calculated. The EW corrections have been calculated for the muonium hyperfine splitting to next-to-leading-order \cite{Asaka:2018qfg}.

We will follow the same approach as Ref.~\cite{Stadnik:2017yge} which derives the shifts to energy level splittings using the expectation value of the position-space potential $V(r)$. Using the experimental and SM-predicted values for the $1S-2S$ and hyperfine splittings of positronium and muonium, we will use the predicted shifts from the exotic neutrino-mediated potentials to put upper bounds on the non-standard coefficients $c^{XY}$, $g^{XY}$ and $h^{XY}$.

\subsection{Shifts to atomic energy levels}

The small shift to an atomic energy level due an exotic force can be calculated to first order in perturbation theory by taking the expectation value of the associated potential $V(r)$,
\begin{align}
\delta E =  -\big\langle V(r)\big\rangle =-\big\langle n\,^{2S+1}L_{J}| V(r)|n\,^{2S+1}L_{J}\big\rangle\,,
\end{align}
where $n \,^{2S+1}L_{J}$ labels the unperturbed atomic state with $n$ the principal quantum number $n$, $S$ the total spin, $L=\{S,P,D,...\}$ the total orbital angular momentum and $J$ the total angular momentum. Shifts to the $1S-nS$ and $n$-hyperfine splittings are respectively
\begin{align}
\label{eq:energyshifts}
\begin{aligned}
\delta E^{1S-nS}&=\delta E(n \,^{3}S_{1}) - \delta E(1 \,^{3}S_{1})\,,\\
\delta E^{n\textendash\mathrm{hfs}}&=\delta E(n \,^{3}S_{1}) - \delta E(n \,^{1}S_{0})\,.
\end{aligned}
\end{align}

The average of the potential over the atomic quantum numbers is the position-space integral
\begin{align}
\label{eq:Vaverage}
\big\langle V(r)\big\rangle_{n,\ell,m} =\int d^{3}\,\boldsymbol{r}\,\Psi^{*}_{n,\ell,m}(\boldsymbol{r})\,V(r)\,\Psi_{n,\ell,m}(\boldsymbol{r})
\end{align}
where $\Psi_{n,\ell,m}(r)$ is the atomic wave-function. For the two-body systems we are considering, $\Psi_{n,\ell,m}(\boldsymbol{r})=R_{n,\ell}(r)Y_{\ell,m}(\theta,\phi)$ is the separable solution of the Schr\"odinger equation with the Coulomb potential $V_C(r) = -Z\alpha/r$. 

We will be comparing the shifts induced by exotic potentials depending differently on $r$. For example, the SM CC and NC induced potential $V^{LL}_{\alpha\beta}(r)$ in Eq.~\eqref{eq:VLL} scales as $1/r^{5}$ in the short-range limit, while the right-handed current induced potential $V^{LR}_{\alpha\beta}(r)$ for Dirac neutrinos scales as $1/r^{3}$. Assuming that $V(r)$ is only a function of $r$ (and not $\theta$ and $\phi$) the integration over the spherical harmonic component $Y_{\ell,m}(\theta,\phi)$ is unity and the average over the hydrogen-like radial wave-function $R_{n,\ell}(r)$ for general $r$-dependence is,
\begin{align}
\label{eq:rint}
\Big\langle \frac{1}{r^d}\Big\rangle_{n,\ell}=\int_{r_c}^{\infty}dr ~r^{2-d}~\big(R_{n,\ell}(r)\big)^2\,,
\end{align}
where $r_{c}$ is a lower cut-off on the distance corresponding to an upper cut-off scale of validity for the four-fermion EFT. For SM CC and NC interactions this distance is around the inverse $Z$ boson mass and we define $r_{c}= 1/m_{Z}= 1.097\times 10^{-11}~\mathrm{eV}^{-1}$. We can write the Fermi coupling in terms of this length scale using
\begin{align}
\label{eq:GFreplace}
G_{F}=\frac{\pi\alpha}{\sqrt{2}s_{W}^2c_{W}^2m_{Z}^2}\equiv A^2r_{c}^2\,,~~A=\left(\frac{\pi\alpha}{\sqrt{2}s_{W}^2c_{W}^2}\right)^{1/2}\,,
\end{align}

This distance scale could be different for a non-standard effective interaction mediated by a particle with a mass above or below the EW scale -- a $Z'$ for example. In this case the distance cut-off is $r'_{c}=1/m_{Z'}$. This mediator may also interact with the SM via a coupling $g'$. Comparing this to the normalisation of the effective interaction to the Fermi coupling,
\begin{align}
G_{F}\,c^{XY}=\frac{g^{\prime 2}}{m_{Z'}^2}\equiv g'^2 r_{c}^{\prime 2}\,.
\end{align}
Depending on whether the new physics is above or below the EW scale, or strongly or weakly coupled, the lower distance scale of validity $r_{c}'$ compares to the SM Fermi cut-off $r_c$ as
\begin{align}
r_{c}^{\prime 2}=\frac{A^2}{g'^2}\,c^{XY}\,r_{c}^2 = \frac{M^2_{Z}}{M^2_{Z'}}\,r_{c}^2\,.
\end{align}
While this discussion is valid for an EFT with point-like particles, for a semi-leptonic system the cut-off $r_{c}$ must take into account the finite size of the nucleon or nucleus -- e.g. for a proton $r_{c} = r_{0}A^{1/3}$ with $r_{0}\approx 1.2$ fm.

We can now integrate Eq.~\eqref{eq:rint} using the hydrogen-like radial wave-function,
\begin{align}
R_{n, l}(r)=\sqrt{\frac{(n-l-1) !}{2 n(n+l) !}\left(\frac{2Z}{n \tilde{a}_{0}}\right)^{3}} e^{-\frac{2Zr}{n \tilde{a}_{0}}}\left(\frac{2Z r}{n \tilde{a}_{0}}\right)^{l} L_{n-l-1}^{2 l+1}\left(\frac{2 Zr}{n \tilde{a}_{0}}\right)\,,
\end{align}
where $L^{j}_{k}(x)$ is the associated Laguerre function and $\tilde{a}_{0}$ is the reduced Bohr radius of the system with reduced mass $m_{r}$,
\begin{align}
\tilde{a}_{0}=\frac{1}{m_{r}\alpha}=\left(\frac{m_{f_{\alpha}}+m_{f_{\beta}}}{m_{f_{\alpha}}m_{f_{\beta}}}\right)\frac{1}{\alpha}\,.
\end{align}
For hydrogen this is the standard Bohr radius $\tilde{a}_{0}\approx a_{0} = 1/(m_{e}\alpha)$. For different values of $d$ in Eq. \eqref{eq:rint} and expanding in $r_{c}$ we obtain
\begin{align}
\label{eq:r_average}
\begin{aligned}
\Big\langle \frac{1}{r^3}\Big\rangle_{n,\ell=0}&=\frac{4Z^3}{n^3\tilde{a}^3_0}\bigg[A_{n}-\gamma_E-\ln\left(\frac{2Zr_c}{na_0}\right)\bigg]+\mathcal{O}\left(\frac{r_{c}}{\tilde{a}_{0}^4}\right)\,,\\
\Big\langle \frac{1}{r^4}\Big\rangle_{n,\ell=0}&=\frac{4Z^3}{n^3r_c\tilde{a}_0^3}+\mathcal{O}\left(\frac{1}{\tilde{a}_{0}^4}\right)\,,\\
\Big\langle \frac{1}{r^5}\Big\rangle_{n,\ell=0}&=\frac{2Z^3}{n^3r_c^2\tilde{a}_0^3}+\mathcal{O}\left(\frac{1}{r_{c}\tilde{a}_{0}^4}\right)\,.
\end{aligned}
\end{align}
Here the parameter $A_{n}$ is given by
\begin{align}
A_{n}=\sum_{j=1}^{n-1}C_{jj}^n~(2j-1)!+\sum_{k>j=0}^{n-1}C_{jk}^n~(j+k-1)!\,,
\end{align}
with
\begin{align}
C^n_{jk}=\frac{1}{j!k!}\frac{(-1)^{j+k}[(n-1)!]^2}{(n-1-j)!(1+j)!(n-1-k)!(1+k)!}\,.
\end{align}

To compute the average in Eq.~\eqref{eq:Vaverage} we must also take the angular average of the spin-dependent terms in $V(r)$ -- for example the factors $\boldsymbol{\sigma}_{\alpha}\cdot\boldsymbol{\sigma}_{\beta}$ and $\left(\boldsymbol{\sigma}_{\alpha} \cdot \hat{\boldsymbol{r}}\right)\left(\boldsymbol{\sigma}_{\beta} \cdot \hat{\boldsymbol{r}}\right)$ in $V^{LL}_{\alpha\beta}(r)$ and $V^{LR}_{\alpha\beta}(r)$. Firstly, as we will be only considering $n\,^{2S+1}S_{J}$ states for the $1S-nS$ and $n$-hyperfine splittings, the following equality holds for $\ell=0$
\begin{align}
\big\langle \left(\boldsymbol{\sigma}_{\alpha} \cdot \hat{\boldsymbol{r}}\right)\left(\boldsymbol{\sigma}_{\beta} \cdot \hat{\boldsymbol{r}}\right)\big\rangle_{\ell=0}=\frac{1}{3}\big\langle\boldsymbol{\sigma}_{\alpha}\cdot\boldsymbol{\sigma}_{\beta}\big\rangle_{\ell=0}\,.
\end{align}
In order to determine the hyperfine splitting between singlet and triplet configurations of external particle spins we must also evaluate the spin dot-product $\big\langle\boldsymbol{\sigma}_{\alpha}\cdot\boldsymbol{\sigma}_{\beta}\big\rangle_{s}$ in these cases. These are $\big\langle\boldsymbol{\sigma}_{\alpha}\cdot\boldsymbol{\sigma}_{\beta}\big\rangle_{s=0} = -3$ (singlet) and $\big\langle\boldsymbol{\sigma}_{\alpha}\cdot\boldsymbol{\sigma}_{\beta}\big\rangle_{s=1} = 1$ (triplet).

The averages of the parity-odd potentials $V^{VS}_{\alpha\beta}(r)$ and $V^{VT}_{\alpha\beta}(r)$ -- which depend on the spin operators $\boldsymbol{\sigma}_{\alpha}\cdot\hat{\boldsymbol{r}}$, $\boldsymbol{\sigma}_{\beta}\cdot\hat{\boldsymbol{r}}$ and $(\boldsymbol{\sigma}_{\alpha}\times\boldsymbol{\sigma}_{\beta})\cdot\hat{\boldsymbol{r}}$ -- vanish. However, the potentials can induce transitions between different $\ell$ states similar to an electric dipole moment. While not the focus of this section, atomic and molecular EDM experiments have been used to constrain spin-dependent, $P$- and $T$-violating potentials induced by axion exchange in Ref.~\cite{Stadnik:2017hpa}. In the context of the neutrino-mediated force, Ref.~\cite{Ghosh:2019dmi} has suggested probing atomic parity violation by measuring the optical rotation of light as it passes through vaporised atoms.
 
The expectation value of the SM-induced potential $V^{LL}_{\alpha\beta}(r)$ can now be written as
\begin{align}
\label{eq:VLLexp}
\big\langle V_{\alpha\beta}^{LL}(r)\big\rangle = \frac{G_{F}^2}{4\pi^3}&\sum_{i,j=1}^{N}\bigg\{X^{LL}_{\alpha\beta;ij}\,\bigg\langle\frac{I_{ij}^{D(M)}(r)}{r^5}\bigg\rangle\nonumber\\
&~\,-Y^{LL}_{\alpha\beta;ij}\left[ \left\langle\frac{(\boldsymbol{\sigma}_{\alpha}\cdot\boldsymbol{\sigma}_{\beta})\,J_{ij}^{D(M)}(r)}{r^5}\right\rangle-\left\langle\frac{(\boldsymbol{\sigma}_{\alpha}\cdot\hat{\boldsymbol{r}})(\boldsymbol{\sigma}_{\beta}\cdot\hat{\boldsymbol{r}}) \,J_{ij}^{V}(r)}{r^5}\right\rangle\right]\bigg\}\,.
\end{align}
Recall that the functions $I_{ij}^{D(M)}(r)$, $J_{ij}^{D(M)}(r)$ and $J_{ij}^{V}(r)$ are exponentially suppressed for distances greater than the Compton wavelength of the neutrinos $r\gg 1/(2m_{i})$. For $r\ll 1/(2m_{i})$ on the other hand the neutrino masses can be neglected $m_{i}\approx m_{j}\approx 0$ and the functions take constant values. For atomic spectroscopy measurements the relevant distance scale (the reduced Bohr radius $\tilde{a}_{0}$) is safely in this regime. In this limit the averages in Eq.~\eqref{eq:VLLexp} become
\begin{align}
\bigg\langle\frac{I_{ij}^{D(M)}(r)}{r^5}\bigg\rangle_{n,\ell=0}&\approx\frac{2Z^3}{n^3r_c^2\tilde{a}_0^3}\,,\\
\left\langle\frac{(\boldsymbol{\sigma}_{\alpha}\cdot\boldsymbol{\sigma}_{\beta})\,J_{ij}^{D(M)}(r)}{r^5}\right\rangle_{n,\ell=0} &\approx \frac{3}{2}\frac{2Z^3}{n^3r_c^2\tilde{a}_0^3}\,\big\langle\boldsymbol{\sigma}_{\alpha}\cdot\boldsymbol{\sigma}_{\beta}\big\rangle\,,\\
\left\langle\frac{(\boldsymbol{\sigma}_{\alpha}\cdot\hat{\boldsymbol{r}})(\boldsymbol{\sigma}_{\beta}\cdot\hat{\boldsymbol{r}}) \,J_{ij}^{V}(r)}{r^5}\right\rangle_{n,\ell=0}&\approx\frac{5}{6}\frac{2Z^3}{n^3r_c^2\tilde{a}_0^3}\,\big\langle\boldsymbol{\sigma}_{\alpha}\cdot\boldsymbol{\sigma}_{\beta}\big\rangle\,,
\end{align}
giving the average for the potential
\begin{align}
\big\langle V_{\alpha\beta}^{LL}(r)\big\rangle_{n,\ell=0} &\approx \frac{G_{F}^2}{2\pi^3}\frac{Z^3}{n^3r_c^2\tilde{a}_0^3}\left\{X_{\alpha\beta;\nu\nu}^{LL}-\frac{2}{3}Y_{\alpha\beta;\nu\nu}^{LL}\big\langle\boldsymbol{\sigma}_{\alpha}\cdot\boldsymbol{\sigma}_{\beta}\big\rangle\right\}\,,
\end{align}
where $X^{LL}_{\alpha\beta;\nu\nu}\equiv\sum\limits^{N}_{i,j}X^{LL}_{\alpha\beta;ij}$ and $Y^{LL}_{\alpha\beta;\nu\nu}\equiv\sum\limits^{N}_{i,j}Y^{LL}_{\alpha\beta;ij}$. The same expression would be obtained in the single neutrino simplification.

Computing the average of the potential $V^{LR}_{\alpha\beta}(r)$ for Dirac neutrinos in Eq.~\eqref{eq:VLRdirac} requires evaluating the average of the factor $m_{i}m_{j}I_{ij}^{LR}(r)/r^3$. For distances $r\ll1/(2m_{i})$,
\begin{align}
\bigg\langle\frac{m_{i}m_{j}I_{ij}^{LR}(r)}{r^3}\bigg\rangle_{n,\ell=0}&\ll\bigg\langle\frac{I_{ij}^{D(M)}(r)}{r^5}\bigg\rangle_{n,\ell=0}\,,
\end{align}
which illustrates that the potential is too suppressed to be a useful probe of the non-standard coefficients $c^{LR}$ and $c^{RR}$. In the Majorana case the potential $V^{LR}_{\alpha\beta}(r)$ has the same $r$-dependence as $V^{LL}_{\alpha\beta}(r)$ and so
\begin{align}
\label{eq:VLRave}
\big\langle V_{\alpha\beta}^{LR}(r)\big\rangle_{n,\ell=0} &\approx -\frac{G_{F}^2}{2\pi^3}\frac{Z^3}{n^3r_c^2\tilde{a}_0^3}\left\{X_{\alpha\beta;\nu\nu}^{LR}-\frac{2}{3}Y_{\alpha\beta;\nu\nu}^{LR}\big\langle\boldsymbol{\sigma}_{\alpha}\cdot\boldsymbol{\sigma}_{\beta}\big\rangle\right\}\,,
\end{align}
where $X^{LR}_{\alpha\beta;\nu\nu}\equiv\sum\limits^{N}_{i,j}X^{LR}_{\alpha\beta;ij}$ and $Y^{LR}_{\alpha\beta;\nu\nu}\equiv\sum\limits^{N}_{i,j}Y^{LR}_{\alpha\beta;ij}$. We remind the reader that $X^{LR}_{\alpha\beta;ij}=-X^{LL}_{\alpha\beta;ij}$ and $Y^{LR}_{\alpha\beta;ij}=-Y^{LL}_{\alpha\beta;ij}$ for Majorana neutrinos, so any new physics contribution to $c_{\alpha\beta;ij}^{LR}=-c_{\alpha\beta;ji}^{LL}$ is added on top of the SM contribution as in Eq. \eqref{eq:SMcorrection}.

\begin{table}[]
	\setlength{\tabcolsep}{2pt}
	\begin{tabular}{ccccc}
		\hline\hline
		&\multicolumn{2}{c}{$\delta E_{\alpha\beta}^{1S-2S}$ [mHz]} & \multicolumn{2}{c}{$\delta E_{\alpha\beta}^{n\textendash\mathrm{hfs}}$ [mHz]}  \\ \hline
		System \,$(f_{\alpha},f_{\beta})$& $V_{\alpha\beta}^{LL}$~(SM-SM) & $\mathrm{exp}-\mathrm{theory}$&$V_{\alpha\beta}^{LL}$~(SM-SM)&$\mathrm{exp}-\mathrm{theory}$ \\ \hline
		Positronium~$(e,e)$&  $10$ &$-5.8(3.3)\cdot 10^{9}$ $^{(\mathrm{1})}$&$57$&$2.2(1.9)\cdot10^{9}$ $^{(\mathrm{a})}$  \\
		Muonium~$(e,\mu)$& $13$ & $5.2(9.9)\cdot 10^{9}$ $^{(2)}$ &$-150$&$-1.1(5.2)\cdot10^{5}$ $^{(\mathrm{b})}$ \\
		Hydrogen~$(e,p)$& \multirow{2}{*}{$-4.1\cdot10^{-4}$} & \multirow{2}{*}{$-1.4(0.5)\cdot10^{7}$ $^{(3)}$} &$-1.2\cdot 10^{-3}$&$-1.1(0.1)\cdot10^{7}$ $^{(\mathrm{c})}$ \\
		Deuterium~$(e,\mathrm{D}^{+})$&   &  &$-1.7\cdot10^{-6}$&$1.4(0.1)\cdot10^6$ $^{(\mathrm{d})}$\\
		Muonic hydrogen~$(\mu,p)$& $2.2\cdot10^{3}$ & -- &$-1.0\cdot 10^{3}$&$-9.4(1.5)\cdot10^{12}$ $^{(\mathrm{e})}$ \\
		Muonic deuterium~$(\mu,\mathrm{D}^{+})$&$-550$ & -- &$-13$&$-1.1(2.1)\cdot10^{12}$ $^{(\mathrm{f})}$ \\ \hline
		\multicolumn{5}{l}{\footnotesize{$^{(\mathrm{1})}$\cite{1998physics...5020M}, \cite{Czarnecki:1999mw}~, $^{(2)}$\cite{PhysRevLett.84.1136}, \cite{Frugiuele:2019drl},~ 
				$^{(\mathrm{3})}$\cite{PhysRevLett.104.233001} (Deuterium$-$Hydrogen $1S-2S$ Isotope Shift)}} \\
		\multicolumn{5}{l}{\footnotesize{$^{(\mathrm{a})}$\cite{Ishida:2013waa}, \cite{Czarnecki:1999mw} (1S-hfs),~ $^{(\mathrm{b})}$\cite{Tanaka2019}, \cite{Eides:2018rph} (1S-hfs),~ $^{(\mathrm{c})}$\cite{2017NatCo...815749D}, \cite{2016PhRvA..93b2513H} (1S-hfs),~ $^{(\mathrm{d})}$\cite{PhysRevA.5.821,2004PhRvA..70f2503K},~ \cite{Karshenboim:2001xw,Faustov:2003pr,Tomalak:2017lxo} (2S-hfs)}}\\
		\multicolumn{5}{l}{\footnotesize{$^{(\mathrm{e})}$\cite{Antognini2013}, \cite{2013AnPhy.331..127A,Tomalak:2017lxo} (2S-hfs),~ $^{(\mathrm{f})}$\cite{Pohl1:2016xoo}, \cite{Krauth:2015nja} (2S-hfs)}}\\\hline \hline
	\end{tabular}
	\caption{Predicted shifts to the $1S-2S$ and hyperfine splittings of two-body systems ($f_{\alpha},\,f_{\beta}$) due to the SM-induced neutrino-mediated potential $V^{LL}_{\alpha\beta}(r)$. The potential is mediated by three light active neutrinos with $m_{1} = 0.1$ eV and the other masses and mixings determined in the NO case. Where possible we compare these to the differences between the experimentally-measured and theoretically-predicted values for these splittings. Uncertainties in these values are calculated by adding the experimental and theoretical uncertainties in quadrature. References for experimental and theoretical values are given in the footnotes below the table respectively.}
	\label{tab:SMshifts}
\end{table}

Computing the shifts to the $1S-2S$ and $n$-$\mathrm{hfs}$ splittings in Eq. \eqref{eq:energyshifts} due to $V^{LL}_{\alpha\beta}(r)$ gives
\begin{align}
\label{eq:1S-2S}
\delta E^{1S-nS}_{\alpha\beta}&= -\frac{G_{F}^2}{2\pi^3}\frac{Z^3}{r_c^2\tilde{a}_0^3}\left(\frac{1}{n^3}-1\right)\left\{X_{\alpha\beta;\nu\nu}^{LL}-\frac{2}{3}Y_{\alpha\beta;\nu\nu}^{LL}\big\langle\boldsymbol{\sigma}_{\alpha}\cdot\boldsymbol{\sigma}_{\beta}\big\rangle_{s=1}\right\}\,,\\
\label{eq:HFS}
\delta E^{n\textendash\mathrm{hfs}}_{\alpha\beta}&=\frac{G_{F}^2}{3\pi^3}\frac{Z^3 }{n^3r_c^2\tilde{a}_0^3}\,Y^{LL}_{\alpha\beta;\nu\nu}\bigg\{\big\langle\boldsymbol{\sigma}_{\alpha}\cdot\boldsymbol{\sigma}_{\beta}\big\rangle_{s=1}-\big\langle\boldsymbol{\sigma}_{\alpha}\cdot\boldsymbol{\sigma}_{\beta}\big\rangle_{s=0}\bigg\}\,,
\end{align}
which can be written as
\begin{align}
\label{eq:1S-2S2}
\delta E^{1S-nS}_{\alpha\beta}&= -\frac{\alpha G_{F}}{2\sqrt{2}\pi^2 c_{W}^2s_{W}^2}\frac{Z^3}{\tilde{a}_0^3}\left(\frac{1}{n^3}-1\right)\left\{X_{\alpha\beta;\nu\nu}^{LL}-\frac{2}{3}Y_{\alpha\beta;\nu\nu}^{LL}\right\}\,,\\
\label{eq:HFS2}
\delta E^{n\textendash\mathrm{hfs}}_{\alpha\beta}&=\frac{2\sqrt{2}\alpha G_{F}}{3\pi^2c_{W}^2s_{W}^2}\frac{Z^3 }{n^3\tilde{a}_0^3}\,Y^{LL}_{\alpha\beta;\nu\nu}\,.
\end{align}
where we have made use of Eq. \eqref{eq:GFreplace} and $r_{c}=1/m_{Z}$. Recalling that $\tilde{a}_{0}= 1/(m_{r}\alpha)$ we can see that the shifts to the splittings are of order $\alpha^{4} G_{F}m_{r}^3$. As a specific example, the shift to the hyperfine splitting between two charged leptons $\ell_{\alpha}$ and $\ell_{\beta}$ is predicted to be
\begin{align}
\label{eq:leptonhfs}
\delta E^{1\textendash\mathrm{hfs}}_{\ell_{\alpha}\ell_{\beta}}=\frac{2\sqrt{2}\alpha^4 G_{F}m_{r}^{3}}{3\pi^2c_{W}^2s_{W}^2}\sum\limits^3_{i,j}(U_{\alpha i}^*U_{\alpha j}+g_{A}^{\ell}\delta_{ij})(U_{\beta i}U_{\beta j}^*+g_{A}^{\ell}\delta_{ij})\,,
\end{align}
while the hyperfine splitting between a charged lepton $\ell_{\alpha}$ and nucleon or nucleus $\mathcal{N}$ is
\begin{align}
\label{eq:leptonnucleonhfs}
\delta E^{1\textendash\mathrm{hfs}}_{\ell_{\alpha}\mathcal{N}}&=\frac{2\sqrt{2}\alpha^4 G_{F}m^{3}_{r}}{3\pi^2c_{W}^2s_{W}^2}\sum\limits^3_{i}\big(|U_{\alpha i}|^2+g_{A}^{\ell}\big)\,g^{\mathcal{N}}_{A}\,.
\end{align}

Using Eqs. \eqref{eq:leptonhfs} and \eqref{eq:leptonnucleonhfs} we list in Table \ref{tab:SMshifts} the predicted shifts to the $1S-2S$ and $n$-hyperfine splittings due to the SM-induced potential $V^{LL}_{\alpha\beta}(r)$ for a range of leptonic and semi-leptonic two-body systems. For both the $1S-2S$ and $n$-hyperfine splitting we compare the predicted shift in units of mHz to the differences between experimentally measured and theoretically predicted values (from QED, hadronic and first-order weak contributions). 

We see in each case that the expected shift from $V^{LL}_{\alpha\beta}(r)$ is much smaller than the experiment-theory discrepancy. We see that the leptonic systems provide larger shifts in relation to the experiment-theory difference compared to the semi-leptonic systems. This is mainly due to the cut-off $r_{c}=1/m_{Z}$ being two orders of magnitude smaller than the charge radii of the proton and deuteron. Of the leptonic systems we see that the experimental measurements of the muonium splittings are the most precise -- the predicted shift due to neutrino-exchange $\delta E_{\alpha\beta}^{n-\mathrm{hfs}} \approx -150$ mHz is around three orders of magnitude smaller than the experiment-theory difference. The hyperfine splitting of muonium is therefore the most stringent probe.

\begin{table}[]
	\setlength{\tabcolsep}{6.5pt}
	\begin{tabular}{ccc}
		\hline\hline
		System \,$(f_{\alpha},f_{\beta})$&$c^{LR(RR)}$~[$V^{LR}_{\alpha\beta,M}(r)$] &$g^{XY}$~[$V^{SS}_{\alpha\beta,M}(r)$]\\\hline
		Positronium~$(e,e)$&   $c^{LR}_{e}<5.7\cdot10^{7}$& $g^{XY}_{e}<7.2\cdot 10^{3}$ \\
		Muonium~$(e,\mu)$& $c^{LR}_{e},\,c^{LR}_{\mu}<3.6\cdot10^{2}$& $g^{XY}_{e}\cdot g^{XY}_{\mu}<5.9\cdot10^{6}$  \\
		H\,$(e,p)$\,/\,D\,$(e, \mathrm{D}^{+})$&  $c^{LR}_{e}<1.5\cdot10^{9},\,c_{p}^{LR}<5.5\cdot10^{9}$ &$g^{XY}_{e}\cdot\,(g^{XY}_{D}-6.48g^{XY}_{p})<1.6\cdot10^{10}$\\\hline\hline
	\end{tabular}
	\caption{Upper limits on the non-standard coefficients $c^{RL}$ and $c^{RR}$ probed by the right-handed current potential $V^{LR}_{\alpha\beta}(r)$ and coefficients $g^{XY}$ (for $X,Y = L,R$) probed by the scalar-scalar potential $V^{SS}_{\alpha\beta}(r)$. $c^{LR}$ and $c^{RR}$ are constrained from the hyperfine splittings of the systems $(f_{\alpha},\,f_{\beta})$, while the $g^{XY}$ are constrained from the $1S-2S$ splittings.  To avoid a helicity-suppression we assume three light active Majorana neutrinos with $m_{1}=0.1$ eV and NO masses and mixings. For simplification we take $c^{LR(RR)}_{\alpha;ij}\equiv c^{LR(RR)}_{\alpha}\delta_{ij}$ and $g^{XY}_{\alpha;ij}\equiv g^{XY}_{\alpha}\delta_{ij}$.}
	\label{tab:coefficientconstraints}
\end{table}

The shift to the hyperfine splitting from the potential $V^{LR}_{\alpha\beta}(r)$ in the Majorana case can be found from Eq. \eqref{eq:VLRave} to be 
\begin{align}
\label{eq:VLRlepton}
\delta E^{1\textendash\mathrm{hfs}}_{\ell_{\alpha}\ell_{\beta}}&=\frac{1}{1+\delta_{\alpha\beta}}\frac{4G_{F}^2}{3\pi^3r_c^{2}\tilde{a}_0^3}\sum\limits^3_{i,j}\Big\{(U_{\alpha i}^*U_{\alpha j}+g_{A}^{\ell}\delta_{ij})\,c^{LR}_{\beta;ij}\,+\,(\alpha, \beta)\Big\}\,,
\end{align}
which depends linearly on the coefficient $c_{\beta\beta}^{LR}$. This potential relies on two effective interactions -- one from SM CC and NC interactions and the other from a non-standard interaction -- which may possess different cut-offs $r_{c}$ and $r'_{c}$. The cut-off appearing in Eq.~\eqref{eq:VLRlepton} must therefore be the larger of these two scales. For simplicity we assume that the new physics arises around the EW scale $m_{Z}$ and therefore $r_{c}'\approx r_{c}$ regardless of the exotic coupling strength $g'$. This allows us to rewrite Eq.~\eqref{eq:VLRlepton} as 
\begin{align}
\label{eq:VLRlepton2}
\delta E^{1\textendash\mathrm{hfs}}_{\ell_{\alpha}\ell_{\beta}}&=\frac{1}{1+\delta_{\alpha\beta}}\frac{2\sqrt{2}\alpha^{4} G_{F}m_{r}^3}{3\pi^2c_{W}^2s_{W}^2}\sum\limits^3_{i,j}\Big\{(U_{\alpha i}^*U_{\alpha j}+g_{A}^{\ell}\delta_{ij})\,c^{LR}_{\beta;ij}\,+\,(\alpha, \beta)\Big\}\,,
\end{align}
We now use Eq.~\eqref{eq:VLRlepton2} to compute the predicted shift as a function of the non-standard coefficient $c^{LR}$. To simplify the sum over mass eigenstates ($i,j$) we take the coefficients to be diagonal in the mass basis, i.e. $c^{LR}_{\alpha;ij}=c^{LR}_{\alpha}\delta_{ij}$. We now write the inequality relating this predicted shift to the difference between experimental and theoretical values,
\begin{align}
\label{eq:VLRlepton3}
|\delta E^{1\textendash\mathrm{hfs}}_{\ell_{\alpha}\ell_{\beta}}|<|\delta E^{1\textendash\mathrm{hfs},\,\mathrm{exp}}_{\ell_{\alpha}\ell_{\beta}}-\delta E^{1\textendash\mathrm{hfs},\,\mathrm{theory}}_{\ell_{\alpha}\ell_{\beta}}|\,
\end{align}
and rearrange to put an upper bound on the value of $c^{LR}_{\alpha}$. We note that $c^{LR}_{\alpha;ij}$ gets a contribution from the SM for Majorana neutrinos -- however, even if we include this contribution it is too small to affect the upper bound derived for the non-standard coefficient. In Table \ref{tab:coefficientconstraints} we give the constraints from positronium (on $c_{e}^{LR}$), muonium ($c_{e}^{LR}$ and $c_{\mu}^{LR}$) and hydrogen ($c_{e}^{LR}$ and $c_{p}^{LR}$). Muonium gives the most stringent upper bounds while the constraints from positronium and hydrogen are five orders of magnitude worse.

We now consider the scalar-scalar potential $V^{SS}_{\alpha\beta}(r)$ in Eq.~\eqref{eq:VSmajorana} which does not depend on the external particle spins -- we must instead use the $1S-2S$ splitting to derive upper bounds on the coefficients $g^{XY}$. For this potential this splitting is found to be
\begin{align}
\delta E^{1S-2S}_{\ell_{\alpha}\ell_{\beta}}&=\frac{21\alpha^4G_{F}m_{r}^2}{32\sqrt{2}\pi^2s_{W}^2c_{W}^2}\sum\limits_{i,j}^{3}g_{\alpha;ij}^{XY}g_{\beta;ij}^{XY}\,.
\end{align}
Taking again the differences in the experimental and theoretical values for the splittings, we derive the upper bounds on the coefficients in Table \ref{tab:coefficientconstraints}. While positronium can put an upper bound on $g_{e}^{XY}$, muonium can only constrain the product of coefficients $g_{e}^{XY}\cdot g_{\mu}^{XY}$. Moreover, we use the experimentally measured difference between the deuterium and hydrogen $1S-2S$ splittings and therefore compare $\delta E^{1S-2S,\,d}_{\ell_{\alpha}\ell_{\beta}}-\delta E^{1S-2S,\,p}_{\ell_{\alpha}\ell_{\beta}}$. This can only constrain the linear combination $g^{XY}_{e}\cdot\,(g^{XY}_{D}-6.48g^{XY}_{p})$. We now see that the constraints from positronium and muonium are roughly comparable while those from hydrogen/deuterium remain less stringent.

\section{Neutrino electromagnetic properties}
\label{sec:mag}

In this final section we will derive long-range potentials induced by possible non-standard electromagnetic properties of the neutrinos. The long-range potential induced by a neutrino magnetic dipole moment has been studied before, for example in Ref. \cite{Lusignoli:2010gw}.

Interactions between the neutrino mass-eigenstate fields and the electromagnetic field can be written generically as
\begin{align}
\mathcal{L}_{\nu\nu\gamma}(x)=-j^{\mu}_{\mathrm{eff}}(x)A_{\mu}(x)=-\big[\bar{\nu}_{i}(x)\,\Lambda_{ij}^{\mu}\,\nu_j(x)\big]A_{\mu}(x)\,,
\end{align}
where $\Lambda_{ij}^{\mu}$ is a $4\times 4$ matrix in spinor space which may contain space-time derivatives. To calculate an amplitude for the $\nu\nu\gamma$ vertex one must take the matrix element of the neutrino current $j^{\mu}_{\mathrm{eff}}(x)$ between initial and final neutrino states,
\begin{align}
\bra{\nu_{i}(p_i)}j^\mu_{\mathrm{eff}}(x)\ket{\nu_{j}(p_{j})} &= e^{i(p_{j}-p_{i})\cdot x}\bra{\nu_{i}(p_i)}j^\mu_{\mathrm{eff}}(0)\ket{\nu_{j}(p_{j})}\nonumber\\
&=e^{i(p_{i}-p_{j})\cdot x}\,\bar{u}(p_i)\,\Gamma_{ij}^{\mu}(q)\,u(p_{j})\,,
\end{align}
where the vertex function $\Gamma_{ij}^{\mu}$ depends only on the momentum-transfer $q$. It is parametrised as
\begin{align}
\Gamma^{\mu}_{ij}(q)=-i\sigma^{\mu\nu} q_{\nu}\big(f^{M}_{ij}(q^2)+if^{E}_{ij}(q^2)\gamma_{5}\big)+\bigg(\gamma^{\mu}-\frac{q^{\mu} \slashed q}{q^{2}}\bigg)\big(f^{Q}_{ij}(q^2)+f^{A}_{ij}(q^2)q^2\gamma_{5}\big)\,.
\end{align}
The functions $f^{Q}_{ij}(q^2)$, $f^{A}_{ij}(q^2)$, $f^{M}_{ij}(q^2)$ and $f^{E}_{ij}(q^2)$ are the real charge, anapole moment, magnetic and electric dipole moment form factors, respectively. When coupling to a real photon with $q^2=0$, $f^{Q}_{ij}(0)=q_{ij}$, $f^{A}_{ij}(0)=a_{ij}$, $f^{M}_{ij}(0)=\mu_{ij}$ and $f^{E}_{ij}(0)=\epsilon_{ij}$ are the neutrino millicharge, anapole moment, magnetic and electric dipole moments, respectively. The above discussion is valid for Dirac neutrinos -- for Dirac antineutrinos the form factors become $\bar{f}^{Q}_{ij}=-f^{Q}_{ji}$, $\bar{f}^{A}_{ij}=f^{A}_{ji}$, $\bar{f}^{M}_{ij}=-f^{M}_{ji}$, $\bar{f}^{E}_{ij}=-f^{E}_{ji}$. For Majorana neutrinos we remember that the same electromagnetic process is described by two terms in the Lagrangian,
\begin{align}
\mathcal{L}_{\nu\nu\gamma}(x)=-j^{\mu}_{\mathrm{eff}}(x)A_{\mu}(x)=-\big[\bar{\nu}_{i}(x)\,\Lambda_{ij}^{\mu}\,\nu_j(x)+\bar{\nu}^c_{j}(x)\,\Lambda_{ji}^{\mu}\,\nu^c_i(x)\big]A_{\mu}(x)\,,
\end{align}
and therefore the matrix element becomes
\begin{align}
\bra{\nu(p_j)}j^\mu_{\mathrm{eff}}(x)\ket{\nu(p_{i})} =e^{i(p_{j}-p_{i})\cdot x}\bar{u}(p_j)\Big\{\Gamma_{ij}^{\mu}(q)+\mathcal{C}[\Gamma_{ji}^{\mu}(q)]^{T}\mathcal{C}^{\dagger}\Big\}u(p_{i})\,.
\end{align}
This enforces the constraints on the form factors: $f^{Q}_{ij}=-f^{Q}_{ji}$, $f^{A}_{ij}=f^{A}_{ji}$, $f^{M}_{ij}=-f^{M}_{ji}$ and $f^{E}_{ij}=-f^{E}_{ji}$. The diagonal elements of the real charge, magnetic and electric dipole form factors therefore vanish for Majorana neutrinos -- only the anapole moment form factor retains non-zero diagonal elements. All off-diagonal elements, or transition moments, can be non-zero depending on the relative CP phases $\eta_{i}$ of the neutrino mass eigenstates. If $\eta_{i}=\eta_{j}$ then the off-diagonal elements of the real charge and magnetic dipole moment vanish, whereas if $\eta_{i}=-\eta_{j}$ the off-diagonal elements of the anapole moment and electric dipole moment vanish.

In the low-energy effective field theory discussed in Sec. \ref{sec:model} one can generate electromagnetic properties for both Dirac and Majorana neutrinos. For Dirac neutrinos the following operator arises at dimension-six,
\begin{eqnarray}\label{eq:Diracmagmoment}
{\cal{L}}^{\nu\nu\gamma}_{\mathrm{eff}} & = & -\frac{\mu_{\rho\sigma}}{2}(\overline{\nu^{\prime}_{\rho R}}\sigma^{\mu\nu}
\nu_{\sigma L}^{\prime})F_{\mu\nu}+\mathrm{h.c.}\,,
\end{eqnarray}
which is written in the flavour-basis and where $F_{\mu\nu}$ is the electromagnetic field strength tensor. If they are instead Majorana, $\overline{\nu^{\prime}_{\rho R}} = \overline{\nu^{\prime c}_{\rho L}}$ and Eq.~\eqref{eq:Diracmagmoment} must instead arise at dimension-seven, because the two $\nu_{L}$ contained within $SU(2)_{L}$ doublets must contract with Higgs doublets. 

It is nonetheless possible for a magnetic moment term to arise at lower dimension when introducing the right-handed Majorana states $N_{R}$ to the SMEFT. This is the LNV operator at dimension-five in Eq. \eqref{eq:Leff5},
\begin{eqnarray}\label{eq:RHmagmoment}
{\cal{L}}_{\zeta} & = & -\frac{\zeta_{ss'}}{2}(\overline{N_{sR}^{\prime c}}\sigma^{\mu\nu}
N^{\prime}_{s'R})B_{\mu\nu}+\mathrm{h.c.}\,,
\end{eqnarray}
which is written in the EW basis, where $\zeta$ is an $n\times n$ matrix with $n$ the number of sterile states. We can rotate Eqs. \eqref{eq:Diracmagmoment} and \eqref{eq:RHmagmoment} to the mass basis in a similar way to Eq. \eqref{eq:Leff1},
\begin{align}
{\cal{L}}^{\nu\nu\gamma}_{\mathrm{eff}} &= -\frac{\mu_{ij}^{D(M)}}{2}\bar{\nu}_{i}\sigma^{\mu\nu}\mathbb{P}_{L}\nu_j F_{\mu\nu}\,+\,\mathrm{h.c.}\,,\label{eq:massbasis67}\\
{\cal{L}}_{\zeta} &= -\frac{1}{2}\bar{n}_{i}\sigma^{\mu\nu}(\zeta_{ij}\mathbb{P}_{R}+\zeta^*_{ij}\mathbb{P}_{L})n_j (c_{W} F_{\mu \nu}-s_{W} Z_{\mu \nu})\,,\label{eq:massbasis5}
\end{align}
where we re-iterate that $\nu = (\nu_{1},\nu_{2},\nu_{3})$ and $n = (\nu_{1},\nu_{2},\nu_{3},N_{1},N_{2},...)$. In the mass basis, $\mu^{D}$ and $\mu^{M}$ are $3\times 3$ matrices given by
\begin{align}
\mu^{D}_{ij} = \sum^{3}_{\rho,\sigma}\mu_{\rho\sigma}\tilde{U}^*_{\rho i}U_{\sigma j}\,,\quad
\mu^{M}_{ij} = \sum^{3}_{\rho,\sigma}\mu_{\rho\sigma} U^*_{\rho i}U_{\sigma j}\,.
\end{align}
and $\zeta$ is a $(3+n)\times (3+n)$ matrix given by
\begin{align}
\zeta_{ij} = \sum^{n}_{s,s'}\zeta_{ss'}\tilde{U}_{s i}\tilde{U}_{s' j}\,.
\end{align}
Splitting these in general complex dipole moments into real and imaginary parts as
\begin{align}
\mu^{D(M)}_{ij}=\hat{\mu}^{D(M)}_{ij}-i\hat{\epsilon}^{D(M)}_{ij}\,,\quad\zeta_{ij}=\hat{\mu}^{M}_{ij}+i\hat{\epsilon}^{M}_{ij}\,,
\end{align}
we obtain
\begin{align}
{\cal{L}}^{\nu\nu\gamma}_{\mathrm{eff}} &= -\frac{1}{2}\bar{\nu}_{i}\sigma^{\mu\nu}(\hat{\mu}_{ij}^{D(M)}+i\hat{\epsilon}_{ij}^{D(M)}\gamma_5)\nu_j F_{\mu\nu}\,+\,\mathrm{h.c.}\,,\\
{\cal{L}}_{\zeta} &= -\frac{1}{2}\bar{\nu}_{i}\sigma^{\mu\nu}(\hat{\mu}_{ij}^{M}+i\hat{\epsilon}_{ij}^{M}\gamma_5)\nu_j (c_{W} F_{\mu \nu}-s_{W} Z_{\mu \nu})\,,
\end{align}
and we now see that $\hat{\mu}^{D(M)}$ and $\hat{\epsilon}^{D(M)}$ correspond to the magnetic and electric dipole moments for Dirac (Majorana) neutrinos. For simplicity we will re-label the magnetic and electric dipole moments as $\mu^{D(M)}$ and $\epsilon^{D(M)}$. Considering the Fermi statistics in the Majorana case -- whether it be Eq. \eqref{eq:massbasis67} or \eqref{eq:massbasis5} -- it is clear that the matrices $\mu^{M}$ and $\epsilon^{M}$ are antisymmetric and have zero diagonal elements. If for Eqs. \eqref{eq:RHmagmoment} and \eqref{eq:massbasis5} we are again considering the type-I seesaw with $N=3+n$ massive neutrinos, we can split for example the magnetic dipole moment into
\begin{align}
\mu^{M}=\begin{pmatrix}
\mu^{M}_{\nu} & \mu^{M}_{\nu N} \\
-(\mu^{M}_{\nu N})^{T} & \mu^{M}_{N}\\
\end{pmatrix}\,,
\end{align}
where the antisymmetric matrices $\mu^{M}_{\nu}$ and $\mu^{M}_{N}$ contain the transition dipole moments for the light and heavy neutrino mass eigenstates respectively, while $\mu^{M}_{\nu N}$ contains the transition dipole moments between active and sterile states. From the form of the mixing matrix $\tilde{U}$ in Eq. \eqref{eq:seesawrotation} in the seesaw limit we can write $\mathcal{L}_{\zeta}$ explicitly in the mass basis as 
\begin{align}
\mathcal{L}_{\zeta}=\left(\bar{N} U_{N}^{T}-\bar{\nu} U_{\nu}^{T} \varepsilon\right) \sigma^{\mu \nu}\left(\zeta \mathbb{P}_{R}+\zeta^{*} \mathbb{P}_{L}\right)\left(U_{N} N-\varepsilon^{T} U_{\nu} \nu\right)\left(c_{W} F_{\mu \nu}-s_{W} Z_{\mu \nu}\right)\,.
\end{align}
It is clear from this that $\mu^{M}_{\nu}$ and $\mu^{M}_{\nu N}$ are suppressed by the factors $\varepsilon^2$ and $\varepsilon$ compared to $\mu^{M}_{N}$, respectively.

\begin{figure}[t!]
	\centering
	\includegraphics[width=0.24\textwidth]{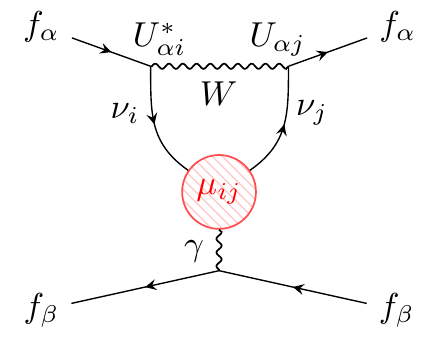}
	\includegraphics[width=0.24\textwidth]{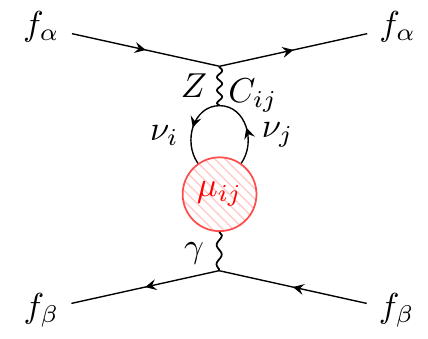}
	\includegraphics[width=0.24\textwidth]{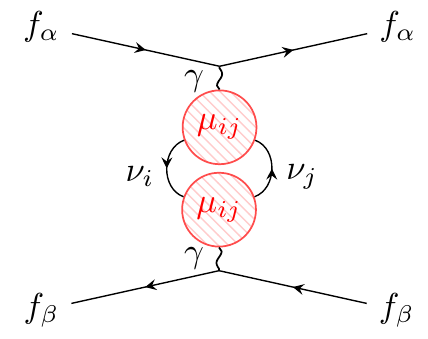}
	\caption{\textit{Left and Centre}: Diagrams depicting the exchange of two massive neutrinos between fermions $f_{\alpha}$ and $f_{\beta}$ with SM CC and NC interactions at one vertex and the exchange of a photon via a neutrino magnetic moment $\mu_{ij}$ at the other, leading to the vector-dipole potential $V^{V\gamma}_{\alpha\beta}(r)$. \textit{Right}: Diagram when a magnetic dipole moment is present at both vertices, resulting in the dipole-dipole potential $V^{\gamma\gamma}_{\alpha\beta}(r)$.}
	\label{fig:magmompotential}
\end{figure}

We now move on to consider the long-range potential for the processes shown in the left two Feynman diagrams of Fig. \ref{fig:magmompotential}. In these diagrams a pair of mass eigenstate neutrinos interacts via a SM CC and NC process at one vertex and via a photon at the other, coupled to the neutrino magnetic or electric dipole moment. The amplitude for this process is given by
\begin{align}
\label{eq:magmomamplitude}
-i\mathcal{M}_{\alpha\beta}=\frac{(-i e)}{4m_{\alpha}m_{\beta}}\left(-i\frac{4G_{F}}{\sqrt{2}}\right)\sum_{i,j=1}^{N}\sum_{X=L,R}\bigg\{c^{XL}_{\alpha\alpha;ij}\,\mathcal{H}^{\alpha\beta}_{\mu\nu}\,\mathcal{N}_{ij}^{\mu\nu}~+~(\alpha, \beta)\bigg\}\,,
\end{align}
where the neutrino loop factor $\mathcal{N}_{ij}^{\mu\nu}$ is given by
\begin{equation}
\mathcal{N}^{\mu \nu}_{ij} =  \frac{1}{q^2}\int {d^4 k \over (2 \pi)^4} {{\rm Tr} [ (-i\sigma^{\mu\rho}q_{\rho}\mu_{ij}^{D(M)}+\sigma^{\mu\rho}q_{\rho}\epsilon_{ij}^{D(M)})
	(\slashed{q}+\slashed{k} +m_j)\gamma^\nu \mathbb{P}_{L}(\slashed{k} +m_i) ] \over {(k^2-m_i^2)\,((q+k)^2-m_j^2)}}\,,
\label{eq:Nmagmom}
\end{equation}
and the product of external fermion bilinears is $\mathcal{H}^{\alpha\beta}_{\mu\nu}=\,[\gamma_{\mu}\,\mathbb{P}_{X}]_{\alpha}[\gamma_{\nu}]_{\beta}$, where the $X=L,R$ depends on the presence of a SM CC or NC interaction.

Taking the discontinuity of Eq. \eqref{eq:magmomamplitude} and using Eq.~\eqref{eq:Vsum}, we obtain
\begin{align}
V^{V\gamma}_{\alpha\beta,D}(r) = {\alpha G_{F} \over {8\sqrt{2}\pi^2r^3}}\frac{1}{m_{e}\mu_{B}}\sum^3_{i,j=1} \bigg\{(m_i+m_j)X^{V\gamma}_{\alpha\beta;ij}I^{S}_{ij}(r) - i(m_i-m_j)Y^{V\gamma}_{\alpha\beta;ij}I^{T}_{ij}(r)\bigg\}\,,
\end{align} 
in the Dirac case, where we have normalised by the Bohr magneton $\mu_{B}=e/2m_{e}$. Here,
\begin{align}
X^{V\gamma}_{\alpha\beta;ij} &= \frac{1}{1+\delta_{\alpha\beta}}\Big\{(c^{LL}+c^{RL})_{\alpha}+(c^{LL}+c^{RL})_{\beta}\Big\}\mu^{D}_{ij}\,\,,\\
Y^{V\gamma}_{\alpha\beta;ij} &=\frac{1}{1+\delta_{\alpha\beta}} \Big\{(c^{LL}+c^{RL})_{\alpha}+(c^{LL}+c^{RL})_{\beta}\Big\}\epsilon^{D}_{ij}\,\,,
\end{align}  
which take into account that the SM current can be at the interaction vertex of fermion $f_{\alpha}$ and magnetic (or electric) dipole moment the interaction vertex of $f_{\beta}$, or vice versa. The $1/(1+\delta_{\alpha\beta})$ factor again takes into account double counting if $\alpha=\beta$.

In the Majorana case we have
\begin{align}
V^{V\gamma}_{\alpha\beta,M}(r) = -{i\alpha G_{F}\over {8\sqrt{2}\pi^2r^3}}\frac{1}{m_e \mu_{B}}\sum^N_{i,j=1} &(m_i-m_j)\,Z^{V\gamma}_{\alpha\beta;ij}\,I^{T}_{ij}(r)\,,
\end{align}  
where
\begin{align}
Z^{V\gamma}_{\alpha\beta;ij} &= \frac{1}{1+\delta_{\alpha\beta}}\Big\{(c^{LL}+c^{RL})_{\alpha;ij}+(c^{LL}+c^{RL})_{\beta;ij}\Big\}\epsilon^{M}_{ij}\,\,.
\end{align}  
It can be seen that the magnetic moment does not contribute to the potential in the Majorana case -- this is simply a case of the whole amplitude vanishing when only the magnetic moment and axial part of $\mathbb{P}_{L}$ contribute. The first thing to observe in these potentials is that the $r$-dependence, $1/r^{3}$, is the same as for the right-handed current potential $V^{LR}_{\alpha\beta}(r)$ in the Dirac case. However, there is now a factor of $\alpha G_{F}$ instead of $G_{F}^2$ and the potential is now proportional to one power of the neutrino masses instead of two. For $\mu_{ij}\sim\mu_{B}$ and noting that $G_{F}^2m_{\nu}^2\ll \alpha G_{F} m_{\nu}/m_{e}$ for $m_{\nu}\sim 0.1$ eV, we see that the potential is far less suppressed than $V^{LR}_{\alpha\beta}(r)$ in the Dirac case.

\begin{figure}[t!]
	\centering
	\includegraphics[width=0.55\textwidth]{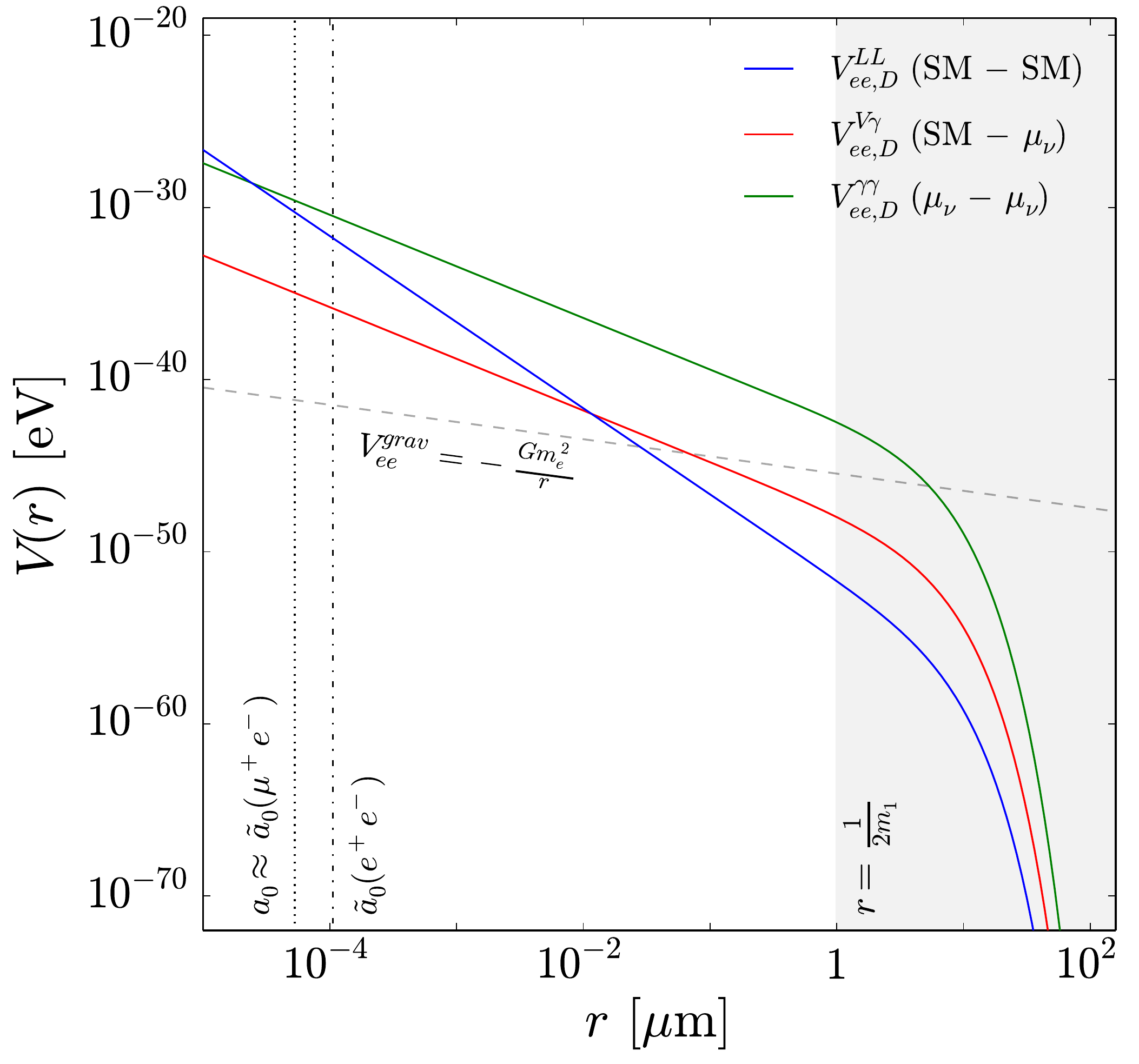}
	\caption{Neutrino-mediated potentials $V^{V\gamma}_{ee}(r)$ and $V^{\gamma\gamma}_{ee}(r)$ compared to the SM potential $V^{LL}_{ee}(r)$, plotted for positronium ($e^{-}e^{+}$) with the exchange of either three light active Dirac neutrinos with $m_{1} = 0.1$ eV and NO mixing parameters and using $\mu_{ij} \equiv \mu_{\nu}= 10^{-12}\,\mu_{B}$. These potentials are compared with the gravitational potential $V^{grav}_{ee}(r)$ between the electron and positron.}
	\label{fig:potentialmuplot}
\end{figure}

We can instead consider the process depicted by the Feynman diagram to the right of Fig.~\ref{fig:magmompotential}, where the two mediated neutrinos are coupled to both the external fermions by their magnetic or electric dipole moment. The dipole-dipole potential obtained in this case (valid for both Dirac and Majorana neutrinos) is
\begin{align}
V^{\gamma\gamma}_{\alpha\beta}(r) = {\alpha^2\over {12\pi r^3}}\frac{1}{m_e^2 \mu^2_{B}}\sum^3_{i,j=1} \bigg\{X^{\gamma\gamma}_{ij}I^{M\gamma}_{ij}(r)-Y^{\gamma\gamma}_{ij}I^{E\gamma}_{ij}(r)\bigg\}\,,
\end{align}  
where $X^{\gamma\gamma}_{ij}=\mu^{D(M)}_{ij}\mu^{D(M)*}_{ij}$ and $Y^{\gamma\gamma}_{ij}=\epsilon^{D(M)}_{ij}\epsilon^{D(M)*}_{ij}$. We see that there are two terms -- one for the presence of two magnetic dipole moments and the other for two electric dipole moments. The cross-term for a magnetic and electric dipole moment vanishes.

In Fig.~\ref{fig:potentialmuplot} we compare (for positronium) the spin-independent potentials $V^{V\gamma}_{\alpha\beta}(r)$ and $V^{\gamma\gamma}_{\alpha\beta}(r)$ to the spin-independent part of SM potential $V^{LL}_{\alpha\beta}(r)$. We take a non-zero value of the magnetic moment, $\mu_{ij}=\mu_{\nu}$ = $10^{-12}\,\mu_{B}$, and let the electric dipole moment vanish. We see, as expected, that the potentials scale as $1/r^{3}$ in the short-range limit $r\ll1/(2m_{i})$. However, unlike the potentials $V^{LR}_{\alpha\beta}(r)$ and $V^{SS}_{\alpha\beta}(r)$ they dominate over the SM potential for a wide range of distances.

\begin{table}[]
	\setlength{\tabcolsep}{7pt}
	\begin{tabular}{ccc}
		\hline\hline
		System \,$(f_{\alpha},f_{\beta})$&$\mu_{\nu}/\mu_{B}$~[$V^{\gamma\gamma}_{\alpha\beta,D}(r)$/$V^{\gamma\gamma}_{\alpha\beta,M}(r)$]&$\mu_{N}/\mu_{B}$ ~[$V^{\gamma\gamma}_{\alpha\beta,M}(r)$, two $N$] \\ \hline
		Positronium~$(e,e)$& $3.6\cdot10^{-2}~(4.4\cdot10^{-2})$&$7.6\cdot10^{-2}$ \\
		Muonium~$(e,\mu)$& $1.3\cdot10^{-2}~(1.5\cdot 10^{-2})$ &$2.6\cdot10^{-2}$ \\
		H\,$(e,p)$\,/\,D\,$(e, \mathrm{D}^{+})$&$2.7\cdot10^{-3}~(3.3\cdot10^{-3})$ &$5.7\cdot10^{-3}$ \\\hline \hline
	\end{tabular}
	\caption{Upper limits on the magnetic moment $\mu_{\nu}$ in units of the Bohr magneton probed by the dipole-dipole potential $V^{\gamma\gamma}_{\alpha\beta}(r)$, derived from the $1S-2S$ splittings of the systems $(f_{\alpha},\,f_{\beta})$. Equivalent limits apply for the electric dipole moment $\epsilon_{\nu}$. We assume three light active neutrinos with $m_{1}=0.1$ eV and NO masses and mixings. For Dirac neutrinos we take $\mu_{ij}\equiv\mu_{\nu}$, $\epsilon_{ij}\equiv\epsilon_{\nu}$, while for Majorana neutrinos $\mu_{ij}\equiv\mu_{\nu}(1-\delta_{ij})$, $\epsilon_{ij}\equiv\epsilon_{\nu}(1-\delta_{ij})$. We also derive upper limits on the heavy sterile neutrino magnetic moments $\mu_{ij}=\mu_{N}(1-\delta_{ij})$ for $i,\,j$ = $4,\,5$, i.e. introducing two heavy Majorana neutrinos in the type-I seesaw. Active neutrino magnetic moments $\mu_{\nu}\propto \varepsilon^2$ and active-sterile transition magnetic moments $\mu_{\nu N}\propto \varepsilon$ are neglected.}
	\label{tab:magconstraints}
\end{table}

Because the vector-dipole potential $V^{V\gamma}_{\alpha\beta}(r)$ is proportional to the neutrino masses and therefore is suppressed in the short-range limit, we focus instead on the shifts induced by the dipole-dipole potential $V^{\gamma\gamma}_{\alpha\beta}(r)$. Using the same procedure outlined as in Sec. \ref{sec:spec} to calculate the expectation value of the potential, we find
\begin{align}
\label{eq:Vgammagamma}
\big\langle V_{\alpha\beta}^{\gamma\gamma}\big\rangle_{n,\,\ell=0} &\approx -\frac{\alpha^2}{12\pi }\frac{1}{m_{e}^2\mu_{B}^2}\frac{4Z^3}{n^3\tilde{a}^3_0}\bigg[A_{n}-\gamma_E-\ln\left(\frac{2Zr_c}{na_0}\right)\bigg]\left\{X_{\alpha\beta;\nu\nu}^{\gamma\gamma}-Y_{\alpha\beta;\nu\nu}^{\gamma\gamma}\right\}\,,
\end{align}
where $X^{\gamma\gamma}_{\nu\nu}=\sum\limits_{i,j}^{3}\mu_{ij}^{D(M)}\mu_{ij}^{D(M)*}$ and $Y^{\gamma\gamma}_{\nu\nu}=\sum\limits_{i,j}^{3}\epsilon_{ij}^{D(M)}\epsilon_{ij}^{D(M)*}$. The potential is spin-independent so we must therefore use the $1S-2S$ splitting to put an upper bound on the neutrino magnetic and electric dipole moments -- this is found from Eq.~\eqref{eq:Vgammagamma} as $\delta E^{1S-2S}_{\alpha\beta} = \big\langle V_{\alpha\beta}^{\gamma\gamma}\big\rangle_{n=1,\,\ell=0}-\big\langle V_{\alpha\beta}^{\gamma\gamma}\big\rangle_{n=2,\,\ell=0}$.

In the second column of Table \ref{tab:magconstraints} we show the upper bounds on the magnetic moments when we assume $\mu_{ij}=\mu_{\nu}$ such that $X^{\gamma\gamma}_{\nu\nu}=9\mu_{\nu}^2$ (for three light Dirac neutrinos) derived from the positronium, muonium and difference in the deuterium and hydrogen $1S-2S$ splittings. In brackets is the upper bound when we assume there to be three light Majorana neutrinos, for which $\mu_{ij}=\mu_{\nu}(1-\delta_{ij})$ and $X^{\gamma\gamma}_{\nu\nu}=6\mu_{\nu}^2$. These limits also apply for the electric dipole moments. In the third column of Table \ref{tab:magconstraints} we consider the scenario where two heavy sterile Majorana neutrinos are introduced in the type-I seesaw. Here the magnetic moments for the light active neutrinos and the active-sterile transition magnetic dipole moments are suppressed as $\varepsilon^2$ and $\varepsilon$ respectively. We therefore take $\mu_{ij}\approx 0$ for all $i,j = 1\ldots 5$ apart from $\mu_{ij}= \mu_{N}(1-\delta_{ij})$, where $\mu_{N}$ is the transition dipole moment between the two sterile states. We again use the $1S-2S$ splittings of the different systems to put an upper bound on this parameter, shown in Table \ref{tab:magconstraints}.

\section{Conclusions}
\label{sec:con}
The exchange of force carriers are the basis of our understanding how structures are formed in nature. On scales larger than nuclei, the SM of particle physics incorporates the photon being largely responsible at the scales of atoms and larger. Beyond the SM, searches for massless or very light exotic mediators are being carried out in a large number of experiments and at different length scales, ranging from nuclear and atomic precision spectroscopy to the effects of fifths forces in astrophysics. Still within the SM, we already have more light particles, neutrinos, that are already known to be lighter than 0.1~eV but as fermions they cannot act as single exchange particles between matter particles. It is nevertheless possible that two neutrinos are simultaneously exchanged between two matter particles. In a Feynman diagram representation, this corresponds to a contribution at the first-loop order, see e.g. Fig.~\ref{fig:SMpotential} for those arising in the SM. Due to the weakness of neutrino interactions and the loop suppression, the effect is small but nevertheless it may be possible to probe the effect of such a SM neutrino exchange in (exotic) atoms, especially muonium \cite{Stadnik:2017yge}. 

We have here considered an EFT approach parametrising the effect of potential New Physics to analyse the long-range potential induced by the exchange of two neutrinos. This includes all possibilities for the relevant four-fermion contact interactions between two neutrinos and two charged leptons or quarks. We have calculated both the spin-independent and spin-dependent long-range potentials as well as for a neutrino magnetic moment. Using our results, we discuss the potential of probing the exchange potentials and the underlying effective operator couplings using state-of-the-art atomic and nuclear spectroscopy experiments high precision QED calculations. Normalising the operator coefficients $G_{\text{eff}}$ of the relevant four-fermion contact interactions relative to the standard Fermi constant $G_{\text{eff}} = G_Fc^{XY}$, we have found that the current precision in atomic spectroscopy is sensitive to coefficients as low as $c^{XY} = \mathcal{O}(10^2)$ for muonium. If the exchange is accompanied by a neutrino magnetic or electric dipole moment $\mu_\nu$, values of order $\mu_\nu = \mathcal{O}(10^{-2})~\mu_B$ are being probed.

We have worked in the low energy effective field theory approach to model both the exchange of SM EW bosons as well as any exotic contributions. In Ref.~\cite{Asaka:2018qfg}, the importance of the second order EW effects have been discussed which is usually ignored in the effective (Fermi theory) approach. Therein, the 1S hyperfine splitting energy shift of muonium was calculated in momentum space and in a gauge invariant fashion, including all relevant EW loop contributions, also those arising from electrons in the loop. For the SM case, they determine the same overall energy shift as in the present work as well as in \cite{Stadnik:2017yge}, and we therefore do not consider any other ultraviolet-complete scenarios. The limits on effective neutrino operators extracted from atomic spectroscopy can be used to constrain new physics scales. For example, the muonium 1S hyperfine splitting energy shift in the SM is of the order
\begin{align}
	|\delta E^{1\textendash\text{hfs}}|
	\approx 0.14 \,\alpha^4 m_e^3 G_F \approx 6\times 10^{-16}~\text{eV}\approx 150 ~\mathrm{mHz},
\end{align}
see Eq.~\eqref{eq:leptonhfs}. This compares with the current sensitivity of $|\delta E^{1\textendash\text{hfs}}| \lesssim 7\times 10^{-14}$~eV \cite{Tanaka2019, Eides:2018rph}. On the other hand, new physics at a scale $\Lambda_\text{NP}$, generating the relevant operators would induce a shift of order
\begin{align}
	|\delta E^{1\textendash\text{hfs}}|
	\approx \frac{\alpha^4 m_e^3}{\Lambda^2_\text{NP}} 
	\approx 10^{-13}\left(\frac{60~\text{GeV}}{\Lambda_\text{NP}}\right)^2~\text{eV},
\end{align}
and new physics scales close to the EW scale are currently being probed. Future advancements in experimental muonium spectroscopy \cite{Asaka:2018qfg} and QED precision calculations \cite{Eides:2013yxa, Eides:2017uoy} are expected to improve the sensitivity to $|\delta E^{1\textendash\text{hfs}}| \approx 10~\text{Hz} \approx 5\times 10^{-15}$~eV.\footnote{The sensitivity in atomic systems involving nuclei is expected to be much weaker as the lower distance cut-off $r \gtrsim 1$~fm means that the dependence on the new physics scale becomes $\propto 1/\Lambda^4_\text{NP}$ for $\Lambda_\text{NP} \gtrsim 100$~MeV.} While this will not improve on the existing limits from other processes as discussed in Sec.~\ref{sec:other-probes}, atomic scale probes have the advantage that the effective operator treatment is valid down to very low energy scales corresponding to the Bohr radius, $\Lambda_\text{NP} \gtrsim \alpha m_e \approx 3$~keV.

\begin{acknowledgments}
	PDB and FFD acknowledge support from the Science and Technology Facilities Council (STFC) via a Consolidated Grant (Reference ST/P00072X/1). CH acknowledges support from the DFG Emmy Noether Grant No. HA 8555/1-1. The authors would like to thank Quan Le Thien Minh and Dennis Krause for illuminating discussions.
\end{acknowledgments}

\begin{appendix}
	
\section{Comparison to other parametrisations}
\label{sec:compare}

In the low-energy effective field theory (LEFT) of the SM a parametrisation for general neutrino interactions similar to this work is given in Ref. \cite{Bischer:2019ttk}. The effective Lagrangian for NC-like operators is written as
\begin{align}
\begin{aligned}
\mathcal{L}^{\bar \nu \nu\bar f f}_{\mathrm{eff}} &=-\frac{G_{F}}{\sqrt{2}} \sum_{j=1}^{10}\ptwiddle{\epsilon}\hspace{-0.8em}\phantom{a}_{\alpha \beta \gamma \delta}\hspace{-2em}\phantom{a}^{j}\hspace{1.1em}\left(\bar{\nu}_{\alpha} \mathcal{O}_{j} \nu_{\beta}\right)\left(\bar{f}_{\gamma} \mathcal{O}_{j}^{\prime} f_{\delta}\right)\,,
\end{aligned}
\end{align}
where $f=\ell,u,d$, the fields are given in the flavour basis and $j$ runs over the ten possible Lorentz-invariant combinations of Dirac matrices in $\mathcal{O}_{j}$ and $\mathcal{O}'_{j}$ for chiral fermions -- analogous to the ten terms in Eq. \eqref{eq:Leff1}. The $\mathcal{O}_{j}$ and $\mathcal{O}'_{j}$ are given in Table II of Ref. \cite{Bischer:2019ttk}.

An alternative basis, frequently used in the literature, is also discussed in Ref. \cite{Bischer:2019ttk}. The effective Lagrangian in this parametrisation is
\begin{align}
\mathcal{L}^{\bar \nu \nu\bar f f}_\mathrm{eff}=-\frac{G_F}{\sqrt{2}}\sum_{a=S,P,V,A,T}\left(\overline{\nu}_{\alpha}\,\Gamma^a\nu_{\beta}\right)
\left(\overline{f_{\gamma}^i}\Gamma^a(C_{\alpha\beta\gamma\delta}^a+\overline{D}_{\alpha\beta\gamma\delta}^a i\gamma^5)f_{\delta}^j\right),
\end{align}
where the five possible independent combinations of Dirac matrices are defined as $\Gamma^a\in \left\{\mathbb{I},i\gamma^5,\gamma^\mu,\gamma^\mu\gamma^5,\sigma^{\mu\nu}\right\}$ for $a = S, P, V, A, T$ and the associated coefficients are denoted by $C^a$ and
\begin{align}
D^{a} \equiv\left\{\begin{array}{ll}
{\bar{D}^{a}} & {(a=S, P, T)} \\
{i \bar{D}^{a}} & {(a=V, A)}\,.
\end{array}\right.
\end{align}
The coefficients $c^{XY}$, $g^{XY}$ and $h^{XX}$ ($X,Y = L,R$) used in this work are simply linear combinations of the $\ptwiddle{\epsilon}\hspace{-0.6em}\phantom{a}^{j}$, $C^{a}$ and $D^{a}$ coefficients,  
\begin{align}
\begin{aligned}
c^{LL} &=\epsilon^{L} = \frac14\left(C^V-D^V+C^A-D^A\right),\\
c^{RL} &= \epsilon^R = \frac14\left(C^V+D^V-C^A-D^A\right),\\
c^{LR} &= \tilde{\epsilon}^{L}= \frac14\left(C^V-D^V-C^A+D^A\right),\\
c^{RR} &= \tilde{\epsilon}^{R}= \frac14\left(C^V+D^V+C^A+D^A\right),\\
g^{LL} &= \epsilon^S+\epsilon^P= \frac12\left(C^S-iD^S-C^P+iD^P\right),\\
g^{RL} &= \epsilon^S-\epsilon^P=\frac12\left(C^S+iD^S+C^P+iD^P\right),\\
g^{LR} &= \tilde{\epsilon}^S+\tilde{\epsilon}^P=\frac12\left(C^S-iD^S+C^P-iD^P\right),\\
g^{RR} &= \tilde{\epsilon}^S-\tilde{\epsilon}^P=\frac12\left(C^S+iD^S-C^P-iD^P\right),\\
h^{LL}&= \epsilon^T=\frac14\left(C^T-iD^T\right),\\
h^{RR}&= \tilde{\epsilon}^{T}=\frac14\left(C^T+iD^T\right),
\end{aligned}
\end{align}
where flavour indices have been suppressed. We note that the flavour indices for fermions and neutrinos are swapped in our convention, i.e. $\epsilon_{\alpha\beta\gamma\delta}^{L}=c_{\gamma\delta;\alpha\beta}^{LL}$.

\section{Spinor identities and non-relativistic limit}
\label{sec:nonrel}

A crucial step to take in deriving the spectral functions or absorptive parts of the invariant scattering amplitudes $\mathcal{M}^{(\alpha,\beta)}$ is taking the non-relativistic limit of the external interacting fermion bilinears. This can be done by expanding the bilinears to first order in both the 3-momentum transfer $\mathbf{q}=\mathbf{p}_{\alpha}-\mathbf{p}'_{\alpha}=\mathbf{p}'_{\beta}-\mathbf{p}_{\beta}$ and the sum of 3-momenta $\mathbf{P}=\mathbf{p}_{\alpha}+\mathbf{p}_{\beta}=\mathbf{p}'_{\alpha}+\mathbf{p}'_{\beta}$,
\begin{align}
[\bar{u}_{s_{\alpha}^{\prime}}\left(\mathbf{p}_{\alpha}^{\prime}\right) \Gamma^a u_{s_{\alpha}}(\mathbf{p}_{\alpha})]\equiv [\Gamma^{a}]_{\alpha} &\approx \xi_{s^{\prime}_{\alpha}}^{\dagger}\left(2m_{f_{\alpha}}\Gamma^{a}-\frac{P_{j}}{2} \left\{\Gamma^{a}, \gamma_{j}\right\}-\frac{q_{j}}{2} \left[\Gamma^{a}, \gamma_{j}\right]\right) \xi_{s_{\alpha}}\,,
\end{align}
where $\Gamma^{a}=\{\mathbb{I},\gamma_{5},\gamma_{\mu},\gamma_{\mu}\gamma_{5},\sigma_{\mu\nu}\}$ is one of the 16 irreducible products of $\gamma$ matrices and $u_{s_{\alpha}}(\mathbf{p}_{\alpha})$ and $\xi_{s_{\alpha}}$ are respectively the 4-component Dirac spinor and 2-component Weyl spinor for a fermion $f_{\alpha}$ with mass $m_{f_{\alpha}}$, 3-momentum $\mathbf{p}_{\alpha}$ and spin $s_{\alpha}$.

This expansion must be made for the external fermion bilinear at each of the interaction vertices. Hence the bilinears only appear as the products $[\Gamma^{a}]_{\alpha}[\Gamma^{b}]_{\beta}$. We retain the higher orders terms in $\mathbf{P}$ and $\mathbf{q}$ arising from this product for comparison with the basis of 16 operators in Ref. \cite{Dobrescu:2006au}, a complete set of scalar operators constructed from two spins and two momenta.

The products of scalar-like fermion bilinears are
\begin{align}
\begin{aligned}
\label{eq:scalarscalar}
[\mathbb{I}]_{\alpha}[\mathbb{I}]_{\beta}&\approx 4m_{\alpha}m_{\beta}\,,\\
[\mathbb{I}]_{\alpha}[\gamma_5]_{\beta}&\approx -2m_{\alpha}(\boldsymbol{\sigma}_{\beta}\cdot\mathbf{q})\,,\\
[\gamma_5]_{\alpha}[\gamma_5]_{\beta}&\approx (\boldsymbol{\sigma}_{\alpha}\cdot\mathbf{q})(\boldsymbol{\sigma}_{\beta}\cdot\mathbf{q})\,,
\end{aligned}
\end{align}
which are proportional to the $\mathcal{O}_1$, $\mathcal{O}_3$ and $\mathcal{O}_9 \pm \mathcal{O}_{10}$ operators in Ref. \cite{Dobrescu:2006au} respectively. Throughout this work however we consider a SM weak vector interaction at one vertex and an arbitrary scalar, vector or tensor-like interaction at the other. These fermion bilinears are therefore not used in this work, but are relevant for axion-mediated long-range potentials \cite{Stadnik:2017hpa, Fadeev:2018rfl, Dzuba:2018anu}.

The products of vector-like fermion bilinears are
\begin{align}
\begin{aligned}
\label{eq:vectorvector}
[\gamma_{\mu}]_{\alpha}[\gamma^{\mu}]_{\beta}&\approx (4m_{\alpha}m_{\beta}-\mathbf{P}^2)+(\boldsymbol{\sigma}_{\alpha}\cdot\boldsymbol{\sigma}_{\beta})\,\mathbf{q}^2-(\boldsymbol{\sigma}_{\alpha}\cdot\mathbf{q})(\boldsymbol{\sigma}_{\beta}\cdot\mathbf{q})\,,\\
&~~~~-i(\boldsymbol{\sigma}_{\alpha}+\boldsymbol{\sigma}_{\beta})\cdot(\mathbf{P}\times\mathbf{q})\,,\\
[\gamma_{\mu}]_{\alpha}[\gamma^{\mu}\gamma_5]_{\beta}&\approx 2im_{\beta}(\boldsymbol{\sigma}_{\alpha}\times\boldsymbol{\sigma}_{\beta})\cdot\mathbf{q}- 2(m_{\alpha}-m_{\beta})(\boldsymbol{\sigma}_{\beta}\cdot\mathbf{P})\,,\\
[\gamma_{\mu}\gamma_5]_{\alpha}[\gamma^{\mu}\gamma_5]_{\beta}&\approx -(4m_{\alpha}m_{\beta}-\mathbf{P}^2) (\boldsymbol{\sigma}_{\alpha}\cdot\boldsymbol{\sigma}_{\beta})\,,\\
[\slashed{q}\gamma_5]_{\alpha}[\slashed{q}\gamma_5]_{\alpha}&\approx 4m_{\alpha}m_{\beta} (\boldsymbol{\sigma}_{\alpha}\cdot\mathbf{q})(\boldsymbol{\sigma}_{\beta}\cdot\mathbf{q})\,.
\end{aligned}
\end{align}
The first of these products contains terms proportional to $\mathcal{O}_{1}$, $\mathcal{O}_{2}$, $\mathcal{O}_{3}$ and $\mathcal{O}_{4}$, the second to $\mathcal{O}_{11}$ and $\mathcal{O}_{12}\pm\mathcal{O}_{13}$, the third to $\mathcal{O}_{2}$ and finally the fourth to $\mathcal{O}_{3}$. These products are relevant in the case of a vector-like current at both interaction vertices. We have not included products containing the bilinear $[\slashed q]_{\alpha}$ which vanishes according to the equations of motion.

The relevant products of scalar-like and vector-like (which must be contracted with the momentum exchange $q^{\mu}$) fermion bilinears are
\begin{align}
\begin{aligned}
\label{eq:vectorscalar}
[\mathbb{I}]_{\alpha}[\slashed{q}\gamma_5]_{\beta}&\approx 4m_{\alpha}m_{\beta}(\boldsymbol{\sigma}_{\beta}\cdot\mathbf{q})\,,\\
[\gamma_5]_{\alpha}[\slashed{q}\gamma_5]_{\beta}&\approx -2m_{\beta}(\boldsymbol{\sigma}_{\alpha}\cdot\mathbf{q})(\boldsymbol{\sigma}_{\beta}\cdot\mathbf{q})\,,
\end{aligned}
\end{align}
proportional to $\mathcal{O}_{2}$ and $\mathcal{O}_{3}$ respectively. These are used for the case of a scalar interaction at one vertex and a CC or NC interaction at the other.

Finally, we list the relevant products of vector-like and tensor-like (where again the free Lorentz index must be contracted with the momentum exchange $q^{\mu}$) fermion bilinears,
\begin{align}
\begin{aligned}
\label{eq:vectortensor}
[\gamma_{\mu}]_{\alpha}[\sigma^{\mu\nu}q_{\nu}]_{\beta}&\approx 2im_{\alpha}\mathbf{q}^2-2im_{\beta}\big[(\boldsymbol{\sigma}_{\alpha}\cdot\boldsymbol{\sigma}_{\beta})\,\mathbf{q}^2-(\boldsymbol{\sigma}_{\alpha}\cdot\mathbf{q})(\boldsymbol{\sigma}_{\beta}\cdot\mathbf{q})\big]\\
& ~~~~+2(m_{\alpha}-m_{\beta})[\boldsymbol{\sigma}_{\beta}\cdot(\mathbf{P}\times\mathbf{q})]\,,\\
[\gamma_{\mu}]_{\alpha}[\sigma^{\mu\nu}q_{\nu}\gamma_5]_{\beta}&\approx i(4m_{\alpha}m_{\beta}-\mathbf{P}^2)(\boldsymbol{\sigma}_{\beta}\cdot\mathbf{q})+[\boldsymbol{\sigma}_{\alpha}\cdot(\mathbf{P}\times\mathbf{q})](\boldsymbol{\sigma}_{\beta}\cdot\mathbf{q})\,,\\
[\gamma_{\mu}\gamma_5]_{\alpha}[\sigma^{\mu\nu}q_{\nu}]_{\beta}&\approx -4m_{\alpha}m_{\beta}(\boldsymbol{\sigma}_{\alpha}\times\boldsymbol{\sigma}_{\beta})\cdot\mathbf{q}-(\boldsymbol{\sigma}_{\alpha}\cdot\mathbf{P})\big[i \,\mathbf{q}^2+\boldsymbol{\sigma}_{\beta}\cdot(\mathbf{P}\times\mathbf{q})\big]\,,\\
[\gamma_{\mu}\gamma_5]_{\alpha}[\sigma^{\mu\nu}q_{\nu}\gamma_5]_{\beta}&\approx2i(m_{\alpha}-m_{\beta})(\boldsymbol{\sigma}_{\alpha}\cdot\mathbf{P})(\boldsymbol{\sigma}_{\beta}\cdot\mathbf{q})-2im_{\alpha}(\boldsymbol{\sigma}_{\alpha}\cdot\boldsymbol{\sigma}_{\beta})(\mathbf{P}\cdot\mathbf{q})\,.
\end{aligned}
\end{align}
The first product contains terms proportional to $\mathcal{O}_{1}$, $\mathcal{O}_{2}$, $\mathcal{O}_{3}$ and  $\mathcal{O}_{4}\pm \mathcal{O}_{5}$, the second to $\mathcal{O}_{9}\pm \mathcal{O}_{10}$ and $\mathcal{O}_{15}$, the third to $\mathcal{O}_{11}$, $\mathcal{O}_{12}\pm \mathcal{O}_{13}$ and $\mathcal{O}_{16}$ and finally the fourth to $\mathcal{O}_{6}\pm \mathcal{O}_{7}$ and $\mathcal{O}_{2}$. These are needed when evaluating the potential for a tensor interaction at one vertex and a CC or NC interaction at the other.

\section{Integrals}
\label{sec:integrals}

The generic form for the dimensionless integrals appearing frequently in this work -- functions of the distance $r$ between the interacting fermions and labelled by the superscript $X$ -- is
\begin{align}
I^{X}_{ij}(r) &=\int\limits^{\infty}_{(m_{i}+m_{j})r} \hspace{-1em}dy~ \Lambda^{1/2}(y^2,m_{i}^2r^2,m_{j}^2r^2) ~G_{ij}^{X}(y,r)~e^{-y}
\,,
\end{align}
where the dimensionless variable $y=r\sqrt{t}$, the indices ($i,j$) run over either the $N$ massive Majorana states or 3 massive Dirac states, and $\Lambda(x,y,z)$ is the K\"all\'{e}n function. The functions $G_{ij}^{X}(y,r)$ are given by
\begin{align}
\begin{aligned}
G^{LR}_{ij}(y,r) &= \frac{1}{y}\,,\\
G^{D}_{ij}(y,r) &= \frac{y}{6}\left\{1-\frac{\overline{m_{ij}^2}r^2}{y^2}
- {(\Delta m_{ij}^2)^2r^4 \over 2y^4}\right\}\,,\\
G^{M}_{ij}(y,r) &= \frac{y}{6}\left\{1-\frac{(\overline{m_{ij}^2}+3m_{i}m_{j})r^2}{y^2} 
- {(\Delta m_{ij}^2)^2r^4 \over 2y^4}\right\}\,,\\
G^{V}_{ij}(y,r) &= \frac{1}{y}\left\{1+ \frac{2\overline{m_{ij}^2}r^2}{y^2}
- {2(\Delta m_{ij}^2)^2r^4 \over y^4}\right\}\,,\\
G^{\Delta}_{ij}(y,r) &= \frac{1}{y}\left\{1- \frac{\Delta m_{ij}^2\, r^2}{y^2}\right\}\,,\\
G^{S}_{ij}(y,r) &=  \frac{1}{y}\left\{1- \frac{(m_{i}-m_{j})^2\, r^2}{y^2}\right\}\,,\\
G^{SD}_{ij}(y,r) &=  \frac{y}{6}\left\{1- \frac{2\overline{m_{ij}^2}\, r^2}{y^2}\right\}\,,\\
G^{SM}_{ij}(y,r) &= \frac{y}{6}\left\{1- \frac{(m_{i}-m_{j})^2\, r^2}{y^2}\right\}\,,\\
G^{T}_{ij}(y,r) &= \frac{1}{y}\left\{1- \frac{(m_{i}+m_{j})^2\, r^2}{y^2}\right\}\,,\\
G^{M\gamma}_{ij}(y,r) &= \frac{1}{y}\left\{1- \frac{(m_{i}-m_{j})^2\, r^2}{y^2}\right\}\left\{1+ \frac{2(m_{i}+m_{j})^2\, r^2}{y^2}\right\}\,,\\
G^{E\gamma}_{ij}(y,r) &= \frac{1}{y}\left\{1- \frac{(m_{i}+m_{j})^2\, r^2}{y^2}\right\}\left\{1+ \frac{2(m_{i}-m_{j})^2\, r^2}{y^2}\right\}\,,
\end{aligned}
\end{align}
where $\overline{m_{ij}^2}=(m_{i}^2+m_{j}^2)/2$ and $\Delta m_{ij}^2=m_{i}^2-m_{j}^2$.

\begin{table}[]
	\setlength{\tabcolsep}{10pt}
	\begin{tabular}{ccccccc}
		\hline\hline$X$	& $D$ & $M$  & $V$ & $\Delta$ & $S$ & $T$ \\ \hline
		$J^X_{ij}(r)$  & $\frac{3}{2}$ & $\frac{3}{2}$ &  $\frac{5}{2}$& $3$  & $3$ & $3$\\\hline\hline
	\end{tabular}
	\caption{Exact values of the dimensionless integrals $J^{X}_{ij}(r)$ for $N =\{D$, $M$, $V$, $\Delta$, $S$, $T\}$ derived from the potentials $I^{X}_{ij}(r)\approx 1$ in the limit of vanishing neutrino masses, $m_{i}\approx 0$. These appear in the potentials $V_{\alpha\beta}^{LL}$, $V_{\alpha\beta}^{LR}$, $V_{\alpha\beta}^{RR}$, $V_{\alpha\beta}^{VS}$ and $V_{\alpha\beta}^{VT}$ in this work.}
	\label{tab:Jlimit}
\end{table}

A second set of dimensionless integrals $J^{N}_{ij}(r)$ is derived from the above by performing the derivative operations
\begin{align}
\begin{aligned}
J_{ij}^{D,M}(r)&=I_{ij}^{D,M}(r)+\left(\frac{1}{2}-\frac{r}{6}\frac{d}{dr}\right)I_{ij}^{V}(r)\,,\\
J_{ij}^{V}(r)&=\left(\frac{5}{2}-\frac{7r}{6}\frac{d}{dr}+\frac{r^2}{6}\frac{d^2}{dr^2}\right)I_{ij}^{V}(r)\,,\\
J_{ij}^{\Delta,S,T}(r)&=\left(3-r\frac{d}{dr}\right)I^{\Delta,S,T}_{ij}(r)\,.
\end{aligned}
\end{align}

The first set of integrals are normalised such that $I^{X}_{ij}(r)\approx 1$ for vanishing neutrino masses $m_{i}\approx 0$ -- applicable in the short-range limit of the potentials in which they appear. The values of the second set $J^{X}_{ij}(r)$ in this limit are given in Table. \ref{tab:Jlimit}.

\end{appendix}

\bibliography{references}

\end{document}